\documentclass{birkjour}

\usepackage{amssymb}
\usepackage{amsfonts}
\usepackage{amsmath}
\usepackage{txfonts}
\usepackage{enumerate}
\usepackage{bm}
\usepackage{mathrsfs}
\usepackage{color}

\usepackage{epstopdf}

 \theoremstyle{definition}
 
 \theoremstyle{remark}

 \numberwithin{equation}{section}

\begin{document}

%
%
%
%
%
%
%
%
%

\title[The GFONLS equation with nonzero boundary conditions]
 {The general fifth-order nonlinear Schr\"{o}dinger equation with nonzero boundary conditions:
Inverse scattering transform and multisoliton solutions}

\author[Xiu-Bin Wang]{Xiu-Bin Wang}

\address{
Department of Mathematics, Harbin Institute of Technology, Harbin 150001, People's Republic of China}
\email{xiubinwang@163.com}

\author{Bo Han}
\address{
Department of Mathematics, Harbin Institute of Technology, Harbin 150001, People's Republic of China }
\email{bohan@hit.edu.cn}

\thanks{This work is supported by the National Natural Science Foundation of China under Grant No.11871180.}


\subjclass{35Q51, 35Q53, 35C99, 68W30, 74J35.}

\keywords{The general fifth-order nonlinear Schr\"{o}dinger equation;
Inverse scattering transform (IST); Soliton solutions; Riemann-Hilbert Problem (RHP).}

\date{January 1, 2004}

\begin{abstract}
Under investigation in this work is the inverse scattering transform of the general fifth-order nonlinear Schr\"{o}dinger equation with nonzero boundary conditions (NZBCs), which can be reduced to several integrable equations.
Firstly, a matrix Riemann-Hilbert problem for the equation with NZBCs at infinity is systematically investigated.
Then the inverse problems are solved through the investigation of the matrix Riemann-Hilbert problem.
Thus, the general solutions for the potentials, and explicit expressions
for the reflection-less potentials are well constructed.
Furthermore, the trace formulae and theta conditions are also presented.
In particular, we analyze the simple-pole and double-pole solutions for the equation with NZBCs.
Finally, the dynamics of the obtained solutions are graphically discussed.
These results provided in this work can be useful to enrich and explain the related nonlinear wave phenomena in nonlinear fields.
\end{abstract}

\maketitle

\section{Introduction}

The fundamental nonlinear Schr\"{o}dinger (NLS) equation
\begin{equation}\label{NLS11}
iq_{t}+q_{xx}+2|q|^2q=0
\end{equation}
is famous to be a key integrable model in the field of mathematical physics.
There are many physical contexts where the NLS equation arises.
For instance, the NLS equation describes the weakly nonlinear surface wave in deep water.
More importantly, the NLS equation models the soliton propagation in optical fibers
where only the group velocity dispersion and the self-phase modulation effects
are taken into account. However, for ultrashort pulse in optical fibers,
the effects of the higher-order dispersion, the self-steepening, and the stimulated Raman scattering should be considered.
Besides, the higher-order dispersion terms and non-Kerr nonlinearity effects have been found interesting applications in
optics \cite{GYYR-1984}-\cite{NS-1991}. Thus, more researches about higher-order NLS equations are inevitable and worthwhile.
Due to these effects, the propagation of subpicosecond and femtosecond pulses can be described by
the general integrable four-parameter $(\alpha_{2},\alpha_{3},\alpha_{4},\alpha_{5})$ fifth-order NLS (GFONLS in brief) equation \cite{TKS-1989,acdj-2014,yzy-2018}
\begin{equation}\label{gtc-NLS}
i\psi_{t}(x,t)+\alpha_{2}K_{2}\left[\psi(x,t)\right]-i\alpha_{3}K_{3}\left[\psi(x,t)\right]+\alpha_{4}K_{4}\left[\psi(x,t)\right]
-i\alpha_{5}K_{5}\left[\psi(x,t)\right]=0,
\end{equation}
where
\begin{equation}\label{gtc-1}
\left\{ \begin{aligned}
&K_{2}=\psi_{xx}+2\left(|\psi|^2-\psi_{0}^2\right)\psi,\\
&K_{3}=\psi_{xxx}+6|\psi|^2\psi_{x},\\
&K_{4}=\psi_{xxxx}+8|\psi|^2\psi_{xx}+6|\psi|^4\psi+6\psi^{*}\psi_{x}^2+2\psi^2\psi^{*}_{xx},\\
&K_{5}=\psi_{xxxxx}+10|\psi|^2\psi_{xxx}+10\left(\psi|\psi_{x}|^2\right)_{x}+20\psi^{*}\psi_{x}\psi_{xx}+30|\psi|^4\psi_{x}.
             \end{aligned} \right.
\end{equation}

Recently, the ISTs of integrable nonlinear wave equations with NZBCs have been paid much attention to
based on the solutions of the related RHP.
The approach has been extended to the focusing and defocusing NLS equation,
the focusing and defocusing Hirota equations, the nonlocal modified KdV equation and the derivative NLS equation etc \cite{BP-2018}-\cite{PB-SAM} such that
many types of nonlinear waves were discussed.
Motivated by the above works of Ablowitz \cite{MJ-IP}, Biondini \cite{BP-2014,MP-2014}, Feng \cite{fbf-2020} and Yan \cite{GQ-2020} etc,
in this work, we would like to extend the IST to study the GFONLS equation \eqref{gtc-NLS} with the following NZBCs as infinity
\begin{equation}\label{NZBC1}
\lim_{x\rightarrow\infty}\psi(x,t)=\psi_{\pm},
\end{equation}
with $|\psi_{\pm}|=\psi_{0}\neq 0$.
Eq.\eqref{gtc-NLS} includes plenty of important nonlinear wave equations as its special cases \cite{RH-1973}-\cite{YY-2015}.
Here we give some crucial cases.

\noindent
\textbf{Case (1):} If $\alpha_{3}=\alpha_{4}=\alpha_{5}=0, \alpha_{2}=1$, Eq.\eqref{gtc-NLS}
can be reduced to the fundamental NLS equation \eqref{NLS11} with NZBCs
\begin{equation}\label{Case1}
i\psi_{t}+ \psi_{xx}+2\left(|\psi|^2-\psi_{0}^2\right)\psi=0.
\end{equation}

\noindent
\textbf{Case (2):} If $\alpha_{4}=\alpha_{5}=0$, Eq.\eqref{gtc-NLS} can be reduced to the following Hirota equation with NZBCs \cite{RH-1973}
\begin{equation}\label{Hirota}
i\psi_{t}+\alpha_{2}\left(\psi_{xx}+2\left(|\psi|^2-\psi_{0}^2\right)\right)\psi+i\alpha_{3}\left(\psi_{xxx}+6|\psi|^2\psi_{x}\right)=0,
\end{equation}
\textbf{Case (3):} If $\alpha_{3}=1, \alpha_{2}=\alpha_{4}=\alpha_{5}=0$, Eq.\eqref{gtc-NLS} can be reduced to the complex modified Korteweg-de Vries (mKdV) equation \cite{mkdx-1973,pwq-1973}.
\begin{equation}\label{mKdV}
\psi_{t}+\psi_{xxx}+6|\psi|^2\psi_{x}=0.
\end{equation}
To the best of the authors' knowledge,
although many mathematical physicist have studied the particular cases of equation \eqref{gtc-NLS},
the IST of equation \eqref{gtc-NLS} with NZBCs has not been reported so far.
The GFONLS equation \eqref{gtc-NLS} is completely integrable, its Lax pair reads \cite{yzy-2018}
\begin{equation}\label{RHP-1}
\left\{ \begin{aligned}
&\phi_{x}=U\phi,~~U=ik\sigma_{3}+Q, \\
&\phi_{t}=V\phi,~~V=\alpha_{2}\Delta_{NLS}+\alpha_{3}\Delta_{MKdV}+\alpha_{4}\Delta_{LPD}+\alpha_{5}\Delta_{FOQ}+3i\alpha_{4}\sigma_{3},
             \end{aligned} \right.
\end{equation}
where $\lambda$ is a complex iso-spectral parameter, the eigenfunction $\phi=\phi(x,t,\lambda)$ is a $2\times2$ matrix function, $\sigma_{3}=\mbox{diag}\{1,-1\}$, and matrices $Q$, $V_{0}$, $L$, $M$, $N$ is expressed by
\begin{align}\label{RHP-2}
&\Delta_{NLS}=-2kU+i\sigma_{3}\left(Q_{x}-Q^2-\psi_{0}^2\right),\notag\\
&\Delta_{MKdV}=-2k\left(\Delta_{NLS}+i\psi_{0}^2\sigma_{3}\right)-\left[Q,Q_{x}\right]-Q_{xx}+2Q^3,\notag\\
&\Delta_{LPD}=2k\left[-\left(4ik^3+k^2Q+kV_{0}\right)+L_{0}\right]+M_{0},~~\Delta_{FOQ}=-2k\Delta_{IPD}+N_{0},
\end{align}
with
\begin{align}\label{RHP-3}
&Q=\left(
    \begin{array}{cc}
      0 &  \psi\\
      -\psi^{*}  & 0 \\
    \end{array}
  \right),~~V_{0}=\frac{1}{2}\left(
                               \begin{array}{cc}
                                 -i|\psi|^2 & -\psi_{x}\\
                                 \psi_{x}^{*} & i|\psi|^2 \\
                               \end{array}
                             \right),~~N_{0}=\left(
                                        \begin{array}{cc}
                                          n_{1} & -n_{2}^{*} \\
                                          n_{2} & -n_{1} \\
                                        \end{array}
                                      \right),\notag\\
&L_{0}=\left(
        \begin{array}{cc}
          \psi\psi_{x}^{*}-\psi^{*}\psi_{x} & i\left(\psi_{xx}+2|\psi|^2\psi\right) \\
          i\left(\psi_{xx}^{*}+2|\psi|^2\psi^{*}\right) & \psi^{*}\psi_{x}-\psi\psi_{x}^{*} \\
        \end{array}
      \right),~~M_{0}=\left(
                        \begin{array}{cc}
                          m_{1} & -m_{2}^{*} \\
                          m_{2} & -m_{1} \\
                        \end{array}
                      \right),\notag\\
&m_{1}=-i\left[\left(\psi\psi_{xx}\right)^{*}-|\psi_{x}|^2+3|\psi|^4\right],~~
m_{2}=\psi_{xxx}^{*}+6|\psi|^2\psi_{x}^{*},\notag\\
&n_{1}=\psi_{xxx}\left(\psi-\psi^{*}\right)-\psi_{x}\psi_{xx}^{*}\psi_{x}^{*}\psi_{xx}+6|\psi|^2\left(\psi\psi_{x}^{*}-\psi^{*}\psi_{x}\right),\notag\\
&n_{2}=i\left(\psi_{xxxx}^{*}+2\psi^{*2}\psi_{xx}+4\left|\psi_{x}^2\psi^{*}\right|+6\psi\psi_{x}^{*2}+8|\psi|^2\psi_{xx}^{*}+6|\psi|^4\psi^{*}\right),
\end{align}
where $\psi^{*}(x,t)$ is the complex conjugate of $\psi(x,t)$, and $k$ is a constant spectral parameter.

It is well-known that the IST is a powerful approach to
derive soliton solutions \cite{MJ-1981}-\cite{wds1}.
However, the research in this work, within our knowledge, has not been conducted before.
The chief idea of the present article is to employ the IST to derive the multisoliton solutions of the GFONLS equation \eqref{gtc-NLS}
with NZBCs \eqref{NZBC1}.
Besides, some graphic analysis
are presented to help readers to understand the propagation phenomena of these wave solutions.

The main results of the present paper are the following theorems.

\noindent
\textbf{Theorem 1.1.} The reflectionless potential with simple poles in the GFONLS equation \eqref{gtc-NLS} with NZBCs \eqref{NZBC1} can be found as
\begin{equation}\label{JSP-35}
\psi(x,t)=\psi_{-}+i\frac{\det\left(
                          \begin{array}{cc}
                            G & v \\
                            w^{T} & 0 \\
                          \end{array}
                        \right)
}{\det G},
\end{equation}
where $w=(w_{j})_{2N\times1}$, $v=(v_{j})_{2N\times1}$, $G=(g_{sj})_{2N\times 2N}$, and $y=(y_{n})_{2N\times1}=G^{-1}v$
with $w_{j}=A_{-}\left[\widehat{\xi}_{j}\right]e^{2i\theta(x,t,\widehat{\xi}_{j})}$, $v_{j}=-iq_{-}/\xi_{j}$,
$g_{sj}=\frac{w_{j}}{\xi_{s}-\widehat{\xi}_{j}}+v_{s}\delta_{sj}$, and $y_{n}=\mu_{-11}\left(x,t,\widehat{\xi}_{n}\right)$.\\

\noindent
\textbf{Theorem 1.2.} The reflectionless potential with double poles of the GFONLS equation \eqref{gtc-NLS} with NZBCs \eqref{NZBC1} can be written as
\begin{equation}\label{DP-20}
\psi(x,t)=\psi_{-}+i\frac{\det\left(
                          \begin{array}{cc}
                            H & v \\
                            w^{T} & 0 \\
                          \end{array}
                        \right)
}{\det H},
\end{equation}
where
$H=\left(H^{(sj)}\right)_{2\times2},~~H^{(sj)}=\left(h_{kn}^{(sj)}\right)_{2N\times2N}$,
$h_{kn}^{(11)}=\widehat{C}_{n}(\xi_{k})\left(D_{n}+\frac{1}{\xi_{k}-\widehat{\xi}_{n}}\right)-\frac{i\psi_{-}}{\xi_{k}}\delta_{kn}$,
$h_{kn}^{(12)}=\widehat{C}_{n}(\xi_{k})$,
$h_{kn}^{(21)}=\frac{\widehat{C}_{n}}{\xi_{k}-\widehat{\xi}_{n}}(\xi_{k})\left(D_{n}+\frac{2}{\xi_{k}-\widehat{\xi}_{n}}\right)
-\frac{i\psi_{-}}{\xi_{k}^2}\delta_{kn}$,
$h_{kn}^{(22)}=\frac{\widehat{C}_{n}(\xi_{k})}{\xi_{k}-\xi_{n}}+\frac{i\psi_{-}\psi_{0}^2}{\xi_{k}^3}\delta_{kn}$,
$w_{n}^{(1)}=A_{-}\left[\widehat{\xi}_{n}\right]e^{2i\theta(\xi_{n})}\widehat{D}_{n}$,
$w_{n}^{(2)}=A_{-}\left[\widehat{\xi}_{n}\right]e^{2i\theta(\xi_{n})}$, $v_{n}^{(1)}=-\frac{\psi_{-}}{\xi_{n}}$,
$v_{n}^{(2)}=-\frac{\psi_{-}}{\xi^2_{n}}$.

The outline of this paper is organized as follows:
In section 2, we consider the direct scattering problem for the GFONLS equation \eqref{gtc-NLS} with NZBCs \eqref{NZBC1}
starting from its spectral problem.
In Section 3, we discuss
the GFONLS equation \eqref{gtc-NLS} with NZBCs \eqref{NZBC1} such that its simple-pole solution are found via solving the
matrix RHP with the reflectionless potentials, and their trace formulae and theta condition are also presented.
Following a similar way, in section 4 we investigate
the GFONLS equation \eqref{gtc-NLS} with NZBCs \eqref{NZBC1} such that its double-pole solutions are found via solving the
matrix RHP. Finally, the conclusion and discussion are provided in the last section.


\section{Direct scattering problem with NZBCs}
%
%
%

$~~~~~~$
{\rotatebox{0}{\includegraphics[width=5.8cm,height=5.0cm,angle=0]{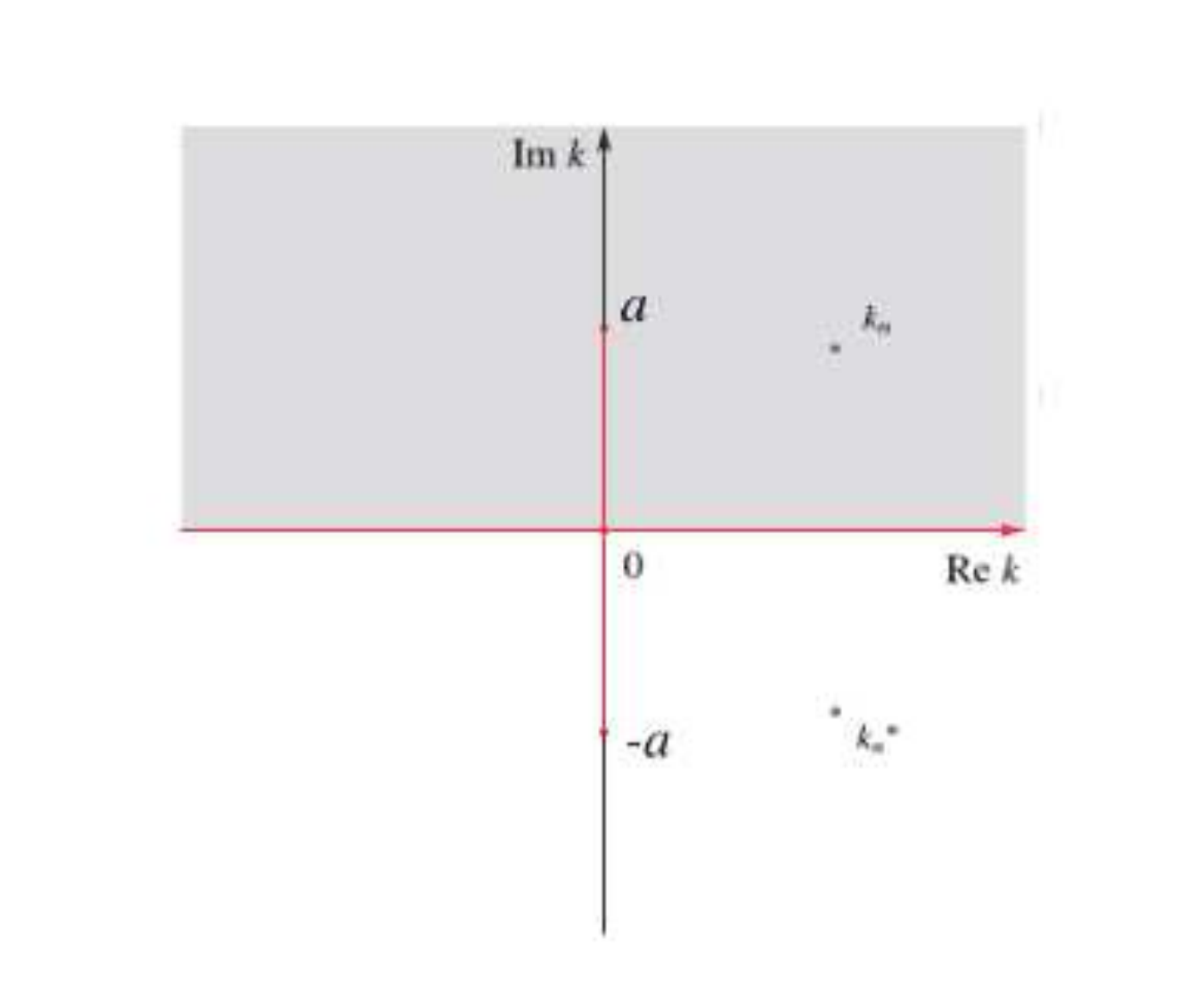}}}
~~~~~~~~~~~~~~~
{\rotatebox{0}{\includegraphics[width=5.8cm,height=5.0cm,angle=0]{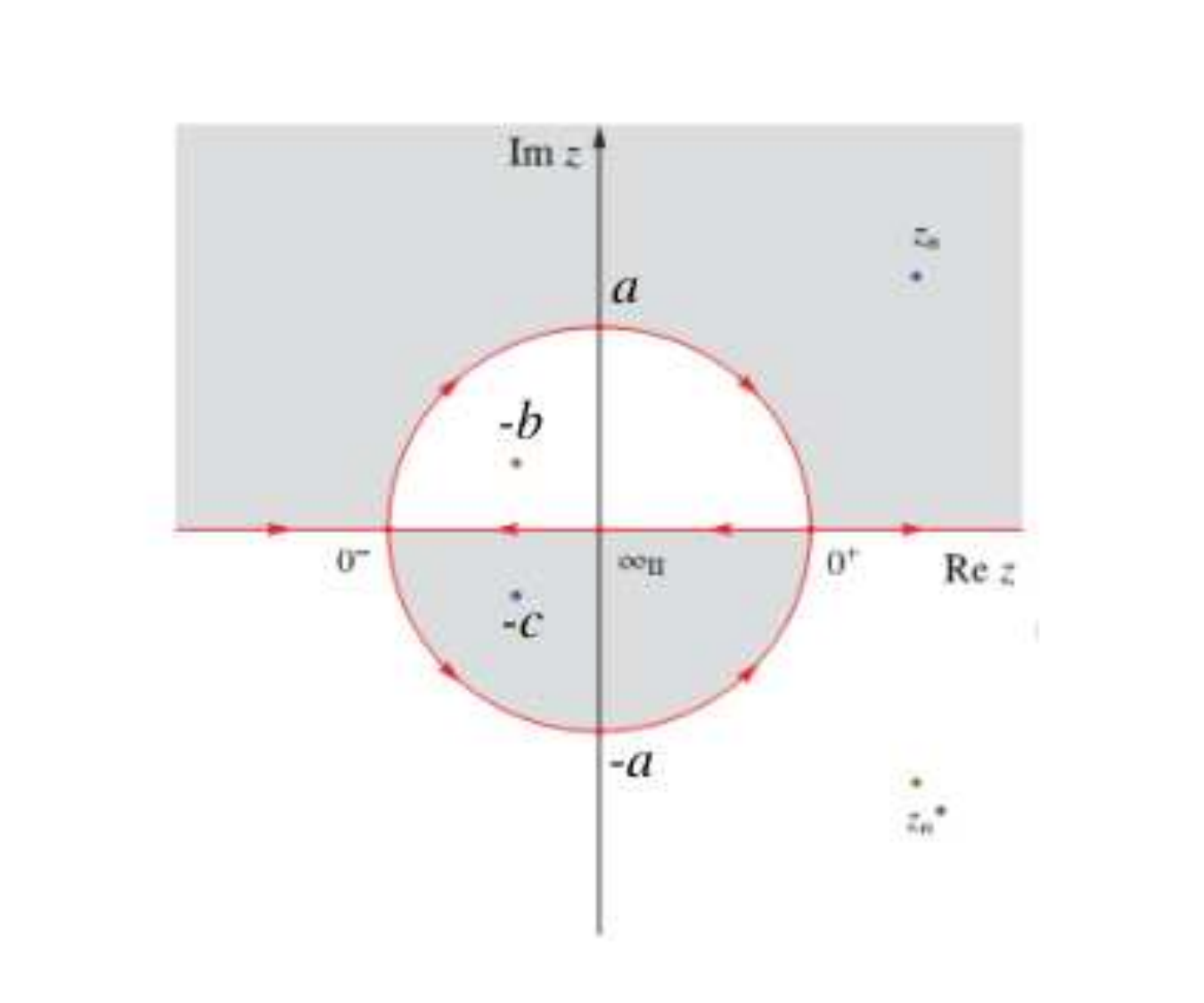}}}

$\qquad\qquad\qquad\textbf{(a)}
\qquad\qquad\qquad\qquad\qquad\qquad\qquad\qquad\textbf{(b)}
$\\
\noindent {\small \textbf{Figure 1.} The grey (white) region for $\mbox{Im}\lambda>0 (\mbox{Im}\lambda<0)$ in distinct spectral planes of the Lax pair with NZBCs. $\textbf{(a)}$: the first sheet of the Riemann surface, showing the discrete spectrum; $\textbf{(b)}$: the complex z-plane,
showing the discrete spectrum [zeros of $s_{11}(z)$ (blue) in the grey region and those of
$s_{22}(z)$ (red) in the white region], and the orientation of the jump contours for the related RHP.
Besides, $a=i\psi_{0}, b=\psi_{0}^2/z_{n}, c=-\psi_{0}^2/z_{n}^{*}$. \\}

First of all, we discuss the first expression in \eqref{RHP-1} as the scattering problem of equation \eqref{NZBC1}.
As $x\rightarrow\pm\infty$, the scattering problem yields
\begin{equation}\label{Lax-4}
\phi_{x}=U_{\pm}\phi,~~U_{\pm}=\lim_{x\rightarrow\pm\infty}U=ik\sigma_{3}+Q_{\pm},
\end{equation}
and
\begin{equation}\label{Lax-41}
Q_{\pm}=\lim_{x\rightarrow\pm\infty}Q(x,t)=\left(
                                                                                            \begin{array}{cc}
                                                                                              0 & \psi_{\pm} \\
                                                                                              -\psi_{\pm}^{*} & 0 \\
                                                                                            \end{array}
                                                                                          \right).
\end{equation}
Consequently, the standard matrix solutions of equation \eqref{Lax-4} can be defined by
\begin{align}\label{Lax-5}
&\phi_{bg}^{f}(x,t,k)=\notag\\
&\left\{ \begin{aligned}
&E^{f}_{\pm}(k)e^{i\theta(x,t,k)\sigma_{3}},~~~~k\neq \pm i\psi_{0},\\
&I+\left[x-2\left(\alpha_{2} k+3\alpha_{3}\psi_{0}^2\pm6\alpha_{4}i\psi_{0}^3+\alpha_{5}\left(-8\psi_{0}^4-7\psi_{0}^2\right)\right)t\right]U_{\pm},~~k=\pm i\psi_{0},
             \end{aligned} \right.
\end{align}
where $I$ represents a $2\times2$ unit matrix, and
\begin{align}\label{Lax-6}
&E^{f}_{\pm}(k)=\left(
                 \begin{array}{cc}
                   1 & \frac{i\psi_{\pm}}{k+\lambda} \\
                   \frac{i\psi^{*}_{\pm}}{k+\lambda} & 1 \\
                 \end{array}
               \right),\notag\\
&\theta(x,t,k)=\notag\\
&\lambda(k)\left\{x+\left[\alpha_{5}\left(-16k^4+8k^2-6\psi_{0}^2\right)
+\alpha_{4}\left(8k^3-4k\psi_{0}^2\right)+\alpha_{3}\left(4k^2-2\psi_{0}^2\right)-2\alpha_{2}k\right]t\right\},
\end{align}
with
\begin{equation}\label{Lax-7}
\lambda^2=k^2+\psi_{0}^2.
\end{equation}
To further study the analyticity of the Jost solutions of the spectral problem \eqref{RHP-1},
we have to consider the regions of $\mbox{Im} \lambda(k)>0 (<0)$ in the function $\theta(x,t,k)$ (also see \cite{BP-2014}).
Taking $\psi_{0}\neq0$, i.e., the boundary conditions are NZBCs, $\lambda(k)$ satisfying \eqref{Lax-7}
in the complex plane is a doubly branched function of $k$ with two branch points being $k\neq\pm i\psi_{0}$
and the branch cut being the segment $i\psi_{0}[-1,1]$.
Take $k\pm i\psi_{0}=r_{\pm}e^{i\theta_{\pm}+2im_{\pm}\pi}(r_{\pm}>0,\theta_{\pm}\in[-\pi/2,3\pi/2], m\in\mathbb{Z})$.
Then two single-valued analytical branches of the complex k-plane are expressed by Sheet-I: $\lambda_{I}(k)=\sqrt{r_{+}r_{-}}e^{i(\theta_{+}+\theta_{-})/2}$
and Sheet-II: $\lambda_{II}(k)=-\lambda_{I}(k)$.
To achieve this aim, one can introduce a uniformization variable $z$ given by the conformal mapping: $z=k+\lambda$,
whose inverse mapping is
\begin{equation}\label{Lax-8}
k(z)=\frac{1}{2}\left(z-\frac{\psi_{0}^2}{z}\right),~~\lambda(z)=\frac{1}{2}\left(z+\frac{\psi_{0}^2}{z}\right).
\end{equation}
In particular, if $\psi_{0}=0$, the NZBCs can be reduced to the ZBC. 

Take $\mathbb{A}=i\psi_{0}[-1,1]$ with 
$C_{0}=\left\{z\in\mathbb{C}:|z|=\psi_{0}\right\}$, and (see Fig.1)
\begin{equation}\label{Lax-9}
D_{+}^{f}=\left\{z\in\mathbb{C}:\left(|z|^2-\psi_{0}^2\right)\mbox{Im}z>0\right\},
~~D_{-}^{f}=\left\{z\in\mathbb{C}:\left(|z|^2-\psi_{0}^2\right)\mbox{Im}z<0\right\}.
\end{equation}

The continuous spectrum of $U_{\pm}=\lim\limits_{x\rightarrow\pm\infty}U$ stands for the set of all values of z satisfying
$\lambda(z)\in\mathbb{R}$, i.e., $z\in\Sigma^{f}=\mathbb{R}\bigcup C_{0}$,
which are the jump contours for the related RHP.
Similar to \cite{BP-2014}, it follows from $[U_{\pm},V_{\pm}]=0$ that the Jost
solutions $\phi_{\pm}(x,t,z)$ of both parts of \eqref{RHP-1} satisfying the boundary conditions
\begin{equation}\label{Lax-10}
\phi_{\pm}(x,t,z)=E_{\pm}^{f}(z)e^{i\theta(x,t,z)\sigma_{3}}+O(1),~~\forall z\in\Sigma^{f},~~x\rightarrow\pm\infty.
\end{equation}
In view of $\phi_{x}=U_{\pm}\phi+\Delta Q_{\pm}\phi$, $\Delta Q_{\pm}(x,t)=Q(x,t)-Q_{\pm}$
with $Q_{\pm}=\lim\limits_{x\rightarrow\pm\infty}Q$, the modified Jost solutions
\begin{equation}\label{Lax-11}
\mu_{\pm}(x,t,z)=\phi_{\pm}(x,t,z)e^{-i\theta(x,t,z)\sigma_{3}}\rightarrow E_{\pm}^{f}(z),~~x\rightarrow\pm\infty,
\end{equation}
possess the following results
\begin{align}\label{Lax-12}
\mu_{\pm}=
\left\{ \begin{aligned}
&E_{\pm}^{f}(z)\left\{I+\int_{\pm\infty}^{x}e^{i\lambda(x-y)\widehat{\sigma}_{3}}\left[(E_{\pm}^{f}(z))^{-1}\Delta Q_{\pm}(y,t)\mu_{\pm}(y,t,z)\right]dy\right\},\\
&~~~~~~~~~z\neq\pm i\psi_{0},~~\psi-\psi_{\pm}\in L^{1}\left(\mathbb{R}^{\pm}\right),\\
&E_{\pm}^{f}(z)+\int_{\pm\infty}^{x}\left[I+(x-y)\left(Q_{\pm}\mp \psi_{0}\sigma_{3}\right)\right]\Delta Q_{\pm}(y,t)\mu_{\pm}(y,t,z)dy,\\
&~~~~~~~~~z=\pm i\psi_{0},~~\left(1+|x|\right)\left(\psi-\psi_{\pm}\right)\in L^{1}\left(\mathbb{R}^{\pm}\right),
             \end{aligned} \right.
\end{align}
where $e^{\widehat{\sigma}_{3}}A:=e^{\sigma_{3}}Ae^{-\sigma_{3}}$.

Let $\Sigma_{0}^{f}:=\Sigma^{f}\setminus\{\pm i\psi_{0}\}$, $\mu_{\pm}(x,t,z)=(\mu_{\pm,1},\mu_{\pm,2})$,
and $\phi_{\pm}(x,t,z)=(\phi_{\pm1},\phi_{\pm2})$.
Since the above expression \eqref{Lax-12} contains $e^{\pm i(x-y)}$,
as a result in terms of the properties of these functions in distinct domains and the definition \eqref{Lax-12}
of $\mu_{\pm}(x,t,z)$ as well as the relationship \eqref{Lax-11} between $\mu_{\pm}(x,t,z)$ and $\phi(x,t,z)$,
the following proposition 2.1 is easily obtained (also refer to \cite{BP-2014} for the detailed derivation).\\

\noindent
\textbf{Proposition 2.1.} Take $\psi-\psi_{\pm}\in L^{1}(\mathbb{R}^{\pm})$, the modified expressions $\mu_{\pm2}(x,t,z)$
and the Jost functions $\phi_{\pm2}$ provided by \eqref{Lax-11} and \eqref{Lax-12} admit unique solutions in $\Sigma_{0}^{f}$.
In addition, $\mu_{+1}(x,t,z)$, $\mu_{-2}(x,t,z)$, $\phi_{+1}(x,t,z)$ and $\phi_{-2}(x,t,z)$
can be continuously to $D_{+}^{f}\bigcup\Sigma_{0}^{f}$, and analytically extended to $D_{+}^{f}$,,
while $\mu_{-1}(x,t,z)$, $\mu_{+2}(x,t,z)$, $\phi_{-1}(x,t,z)$ and $\phi_{+2}(x,t,z)$,
can be continuously extended to $D_{-}^{f}\bigcup\Sigma_{0}^{f}$, and analytically extended to $D_{-}^{f}$.

Since $\mbox{tr}U(x,t,z)=\mbox{tr}V(x,t,z)=0$, we have $(\det\phi_{\pm})_{x}=(\det\phi_{\pm})_{t}=0$.
Besides we can find
\begin{equation}\label{Lax-13}
\det \phi_{\pm}=\lim_{x\rightarrow\pm\infty}\mu_{\pm}=\det E_{\pm}^{f}(z)=\gamma_{f}(z)=1+\psi_{0}^2/z^2\neq0,~~z\neq\pm i\psi_{0},
\end{equation}
on the basic of Liouville's formula.
Since $\phi_{\pm}(x,t,z)$ are both primary matrix solutions of the spectral problem \eqref{RHP-1},
Thus, we find a constant matrix $S(z)$ so that
\begin{equation}\label{Lax-14}
\phi_{+}(x,t,z)=\phi_{-}(x,t,z)S(z),~~z\in\Sigma_{0}^{f},
\end{equation}
where $S(z)=(s_{ij}(z))_{2\times2}$ are scattering coefficients.
According to the relation \eqref{Lax-14}, we have
\begin{equation}\label{Lax-15}
\left\{ \begin{aligned}
&s_{11}(z)=\gamma_{f}^{-1}(z)|\phi_{+1}(x,t,z),\phi_{-2}(x,t,z)|,~~s_{12}(z)=\gamma_{f}^{-1}(z)|\phi_{+1}(x,t,z),\phi_{+2}(x,t,z)|,\\
&s_{21}(z)=\gamma_{f}^{+2}(z)|\phi_{+1}(x,t,z),\phi_{-2}(x,t,z)|,~~s_{22}(z)=\gamma_{f}^{-1}(z)|\phi_{+1}(x,t,z),\phi_{+1}(x,t,z)|,
             \end{aligned} \right.
\end{equation}
and $\det S(z)=1$.

Form Proposition 2.1, it is not hard to see that the scattering coefficients $s_{11}(z)$ and $s_{22}(z)$ in $z\in\Sigma_{0}^{f}$
can be continuously extended to
$D_{+}^{f}\bigcup\Sigma_{0}^{f}$ and $D_{-}^{f}\bigcup\Sigma_{0}^{f}$, and analytically extended to $D_{+}^{f}$ and $D_{-}^{f}$, respectively.

To further tackle the matrix RHP in the next section, we take $s_{11}(z)s_{22}(z)\neq 0$ for $z\in\Sigma^{f}$,
and $S(z)$ is continuous for $z=i\psi_{0}$.
Therefore, we can obtain the so-called reflection coefficients
\begin{equation}\label{Lax-16}
\rho(z)=\frac{s_{21}(z)}{s_{11}(z)},~~\hat{\rho}(z)=\frac{s_{21}(z)}{s_{22}(z)},~~\forall z\in\Sigma^{f}.
\end{equation}

\section{Inverse scattering problem with NZBCs: Simple pole}
To construct the residue conditions and discrete spectrum in the section,
here we introduce the symmetries of the scattering matrix $S(k)$.
Similar to the results of \cite{BP-2014}, we have $k(z)=k^{*}(z^{*})$, $k(z)=k\left(-\psi_{0}^2/z\right)$, $\lambda(z)=\bar{\lambda}(z^{*})$,
and $\lambda(z)=-\lambda\left(-\psi_{0}^2/z\right)$, as a result the symmetries of $U$, $V$, and $\theta$ reach to
\begin{equation}\label{ISP-1}
\left\{ \begin{aligned}
&U(x,t,z)=\sigma_{2}U^{*}(x,t,z^{*})\sigma_{2},~~U(x,t,z)=U\left(x,t,-\psi_{0}^2/z\right),\\
&V(x,t,z)=\sigma_{2}V(x,t,z^{*})^{*}\sigma_{2},~~V(x,t,z)=V\left(x,t,-\psi_{0}^2/z\right),\\
&\theta(x,t,z)=\theta^{*}(x,t,z^{*}),~~\theta(x,t,z)=-\theta\left(x,t,-\psi_{0}^2/z\right),
           \end{aligned} \right.
\end{equation}
with $\sigma_{2}=\left(
                   \begin{array}{cc}
                     0 & -i \\
                     i & 0 \\
                   \end{array}
                 \right)$.

In view of the above-mentioned symmetries, equations \eqref{RHP-1} and \eqref{Lax-11} yield
\begin{equation}\label{ISP-2}
\left\{ \begin{aligned}
&\phi_{\pm}(x,t,z)=\sigma_{2}\phi^{*}_{\pm}(x,t,z^{*})\sigma_{2},~~\phi_{\pm}(x,t,z)=\frac{i}{z}\phi_{\pm}\left(x,t,-\psi_{0}^2/z\right)\sigma_{3}Q_{\pm},\\
&\mu_{\pm}(x,t,z)=\sigma_{2}\mu^{*}_{\pm}(x,t,z^{*})\sigma_{2},~~\mu_{\pm}(x,t,z)=\frac{i}{z}\mu_{\pm}\left(x,t,-\psi_{0}^2/z\right)\sigma_{3}Q_{\pm}.
           \end{aligned} \right.
\end{equation}
It follows from equations \eqref{ISP-2} and \eqref{Lax-14} that
\begin{equation}\label{ISP-3}
S(z)=\sigma_{2}S^{*}(z^{*})\sigma_{2},~~S(z)=\left(\sigma_{3}Q_{-}\right)^{-1}S\left(-\psi_{0}^2/z\right)\sigma_{3}Q_{+},
\end{equation}
which lead to the symmetries between $\rho(z)$ and $\hat{\rho}(z)$ as
\begin{equation}\label{ISP-4}
\rho(z)=-\hat{\rho}^{*}(z^{*}),~~\rho(z)=\frac{q^{*}_{-}}{q_{-}}\rho\left(-\psi_{0}^2/z\right).
\end{equation}
The discrete spectrum is the set of all values
$z\in\mathbb{C}\setminus\Sigma^{f}$ such that they admit eigenfunctions in $L^2(\mathbb{R})$.
Similar to the works of \cite{BP-2014}, they satisfy $s_{11}(z)=0$ for $z\in D_{+}^{f}$
and $s_{22}(z)=0$ for $z\in D_{-}^{f}$ such that the corresponding eigenfunctions are in $L^2(\mathbb{R})$
from \eqref{Lax-15} and the expression \eqref{Lax-11} of $\phi_{\pm}$.

Next, we require that $s_{11}(z)$ admits $N$ simple zeros in
\begin{equation*}
D_{+}^{f}\cap\left\{z\in\mathbb{C}:|z|>\psi_{0}, \mbox{Im}z>0\right\}
\end{equation*}
given by $z_{n}$, $n=1,2,\ldots,N$, i.e., $s_{11}(z_{n})=0$ and $s'_{11}(z_{n})\neq0(n=1,2,\ldots,N)$.
If $s_{11}(z_{n})=0$,
we have $s_{22}(z^{*}_{n})=s_{22}\left(-\psi_{0}^2/z_{n}\right)=s_{11}\left(-\psi_{0}^2/z^{*}_{n}\right)=0$,
Thus the set of discrete spectrum yields
\begin{equation}\label{ISP-5}
Z^{f}=\left\{z_{n},-\frac{\psi_{0}^2}{z^{*}_{n}},z^{*}_{n},-\frac{\psi_{0}^2}{z_{n}}\right\}_{n=1}^{N},
~~s_{11}(z_{n})=0,~~z_{n}\in D_{+}^{f}\cap\left\{z\in\mathbb{C}:|z|>\psi_{0},\mbox{Im}z>0\right\}.
\end{equation}
Since $s_{11}(z_{0})=0$ and $s'_{11}(z_{0})\neq0$ are taken for $z_{0}\in Z^{f}\bigcap D_{+}^{f}$,
then according to the first expression in \eqref{Lax-15}, a norming constant $b_{+}(z_{0})$ satisfies
\begin{equation}\label{ISP-6}
\phi_{+1}(x,t,z_{0})=b_{+}(z_{0})\phi_{-2}(x,t,z_{0}).
\end{equation}
The residue condition of $\phi_{+1}(x,t,z)/s_{11}(z)$ in $z_{0}\in Z^{f}\bigcap D_{+}^{f}$ reaches to
\begin{equation}\label{ISP-7}
\mathop{\mbox{Res}}\limits_{z=z^{*}_{0}}\left[\frac{\phi_{+1}(x,t,z)}{s_{11}(z)}\right]
=\frac{\phi_{+1}(x,t,z_{0})}{s'_{11}(z_{0})}=\frac{b_{+}(z_{0})}{s'_{11}(z_{0})}\phi_{-2}(x,t,z_{0}).
\end{equation}
Following the similar way, from $s_{22}(z^{*}_{0})=0$ and $s'_{22}(z^{*}_{0})\neq0$ for $z^{*}_{0}\in Z^{f}\bigcap D_{-}^{f}$
and the second expression in \eqref{Lax-15},
we see that a norming constant $b_{-}(z^{*}_{0})$ yields
\begin{equation}\label{ISP-8}
\phi_{-2}(x,t,z^{*}_{0})=b_{-}(z^{*}_{0})\phi_{-1}(x,t,z^{*}_{0}).
\end{equation}
The residue condition $\phi_{+2}(x,t,z)/s_{22}(z)$ in $z^{*}_{0}\in Z^{f}\bigcap D_{-}^{f}$ arrives at
\begin{equation}\label{ISP-9}
\mathop{\mbox{Res}}\limits_{z=z^{*}_{0}}\left[\frac{\phi_{+2}(x,t,z)}{s_{22}(z)}\right]=\frac{\phi_{+2}(x,t,z^{*}_{0})}{s'_{22}(z^{*}_{0})}
=\frac{b_{-}(z^{*}_{0})}{s'_{22}(z^{*}_{0})}\phi_{-1}(x,t,z^{*}_{0}).
\end{equation}
For simplicity, equations \eqref{ISP-7} and \eqref{ISP-9} can be rewritten as
\begin{equation}\label{ISP-10}
\left\{ \begin{aligned}
&\mathop{\mbox{Res}}\limits_{z=z_{0}}\left[\frac{\phi_{+1}(x,t,z)}{s_{11}(z)}\right]=
A_{+}[z_{0}]\phi_{-2}(x,t,z_{0}),~~A_{+}[z_{0}]=\frac{b_{+}(z_{0})}{s'_{11}(z_{0})},~~z_{0}\in Z^{f}\cap D_{+}^{f},\\
&\mathop{\mbox{Res}}\limits_{z=z^{*}_{0}}\left[\frac{\phi_{+1}(x,t,z)}{s_{11}(z)}\right]=
A_{+}[z^{*}_{0}]\phi_{-2}(x,t,z^{*}_{0}),~~A_{+}[z^{*}_{0}]=\frac{b_{+}(z^{*}_{0})}{s'_{11}(z^{*}_{0})},~~z^{*}_{0}\in Z^{f}\cap D_{+}^{f}.
           \end{aligned} \right.
\end{equation}
It follows from \eqref{ISP-10} that
\begin{equation}\label{ISP-11}
A_{+}[z_{0}]=-A^{*}_{-}[z^{*}_{0}],~~A_{+}[z_{0}]=\frac{z_{0}^2}{\psi_{-}^2}A_{-}\left[-\frac{\psi_{0}^2}{z_{0}}\right],
~~z_{0}\in Z^{f}\cap D_{+}^{f},
\end{equation}
in terms of symmetries \eqref{ISP-2} and \eqref{ISP-3}, which lead directly to
\begin{equation}\label{ISP-12}
A_{+}[z_{n}]=-A^{*}_{-}[z^{*}_{n}]=\frac{z_{n}^2}{\psi_{-}^2}A_{-}\left[-\frac{\psi_{0}^2}{z_{n}}\right]
=-\frac{z_{n}^2}{\psi_{-}^2}A^{*}_{+}\left[-\frac{\psi_{0}^2}{z^{*}_{n}}\right],
~~z_{n}\in Z^{f}\cap D_{+}^{f}.
\end{equation}
We rewrite the relation $\phi_{+}(x,t,z)=\phi_{-}(x,t,z)S(z)$ as
\begin{equation}\label{ISP-13}
\left\{ \begin{aligned}
&\frac{\phi_{+1}(x,t,z)}{s_{11}(z)}=\phi_{-1}(x,t,z)+\rho(z)\phi_{-2}(x,t,z),\\
&\frac{\phi_{+2}(x,t,z)}{s_{22}(z)}=\hat{\rho}(z)\phi_{-1}(x,t,z)+\phi_{-2}(x,t,z),
           \end{aligned} \right.
\end{equation}
which yield
\begin{equation}\label{ISP-14}
\left[\phi_{-1}(x,t,z),\frac{\phi_{+2}(x,t,z)}{s_{22}(z)}\right]=\left[\frac{\phi_{+1}(x,t,z)}{s_{11}(z)},\phi_{-2}(x,t,z)\right]
\left[I-J_{0}(x,t,\lambda)\right],
\end{equation}
with
\begin{equation}\label{ISP-15}
J_{0}=\left(
        \begin{array}{cc}
          0 & -\hat{\rho}(z) \\
          \rho(z) & \rho(z)\hat{\rho}(z) \\
        \end{array}
      \right).
\end{equation}
Similar to \cite{BP-2014}, the asymptotics for modified Jost solutions and scattering data satisfy
\begin{equation}\label{ISP-16}
\left\{ \begin{aligned}
&\mu_{\pm}(x,t,z)=I+O\left(\frac{1}{z}\right),~~S(z)=I+O\left(\frac{1}{z}\right),~~z\rightarrow\infty,\\
&\mu_{\pm}(x,t,z)=\frac{i}{z}\sigma_{3}Q_{\pm}+O(1),~~S(z)=\frac{\psi_{+}}{\psi_{-}}I+O(z),~~z\rightarrow 0.
           \end{aligned} \right.
\end{equation}
On the basis of the modified Jost functions, let the sectionally meromorphic matrix $M(x,t,z)$ be
\begin{align}\label{ISP-17}
M(x,t,z)=
\left\{ \begin{aligned}
M^{+}(x,t,z)=&\left(\frac{\mu_{+1}(x,t,z)}{s_{11}(z)},\mu_{-2}(x,t,z)\right)\\
&=\left(\frac{\phi_{+1}(x,t,z)}{s_{11}(z)},\phi_{-2}(x,t,z)\right)e^{-i\theta(x,t,z)\sigma_{3}},~~z\in D_{+}^{f},\\
M^{-}(x,t,z)=&\left(\mu_{-1}(x,t,z),\frac{\mu_{+2}(x,t,z)}{s_{22}}\right)\\
&=\left(\phi_{-1}(x,t,z),\frac{\phi_{+2}(x,t,z)}{s_{22}}\right)e^{-i\theta(x,t,z)\sigma_{3}},~~z\in D_{-}^{f}.
           \end{aligned} \right.
\end{align}
Summarizing the above results, the following proposition 3.1 holds.\\

\noindent
\textbf{Proposition 3.1}.
The matrix function $M(x,t,z)$ admits the following matrix RHP\\

\noindent
$\bullet$ Analyticity: $M(x,t,z)$ is analytic in $\left(D_{+}^{f}\bigcup D_{-}^{f}\right)\setminus Z^{f}$;\\
$\bullet$ Jump condition: $M^{-}(x,t,z)=M^{+}(x,t,z)(I-J(x,t,z))$,~~$z\in\Sigma^{f}$ with $J(x,t,z)=e^{i\theta(x,t,z)\widehat{\sigma}_{3}}J_{0}$;\\
$\bullet$ Asymptotic behavior: $M^{\pm}(x,t,z)=I+(1/z)$ for $z\rightarrow\infty$, Besides, $M^{\pm}=\frac{i}{z}\sigma_{3}Q_{-}+O(1)$ for $z\rightarrow0$.

To conveniently solve the above RHP (i.e.,Proposition 3.1), take
\begin{equation}\label{JSP-18}
\xi_{n}=\left\{ \begin{aligned}
&z_{n},~~n=1,2,\ldots,N,\\
&-\frac{\psi_{0}^2}{z^{*}_{n-N}},~~n=N+1,N+2,\ldots,2N,
           \end{aligned} \right.
\end{equation}
and $\widehat{\xi}_{n}=-\psi_{0}^2/\xi_{n}$. Then $Z^{f}=\left\{\xi_{n},\widehat{\xi}_{n}\right\}_{n=1}^{2N}$
with $\xi_{n}\in D_{+}^{f}$ and $\widehat{\xi}_{n}\in D_{-}^{f}$.
Subtracting out the simple pole contributions and the asymptotics, i.e.,
\begin{equation}\label{JSP-19}
M_{sp}(x,t,z)=I+\frac{i}{z}\sigma_{3}Q_{-}+\sum_{n=1}^{2N}\left[\frac{\mathop{\mbox{Res}}\limits_{z=\xi_{n}}M^{+}(x,t,z)}{z-\xi_{n}}+
\frac{\mathop{\mbox{Res}}\limits_{z=\widehat{\xi}_{n}}M^{-}(x,t,z)}{z-\widehat{\xi}_{n}}\right],
\end{equation}
from both sides of the above jump condition $M^{-}=M^{+}(I-J)$ leads to
\begin{equation}\label{JSP-20}
M^{-}(x,t,z)-M_{sp}(x,t,z)=M^{+}(x,t,z)-M_{sp}(x,t,z)-M^{+}(x,t,z)J(x,t,z).
\end{equation}
Here $M^{\pm}(x,t,z)\rightarrow M_{sp}(x,t,z)$ are analytic in $D_{\pm}^{f}$.
Furthermore, the asymptotics are both $O(1/z)$ as $z\rightarrow\infty$ and $O(1)$ as $z\rightarrow0$
and $J(x,t,z)$ is $O(1/z)$ as $z\rightarrow\infty$, and $O(z)$ as $z\rightarrow 0$.
Thus, the Cauchy projectors
\begin{equation}\label{JSP-21}
P^{\pm}[f](z)=\frac{1}{2\pi i}\int_{\Sigma^{f}}\frac{f(\zeta)}{\zeta-(z\pm i0)}d\zeta,
\end{equation}
(where $z\pm i0$ is the limit taken from the left/right of $z$),
and Plemelj's formulae are employed to solve \eqref{JSP-20} to give
\begin{equation}\label{JSP-22}
M(x,t,z)=M_{sp}(x,t,z)+\frac{1}{2\pi i}\int_{\Sigma^{f}}\frac{M^{+}(x,t,\zeta)J(x,t,\zeta)}{\zeta-z}d\zeta,~~z\in\mathbb{C}\setminus\Sigma^{f},
\end{equation}
where $\int_{\Sigma^{f}}$ represents the integral along the oriented contours seen in Fig.1.

From \eqref{ISP-17}, we find that only its first (second) column admits a simple pole at $z=\xi_{n}(z=\widehat{\xi}_{n})$.
As a consequence, by using \eqref{Lax-11} and \eqref{ISP-10}, the residue part in \eqref{JSP-22} can be written as
\begin{align}\label{JSP-23}
\frac{\mathop{\mbox{Res}}\limits_{z=\xi_{n}} M^{+}(x,t,z)}{z-\xi_{n}}&
+\frac{\mathop{\mbox{Res}}\limits_{z=\widehat{\xi}_{n}} M^{-}(x,t,z)}{z-\widehat{\xi}_{n}}\notag\\
&=\left(\frac{A_{+}\left[\xi_{n}\right]e^{-2i\theta(x,t,\xi_{n})}}{z-\xi_{n}}\mu_{-2}\left(x,t,\xi_{n}\right),
\frac{A_{-}\left[\widehat{\xi}_{n}\right]e^{2i\theta\left(x,t,\widehat{\xi}_{n}\right)}}{z-\widehat{\xi}_{n}}\mu_{-1}\left(x,t,\widehat{\xi}_{n}\right)\right).
\end{align}

For $z=\xi_{s} (s=1,2,\ldots,2N)$, from the second column of $M(x,t,z)$ provided by \eqref{JSP-22} with \eqref{JSP-23}, we obtain
\begin{align}\label{JSP-24}
\mu_{-2}(x,t,\xi_{s})=\left(
                         \begin{array}{c}
                           \frac{i\psi_{-}}{\xi_{s}} \\
                           1 \\
                         \end{array}
                       \right)&+\sum_{n=1}^{2N}\frac{A_{-}\left[\widehat{\xi}_{n}\right]e^{2i\theta\left(x,t,\widehat{\xi}_{n}\right)}}
                       {\xi_{s}-\widehat{\xi}_{n}}
                       \mu_{-1}\left(x,t,\widehat{\xi}_{n}\right)\notag\\
                       &+\frac{1}{2\pi i}\int_{\Sigma^{f}}\frac{\left(M^{+}J\right)_{2}\left(x,t,\zeta\right)}{\zeta-\xi_{s}}d\zeta,~~s=1,2,\ldots,2N.
\end{align}
In view of \eqref{ISP-2}, we have
\begin{equation}\label{JSP-25}
\mu_{-2}(x,t,\xi_{s})=\frac{i\psi_{-}}{\xi_{s}}\mu_{-1}\left(x,t,\widehat{\xi}_{s}\right),~~s=1,2,\ldots,2N.
\end{equation}
Substituting \eqref{JSP-25} into \eqref{ISP-2} leads directly to
\begin{align}\label{JSP-26}
\sum_{n=1}^{2N}\left(\frac{A_{-}\left[\widehat{\xi}_{n}\right]e^{2i\theta\left(x,t,\widehat{\xi}_{n}\right)}}{\xi_{s}-\widehat{\xi}_{n}}-\frac{i\psi_{-}}{\xi_{s}}\delta_{sn}\right)
&\mu_{-1}\left(x,t,\widehat{\xi}_{n}\right)+\left(
                                   \begin{array}{c}
                                     \frac{i\psi_{-}}{\xi_{s}} \\
                                     1 \\
                                   \end{array}
                                 \right)\notag\\
                                 &+\frac{1}{2\pi i}\int_{\Sigma^{f}}\frac{\left(M^{+}J\right)_{2}\left(x,t,\zeta\right)}{\zeta-\xi_{s}}d\zeta=0,~~s=1,2,\ldots,2N.
\end{align}
where
\begin{equation*}
\delta_{sn}=\left\{ \begin{aligned}
1,~~s=n,\\
0,~~s\neq n.
           \end{aligned} \right.
\end{equation*}
System \eqref{JSP-26} including $2N$ equations with $2N$ unknowns $\mu_{-1}\left(x,t,\widehat{\xi}_{n}\right)$
can lead to the solutions $\mu_{-1}\left(x,t,\widehat{\xi}_{s}\right)$
such that one can find $\mu_{-2}\left(x,t,\xi_{s}\right)$ from \eqref{JSP-26}.
As a consequence, plugging $\mu_{-1}\left(x,t,\widehat{\xi}_{s}\right)$ and $\mu_{-2}\left(x,t,\xi_{s}\right)$ into \eqref{JSP-23},
and then plugging \eqref{JSP-23} into \eqref{JSP-22} can lead to the $M(x,t,z)$ on the basis of the scattering data.

In view of \eqref{JSP-23} and \eqref{JSP-22}, the asymptotic behavior of $M(x,t,z)$ yields
\begin{equation}\label{JSP-27}
M(x,t,z)=I+\frac{M^{(1)}(x,t)}{z}+O\left(\frac{1}{z^2}\right),~~z\rightarrow\infty,
\end{equation}
where
\begin{align}\label{JSP-28}
M^{(1)}(x,t)&=i\sigma_{3}Q+\sum_{n=1}^{2N}\left[A_{+}\left[\xi_{n}\right]e^{-2i\theta\left(x,t,\xi_{n}\right)},
A_{-}\left[\widehat{\xi}_{n}\right]e^{2i\theta\left(x,t,\widehat{\xi}_{n}\right)}\mu_{-1}\left(x,t,\widehat{\xi}_{n}\right)\right]\notag\\
&-\frac{1}{2\pi i}\int_{\Sigma^{f}}M^{+}\left(x,t,\zeta\right)J(x,t,\zeta)d\zeta,
\end{align}
with $\mu_{-1}\left(x,t,\widehat{\xi}_{s}\right)$ and $\mu_{-2}\left(x,t,\widehat{\xi}_{s}\right)$ given by \eqref{JSP-25} and \eqref{JSP-26}.

It follows from \eqref{ISP-17} that $M(x,t,z)e^{i\theta(x,t,z)\sigma_{3}}$ satisfies \eqref{RHP-1}.
Substitution of $M(x,t,z)e^{i\theta(x,t,z)\sigma_{3}}$ with \eqref{JSP-27} into the x-part of the Lax pair \eqref{RHP-1}
and then choosing the coefficients of $z^{0}$ can arrive at the following proposition 3.2 for the potential $u(x,t)$.\\

\noindent
\textbf{Proposition 3.2.} The potential with simple poles of the GFONLS equation \eqref{gtc-NLS} with NZBCs \eqref{NZBC1} can be found as
\begin{equation}\label{JSP-29}
\psi(x,t)=\psi_{-}-i\sum_{n=1}^{2N}A_{-}\left[\widehat{\xi}_{n}\right]e^{2i\theta\left(x,t,\widehat{\xi}_{n}\right)}\mu_{-11}\left(x,t,\widehat{\xi}_{n}\right)
+\frac{1}{2\pi}\int_{\Sigma^{f}}\left(M^{+}J\right)_{12}\left(x,t,\zeta\right)d\zeta.
\end{equation}
where $\xi_{n}=z_{n}$, $\xi_{n+N}=-\psi_{0}^2/z^{*}_{n-N}$, $n=1,2,\ldots,N$, $\widehat{\xi}_{n}=-\psi_{0}^2/\xi_{n}$, and
$\mu_{-11}\left(x,t,\widehat{\xi}_{n}\right)$ are expressed by
\begin{align}\label{JSP-30}
\sum_{n=1}^{2N}\left(\frac{A_{-}\left[\widehat{\xi}_{n}\right]e^{2i\theta\left(x,t,\widehat{\xi}_{n}\right)}}{\xi_{s}-\widehat{\xi}_{n}}
-\frac{i\psi_{-}}{\xi_{s}}\delta_{sn}\right)
&\mu_{-11}\left(x,t,\widehat{\xi}_{n}\right)+\frac{i\psi_{-}}{\xi_{s}}\notag\\
&+\frac{1}{2\pi i}\int_{\Sigma^{f}}\frac{\left(M^{+}J\right)_{12}\left(x,t,\zeta\right)}{\zeta-\xi_{s}}d\zeta=0,~~s=1,2,\ldots,2N,
\end{align}
which can be obtained from expression \eqref{JSP-26}.

Since $s_{11}(z)$ and $s_{22}(z)$ are analytic in $D_{+}^{f}$ and $D_{-}^{f}$, respectively,
and the discrete spectral points $\xi_{n}'$s and $\widehat{\xi}_{n}$'s are the
simple zeros of $s_{11}(z)$ and $s_{22}(z)$, respectively.
Based on \cite{BP-2014}, the trace formulae for the GFONLS equation \eqref{gtc-NLS} with NZBCs can be found as
\begin{equation}\label{JSP-31}
s_{11}(z)=e^{s(z)}s_{0}(z)~~\mbox{for}~~z\in D_{+}^{f},~~
s_{22}(z)=e^{-s(z)}/s_{0}(z)~~\mbox{for}~~z\in D_{-}^{f},
\end{equation}
where
\begin{equation}\label{JSP-32}
s(z)=-\frac{1}{2\pi i}\int_{\Sigma^{f}}\frac{\log\left[1+\rho\left(\zeta\right)\rho^{*}\left(\zeta^{*}\right)\right]}{\zeta-z}d\zeta,~~
s_{0}(z)=\prod_{n=1}^{N}\frac{(z-z_{n})\left(z+\psi_{0}^2/z^{*}_{n}\right)}{\left(z-z^{*}_{n}\right)\left(z+\psi_{0}^2/z_{n}\right)}.
\end{equation}
See \cite{BP-2014} for the detailed derivation.

In view of the limit $z\rightarrow0$ of $s_{11}(z)$ in \eqref{JSP-32} and \eqref{ISP-16}
we have the following theta condition
\begin{equation}\label{JSP-33}
\mbox{arg}\left(\frac{\psi_{+}}{\psi_{-}}\right)=4\sum_{n=1}^{N}\mbox{arg}z_{n}+\int_{\Sigma^{f}}\frac{\log\left[1
+\rho(\zeta)\rho^{*}\left(\zeta^{*}\right)\right]}{2\pi\zeta}d\zeta,~~z\rightarrow 0.
\end{equation}
In particular, in the case of the reflectionless potential, i.e., $\rho(z)=\hat{\rho}(z)=0$,
which can result in $J=(0)_{2\times2}$.
As result, Eq.\eqref{JSP-30} arrives at
\begin{equation}\label{JSP-34}
\sum_{n=1}^{2N}\left(\frac{A_{-}\left[\widehat{\xi}_{n}\right]e^{2i\theta\left(x,t,\widehat{\xi}_{n}\right)}}{\xi_{s}-\widehat{\xi}_{n}}
-\frac{i\psi_{-}}{\xi_{s}}\delta_{sn}\right)\mu_{-11}\left(x,t,\widehat{\xi}_{n}\right)=\frac{i\psi_{-}}{\xi_{s}},~~s=1,2,\ldots,2N.
\end{equation}
which can be found for $\mu_{-11}(x,t,\widehat{\xi}_{n})$ by making use of the Cramer's rule.
Summarizing the above analysis, the Theorem 1.1 for the potential $\psi(x,t)$ in the case of simple pole holds.\\

In the case of the reflectionless potential $\rho(z)=\hat{\rho}(z)=0$, the above obtained trace formulae and theta condition become
\begin{equation}\label{JSP-36}
\left\{ \begin{aligned}
&s_{11}=\prod_{n=1}^{N}\frac{\left(z-z_{n}\right)\left(z+\psi_{0}^2/z^{*}_{n}\right)}{\left(z-z^{*}_{n}\right)\left(z+\psi_{0}^2/z_{n}\right)},~~\mbox{for}~~z\in D_{+}^{f},\\
&s_{22}=\prod_{n=1}^{N}\frac{\left(z-z^{*}_{n}\right)\left(z+\psi_{0}^2/z_{n}\right)}{\left(z-z_{n}\right)\left(z+\psi_{0}^2/z^{*}_{n}\right)},~~\mbox{for}~~z\in D_{-}^{f},
           \end{aligned} \right.
\end{equation}
and
\begin{equation}\label{JSP-37}
\mbox{arg}\left(\frac{\psi_{+}}{\psi_{-}}\right)=\mbox{arg}(\psi_{+})-\mbox{arg}(\psi_{-})=4\sum_{n=1}^{N}\mbox{arg}(z_{n}),
\end{equation}
respectively.

\textbf{Case (I):} For $N=1, z_{1}=1.5i$ in Theorem 1.1, equation \eqref{JSP-37} indicates that the asymptotic phase difference is $2\pi$.
As shown in Fig.2, the solution  can represent the Kuznetzon-Ma (KM) breather that is spatially localized and temporally breathing.
As shown in Figs.3(a)-3(c), as $\psi_{-}$ becomes smaller, the periodic behavior of the breather wave
only appears in its top part, and the maximal amplitude under the background gradually decreases.
Particularly, as seen in Fig.3(d), as $\psi_{-}\rightarrow0$, the breather wave of the GFONLS equation \eqref{gtc-NLS} with NZBCs
yields the bright soliton of the the GFONLS equation \eqref{gtc-NLS} with the ZBC.

\textbf{Case (II):} For $N=1, z_{1}=ae^{\pi/4}$ in Theorem 1.1, the asymptotic phase difference is $\pi$.
The GFONLS equation \eqref{gtc-NLS} with the NZBCs possesses the non-stationary solitons.
Fig.3 is plotted for the non-stationary solitons for the GFONLS equation \eqref{gtc-NLS} with NZBCs,
which are localized both in time and space, thus revealing the usual Akhmediev breather wave features.

\textbf{Case (III):} For $N=2$ in Theorem 1.1, we have the interactions of breather-breather solutions of
the GFONLS equation \eqref{gtc-NLS} with NZBCs.  As shown in Fig.4, the interaction phenomenon is strong.
In particular, as $\psi_{-}\rightarrow0$, we have the strong interactions of the bright-bright solitons of
the GFONLS equation \eqref{gtc-NLS} with the ZBC.
However, as shown Fig.5, if we take two appropriate eigenvalues,
then we have the weak interactions of breather-breather solutions of the GFONLS equation \eqref{gtc-NLS} with NZBCs.
Likewise, as $\psi_{-}\rightarrow0$, we have the weak interactions of the simple-pole bright-bright solitons of the GFONLS equation \eqref{gtc-NLS}
with ZBC (see Figs.5(a)-(5)).

\textbf{Case (IV):} One interesting example of the breather-breather waves is seen in
Fig.6(a), where both (two breather waves) have different modulation frequencies.
In particular, as $z_{1}=z_{2}$, they appear as a first-order Akhmediev breather (see Fig.6(b)).
As shown Fig.6(c), as $z_{1}=-z_{2}=0.1=1.5i$, we can see the interactions of simple-pole breather-breather solutions of the GFONLS equation \eqref{gtc-NLS} with NZBCs.

\textbf{Case (V):} For $N=2$ in Theorem 1.1, we present another interesting example of the breather-breather waves.
As seen in Fig.7, the result is a simply periodic solution.
Specifically, as $\psi_{-}\rightarrow0$, we have the simple-pole bright-bright solitons of the GFONLS equation \eqref{gtc-NLS} with NZBCs.
To our surprise, the bright-bright solitons is also a simply periodic solution (see Fig.7(d)).

$~~~$
{\rotatebox{0}{\includegraphics[width=5.2cm,height=3.6cm,angle=0]{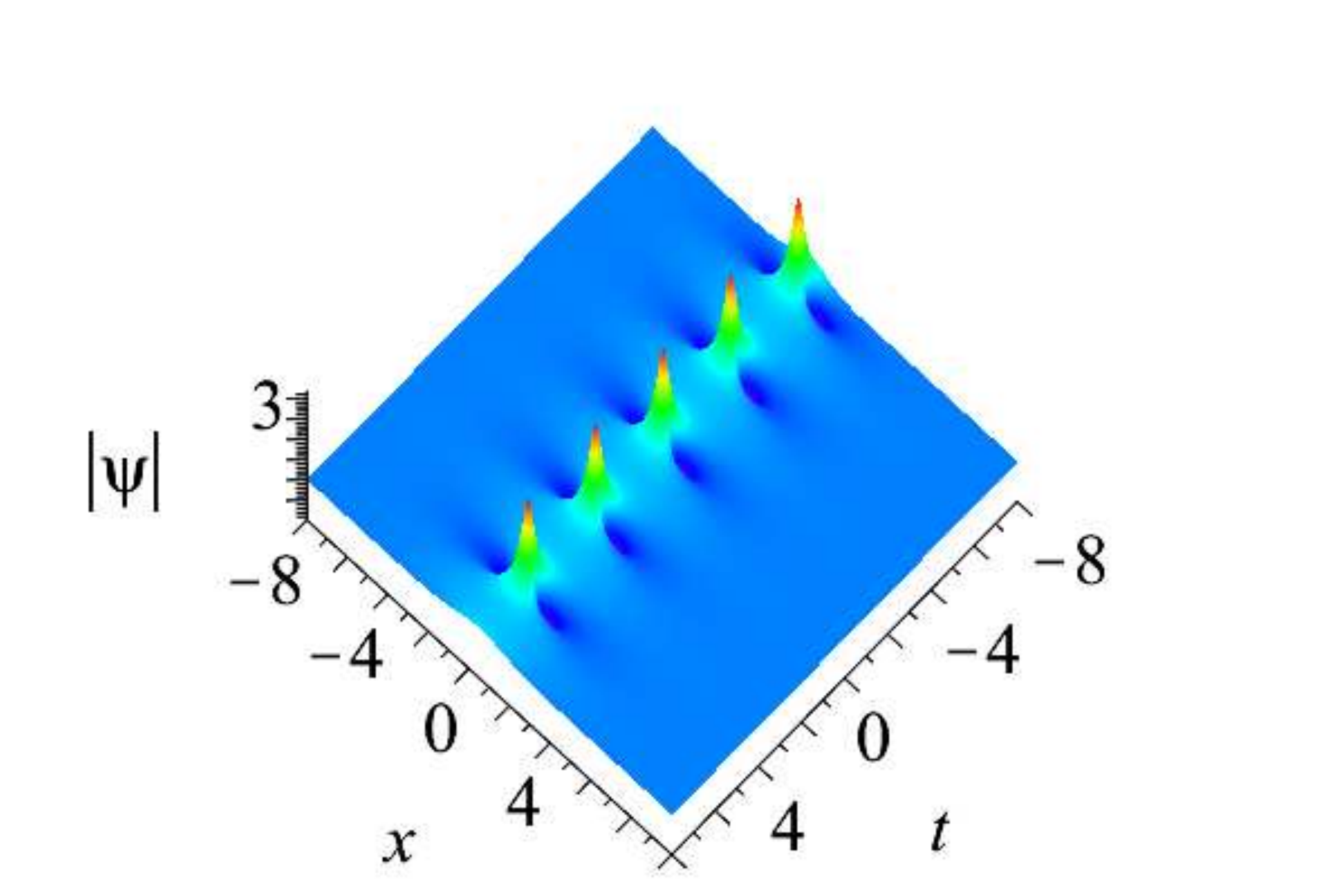}}}
~~~~~~~~~~~~~~~
{\rotatebox{0}{\includegraphics[width=5.2cm,height=3.6cm,angle=0]{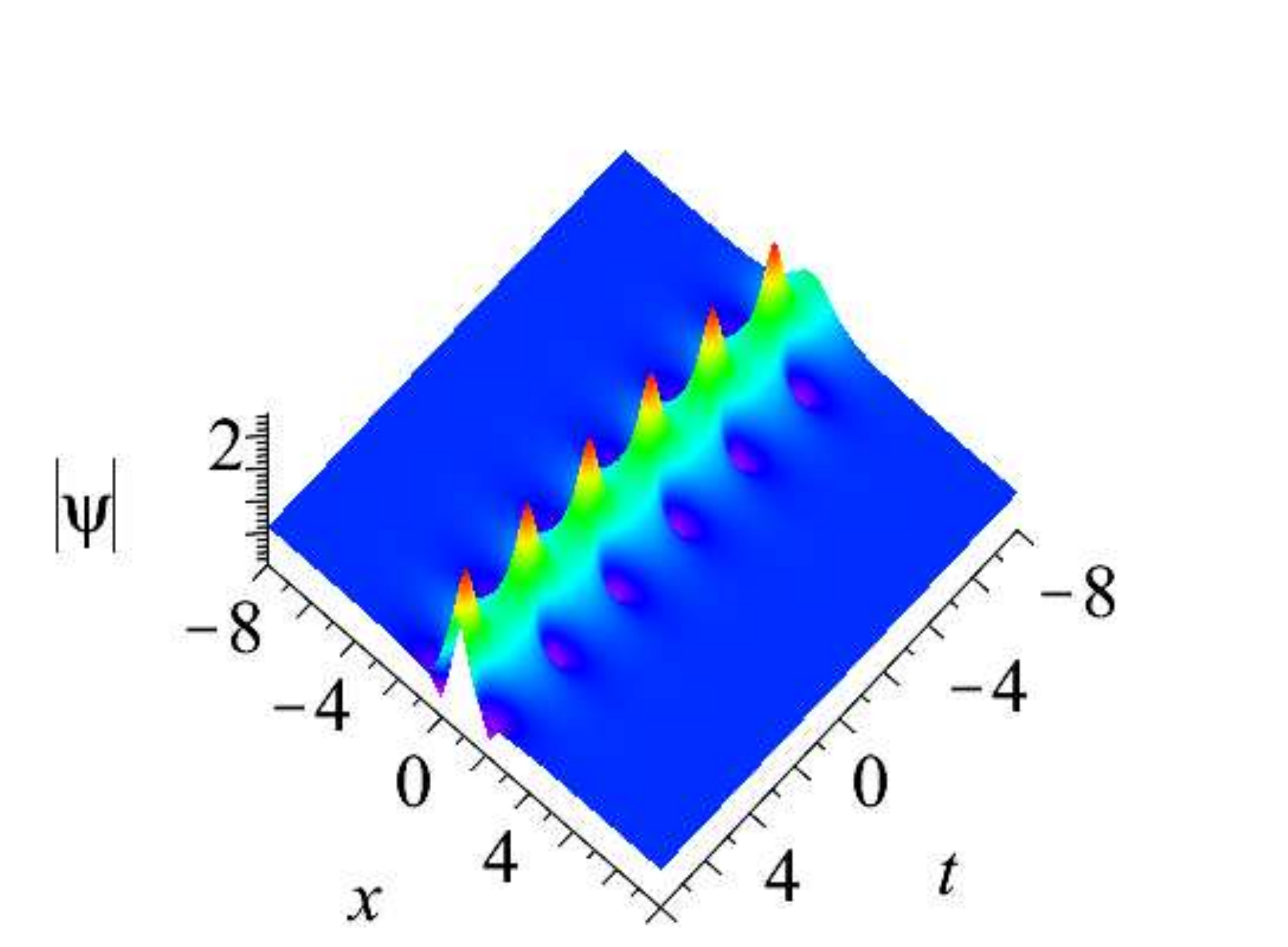}}}

$\qquad\qquad\qquad\qquad\textbf{(a)}
\qquad\qquad\qquad\qquad\qquad\qquad\qquad\textbf{(b)}
$\\

$~~~$
{\rotatebox{0}{\includegraphics[width=5.2cm,height=3.6cm,angle=0]{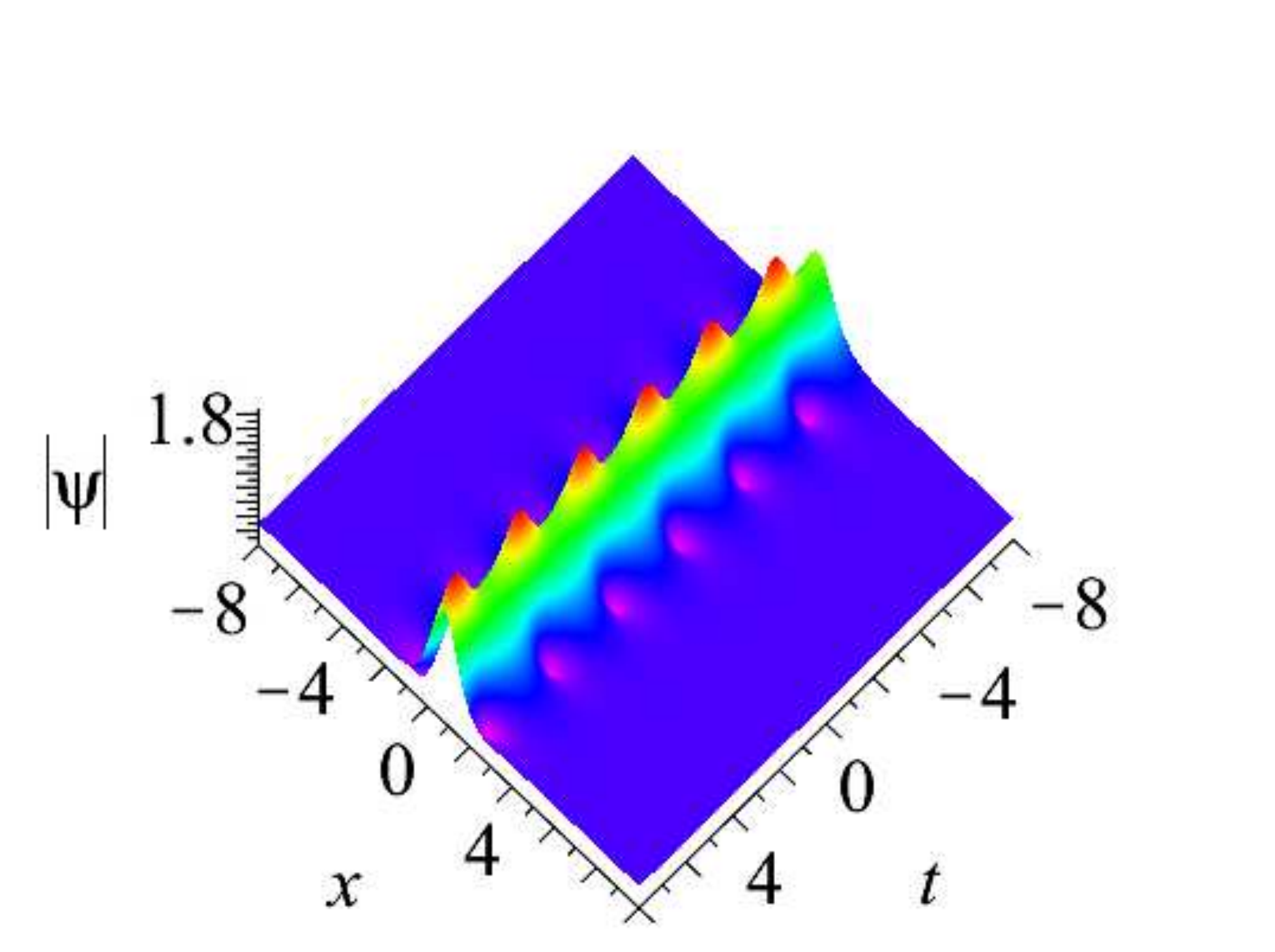}}}
~~~~~~~~~~~~~~~
{\rotatebox{0}{\includegraphics[width=5.2cm,height=3.6cm,angle=0]{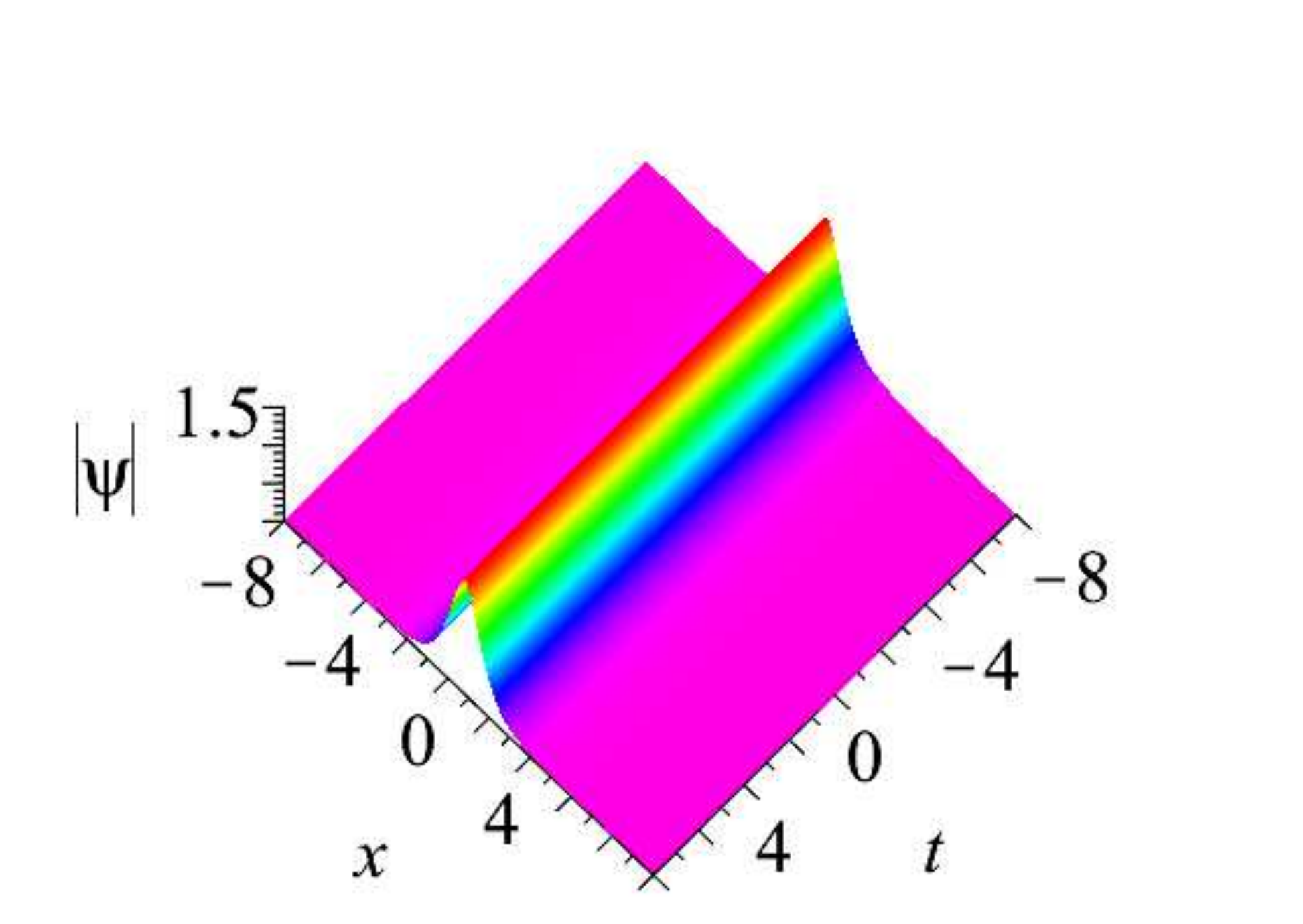}}}

$\qquad\qquad\qquad\qquad\textbf{(c)}
\qquad\qquad\qquad\qquad\qquad\qquad\qquad\textbf{(d)}
$\\
\noindent { \small \textbf{Figure 2.} (Color online) Breather waves via solution \eqref{JSP-35} with parameters
$N=1, z_{1}=1.5i, \alpha_{3}=\alpha_{4}=\alpha_{5}=0.01, A_{+}[z_{1}]=1$:
$\textbf{(a)}$: $\psi_{-}=1$; $\textbf{(b)}$: $\psi_{-}=0.6$; $\textbf{(c)}$: $\psi_{-}=0.3$;
$\textbf{(d)}$: bright-soliton solution with $\psi_{-}\rightarrow0$.\\}

$~~~$
{\rotatebox{0}{\includegraphics[width=5.2cm,height=3.6cm,angle=0]{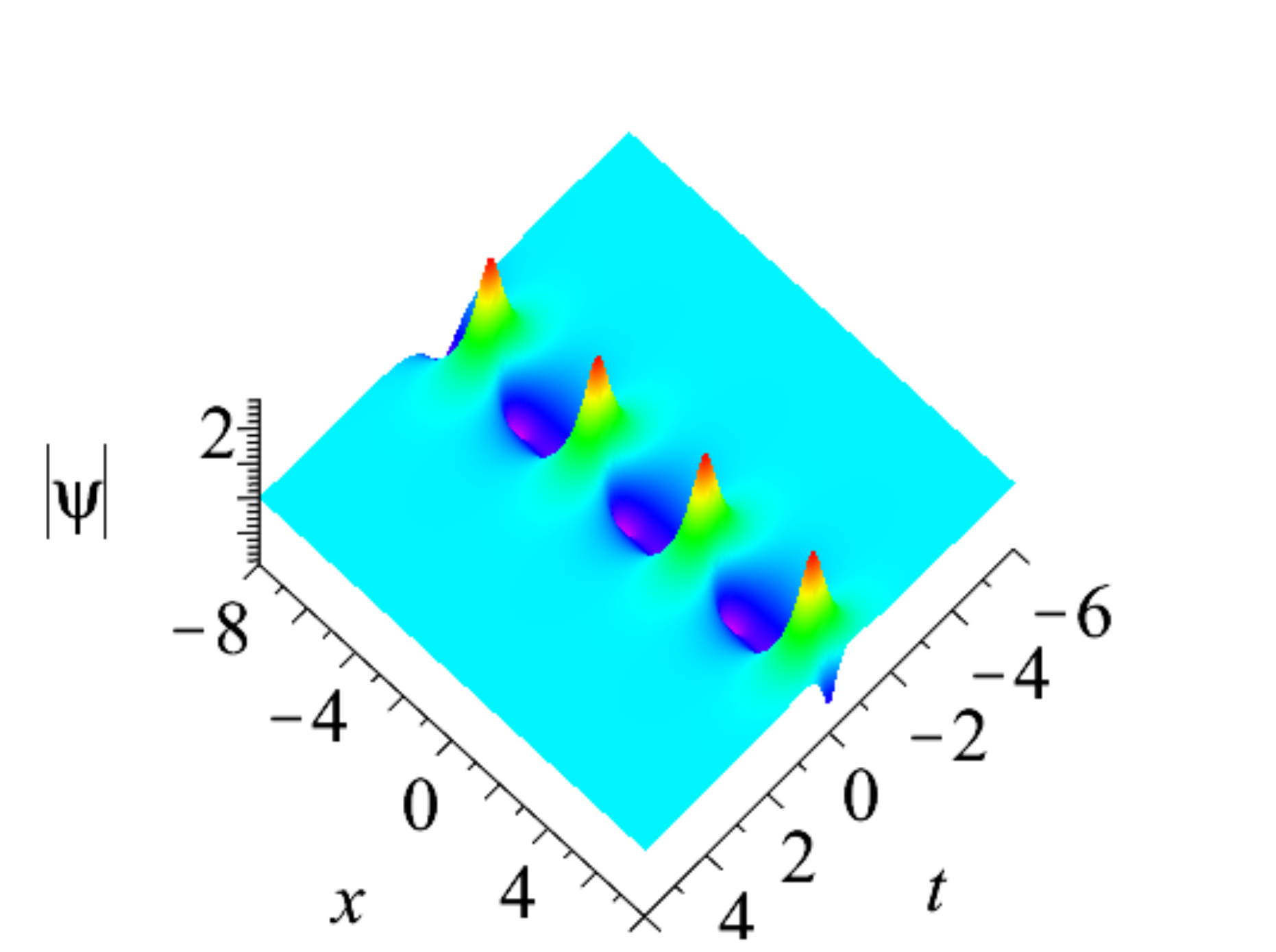}}}
~~~~~~~~~~~~~~~
{\rotatebox{0}{\includegraphics[width=5.2cm,height=3.6cm,angle=0]{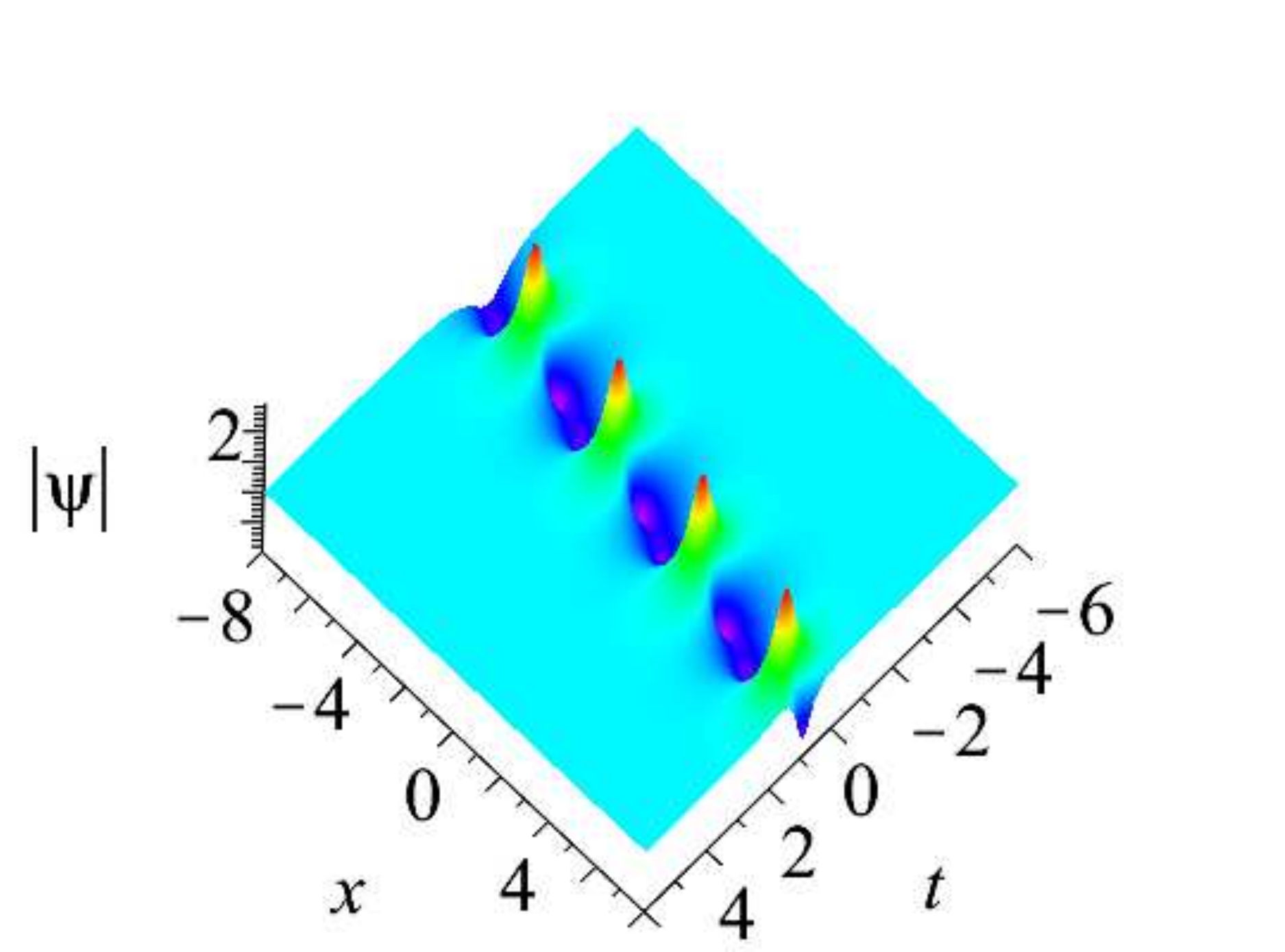}}}

$\qquad\qquad\qquad\qquad\textbf{(a)}
\qquad\qquad\qquad\qquad\qquad\qquad\qquad\textbf{(b)}
$\\
\noindent { \small \textbf{Figure 3.} (Color online) Breather waves via solution \eqref{JSP-35} with parameters
$N=1, A_{+}[z_{1}]=1, \alpha_{3}=\alpha_{4}=\alpha_{5}=0.01, \psi_{-}=1$:
$\textbf{(a)}$: $z_{1}=e^{i\pi/4}$; $\textbf{(b)}$: $z_{1}=0.8e^{i\pi/4}$.\\}

$~~~$
{\rotatebox{0}{\includegraphics[width=5.2cm,height=3.6cm,angle=0]{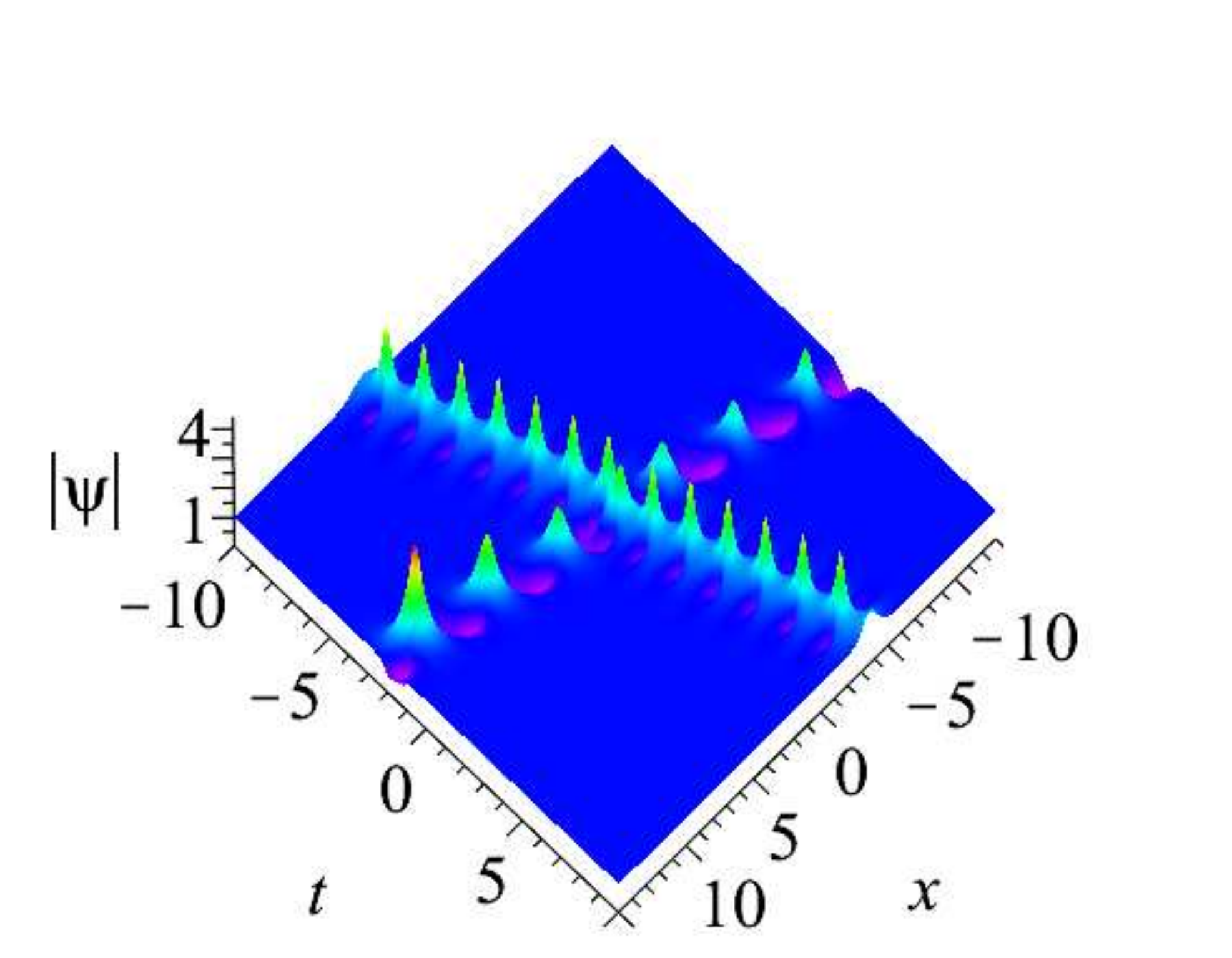}}}
~~~~~~~~~~~~~~~
{\rotatebox{0}{\includegraphics[width=5.2cm,height=3.6cm,angle=0]{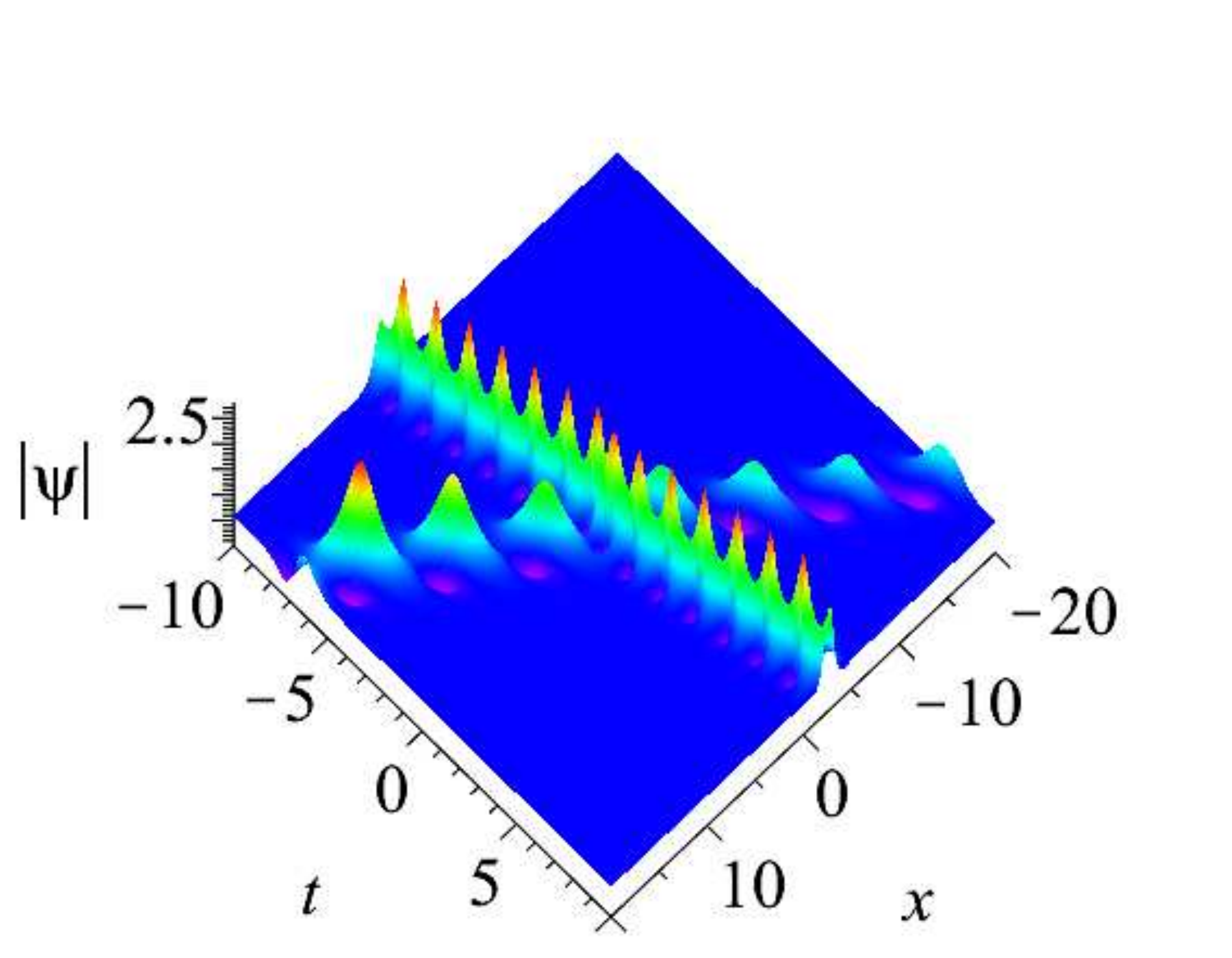}}}

$\qquad\qquad\qquad\qquad\textbf{(a)}
\qquad\qquad\qquad\qquad\qquad\qquad\qquad\textbf{(b)}
$\\

$~~~$
{\rotatebox{0}{\includegraphics[width=5.2cm,height=3.6cm,angle=0]{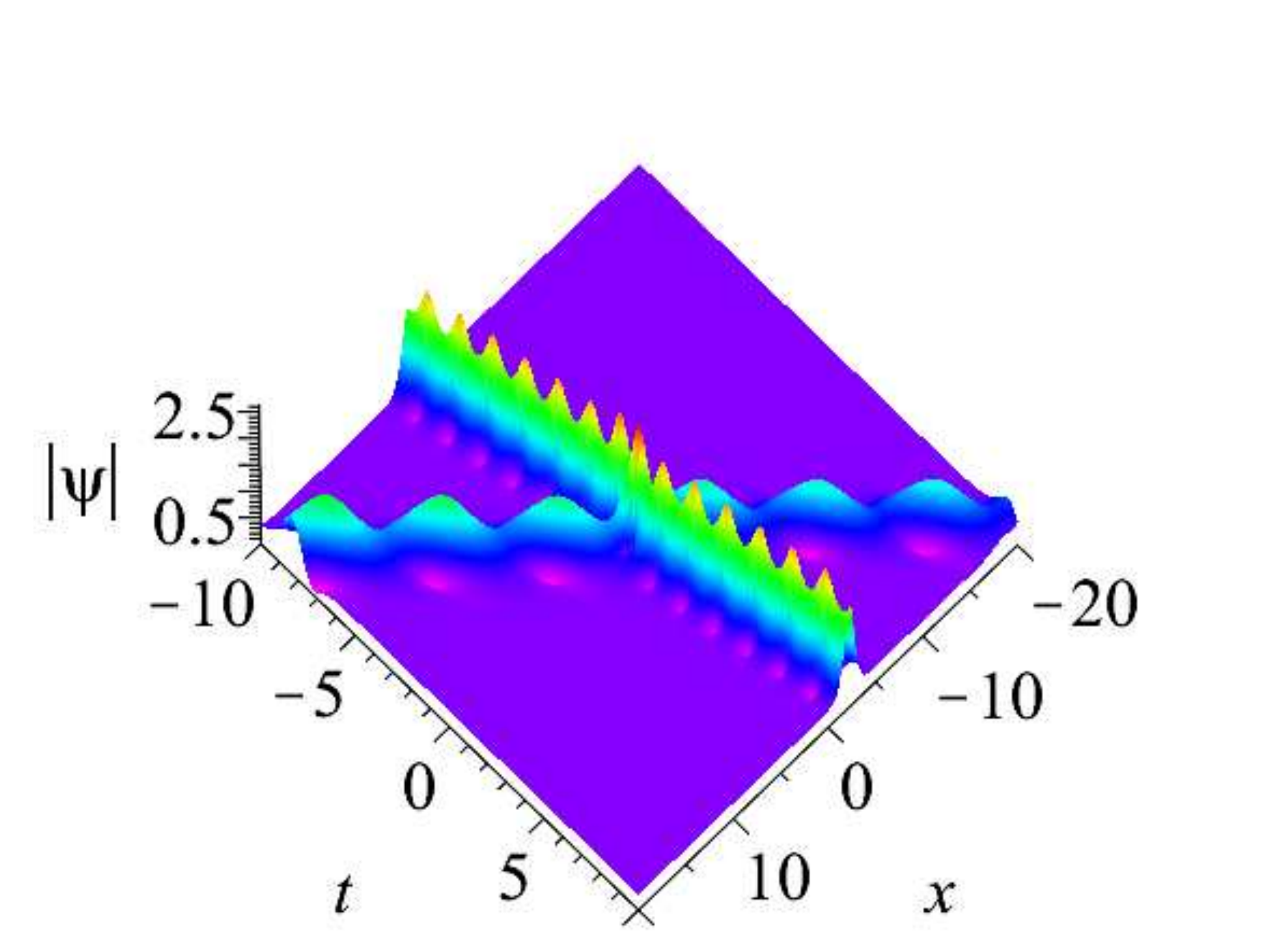}}}
~~~~~~~~~~~~~~~
{\rotatebox{0}{\includegraphics[width=5.2cm,height=3.6cm,angle=0]{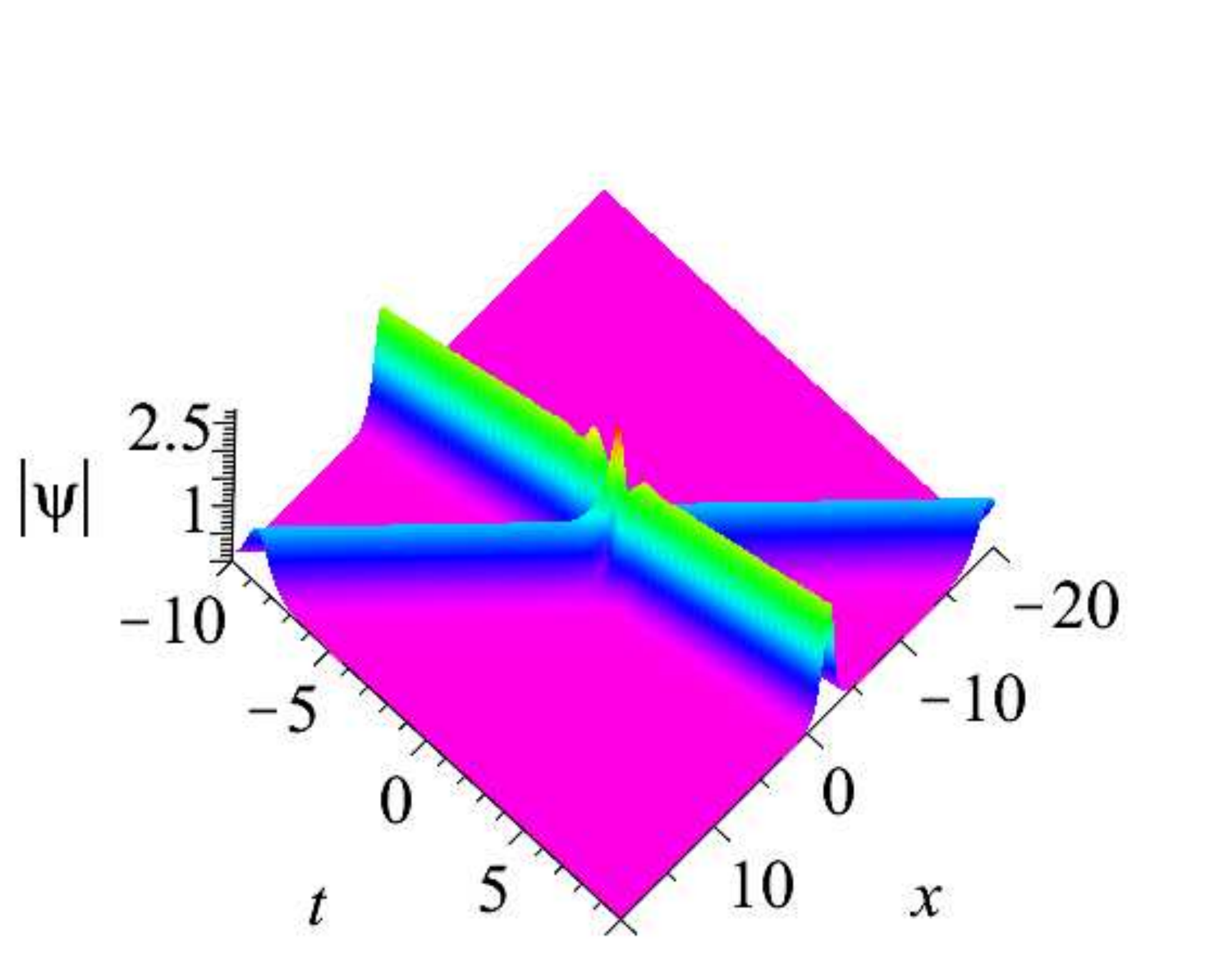}}}

$\qquad\qquad\qquad\qquad\textbf{(c)}
\qquad\qquad\qquad\qquad\qquad\qquad\qquad\textbf{(d)}
$\\
\noindent { \small \textbf{Figure 4.} (Color online) Breather waves via solution \eqref{JSP-35} with parameters
$N=2, z_{1}=0.2+2i, z_{2}=1+i, \alpha_{3}=\alpha_{4}=\alpha_{5}=0.01, A_{+}[z_{1}]=A_{+}[z_{2}]=1$:
$\textbf{(a)}$: breather-breather solutions with $\psi_{-}=1$;
$\textbf{(b)}$: breather-breather solutions with $\psi_{-}=0.6$;
$\textbf{(c)}$: breather-breather solutions with $\psi_{-}=0.3$;
$\textbf{(d)}$: bright-bright solitons with $\psi_{-}\rightarrow0$.\\}

$~~~$
{\rotatebox{0}{\includegraphics[width=5.2cm,height=3.6cm,angle=0]{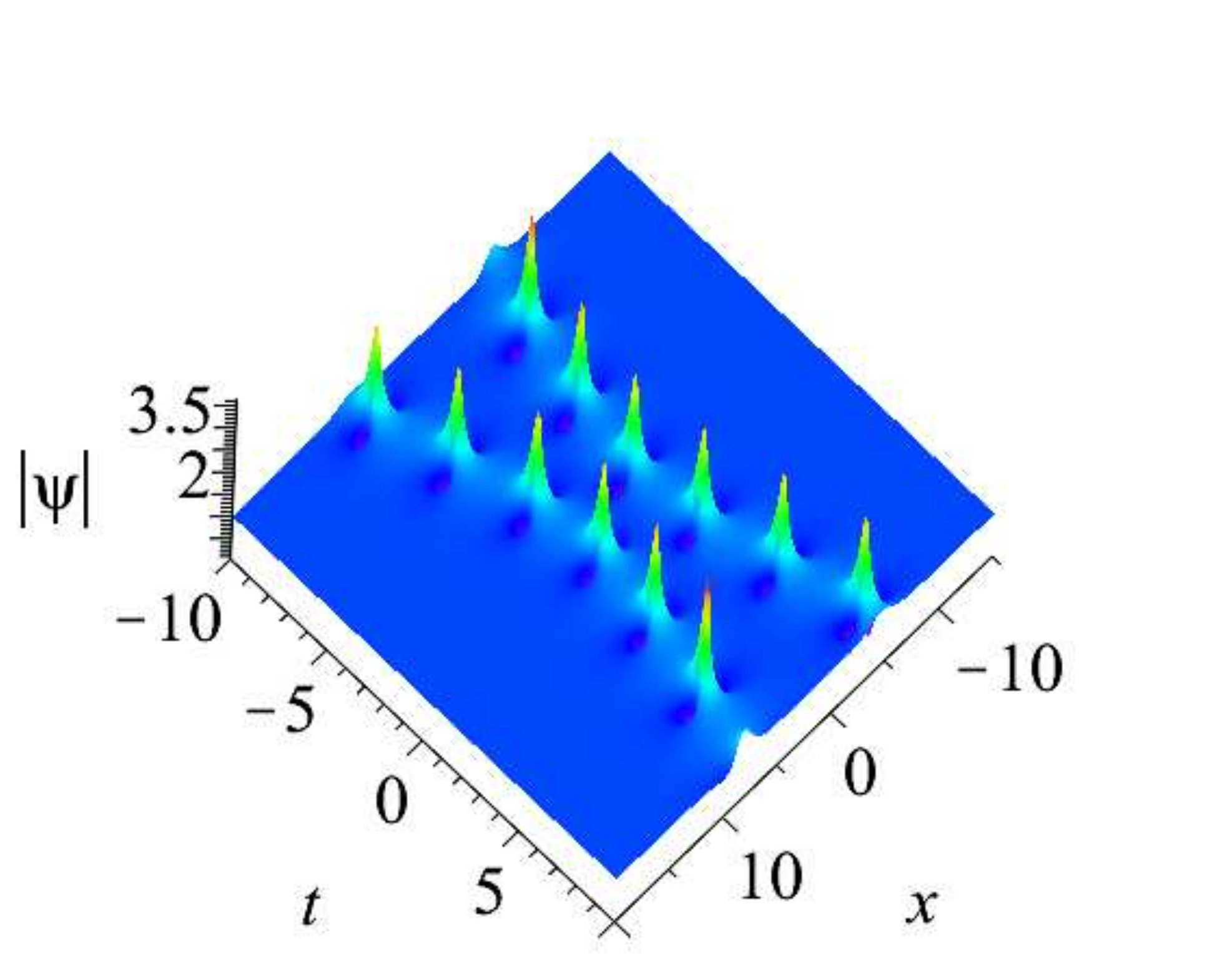}}}
~~~~~~~~~~~~~~~
{\rotatebox{0}{\includegraphics[width=5.2cm,height=3.6cm,angle=0]{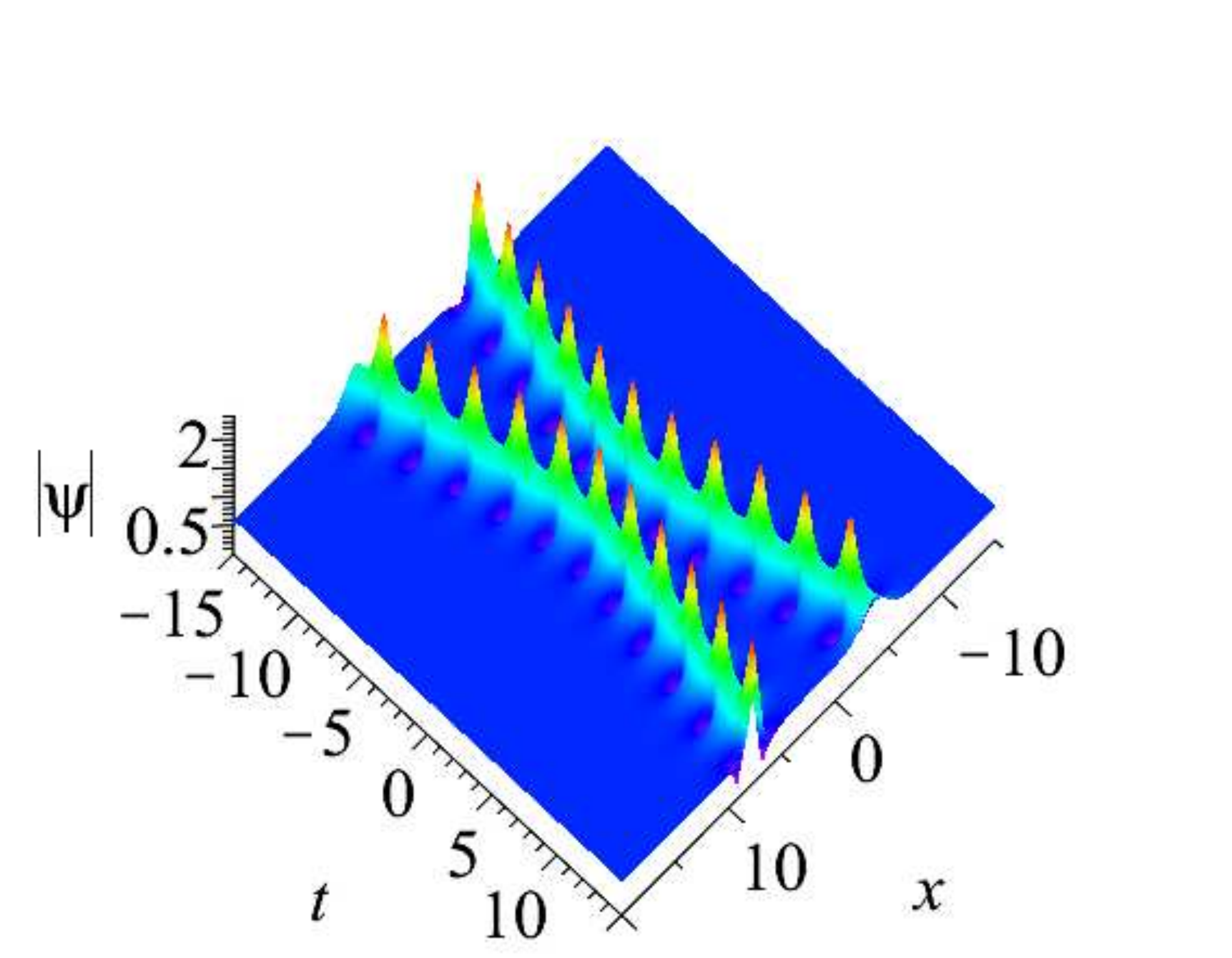}}}

$\qquad\qquad\qquad\qquad\textbf{(a)}
\qquad\qquad\qquad\qquad\qquad\qquad\qquad\textbf{(b)}
$\\

$~~~$
{\rotatebox{0}{\includegraphics[width=5.2cm,height=3.6cm,angle=0]{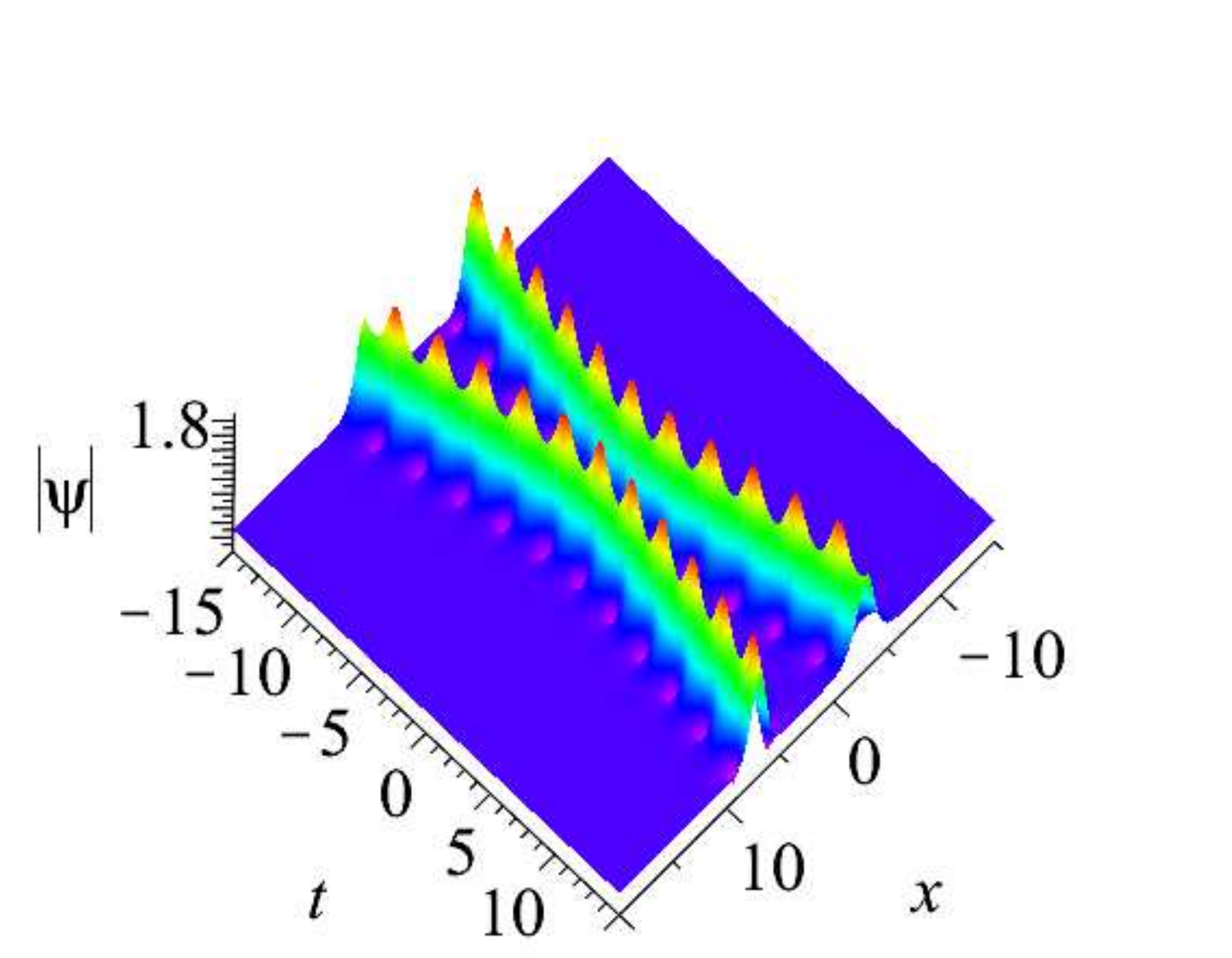}}}
~~~~~~~~~~~~~~~
{\rotatebox{0}{\includegraphics[width=5.2cm,height=3.6cm,angle=0]{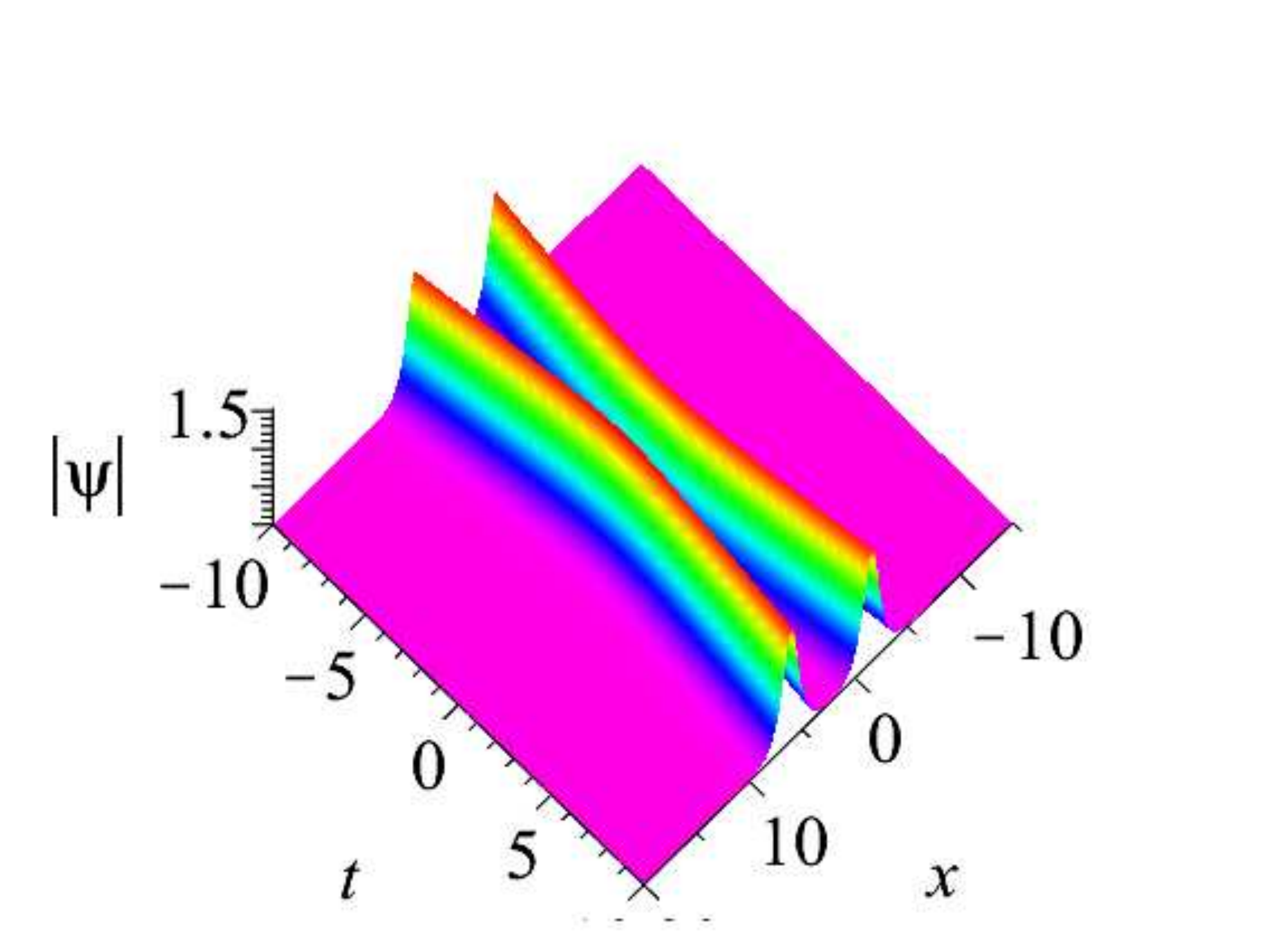}}}

$\qquad\qquad\qquad\qquad\textbf{(c)}
\qquad\qquad\qquad\qquad\qquad\qquad\qquad\textbf{(d)}
$\\
\noindent { \small \textbf{Figure 5.} (Color online) Breather waves via solution \eqref{JSP-35} with parameters
$N=2, z_{1}=0.1+1.5i, z_{2}=-0.1+1.5i, \alpha_{2}=1, \alpha_{3}=\alpha_{4}=\alpha_{5}=0.01,A_{+}[z_{1}]=A_{+}[z_{2}]=1$:
$\textbf{(a)}$: breather-breather solutions with $\psi_{-}=1$;
$\textbf{(b)}$: breather-breather solutions with $\psi_{-}=0.6$;
$\textbf{(c)}$: breather-breather solutions with $\psi_{-}=0.3$;
$\textbf{(d)}$: bright-bright solitons with $\psi_{-}\rightarrow0$.\\}

\noindent
{\rotatebox{0}{\includegraphics[width=4.0cm,height=3.2cm,angle=0]{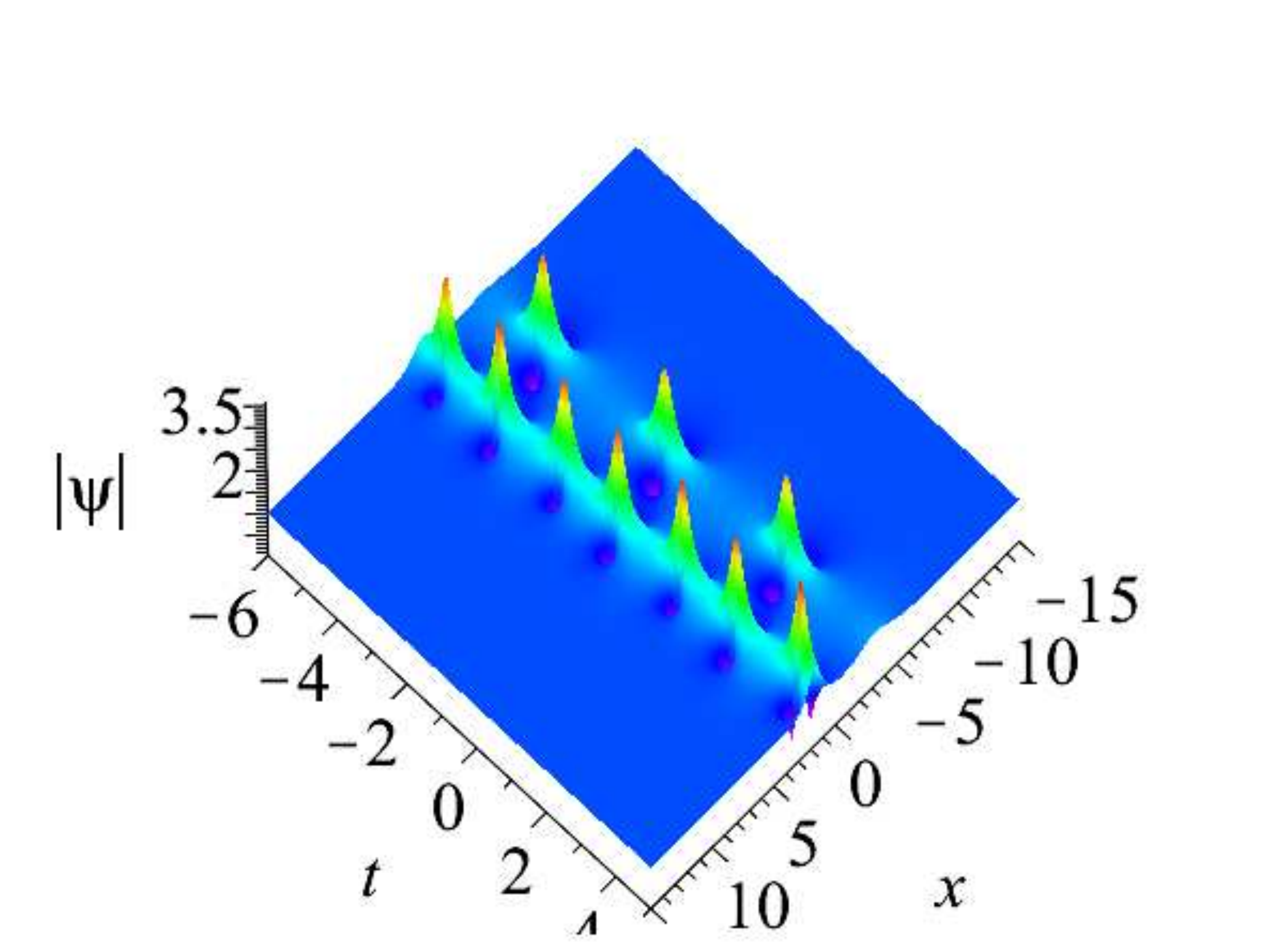}}}
{\rotatebox{0}{\includegraphics[width=4.0cm,height=3.2cm,angle=0]{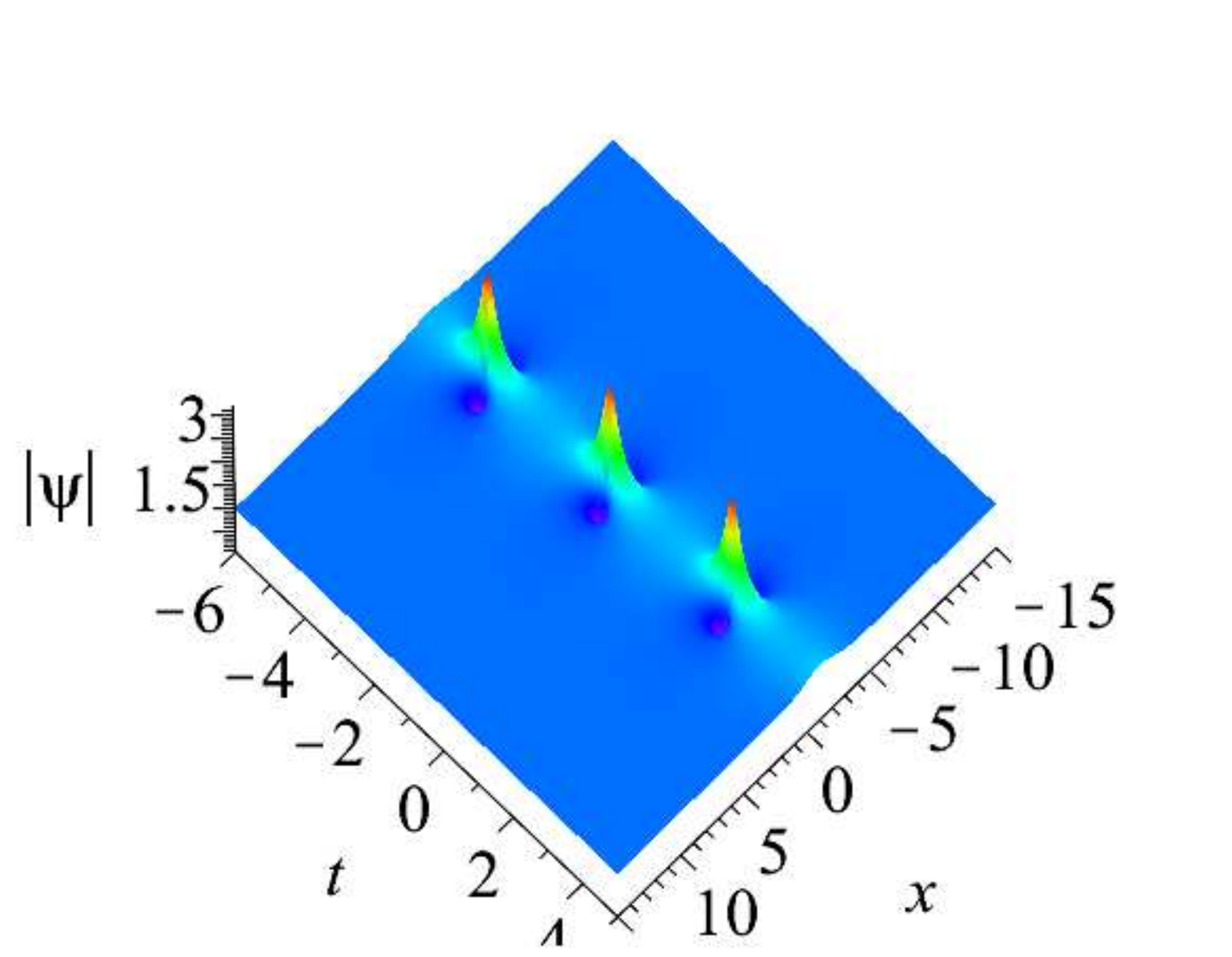}}}
{\rotatebox{0}{\includegraphics[width=4.0cm,height=3.2cm,angle=0]{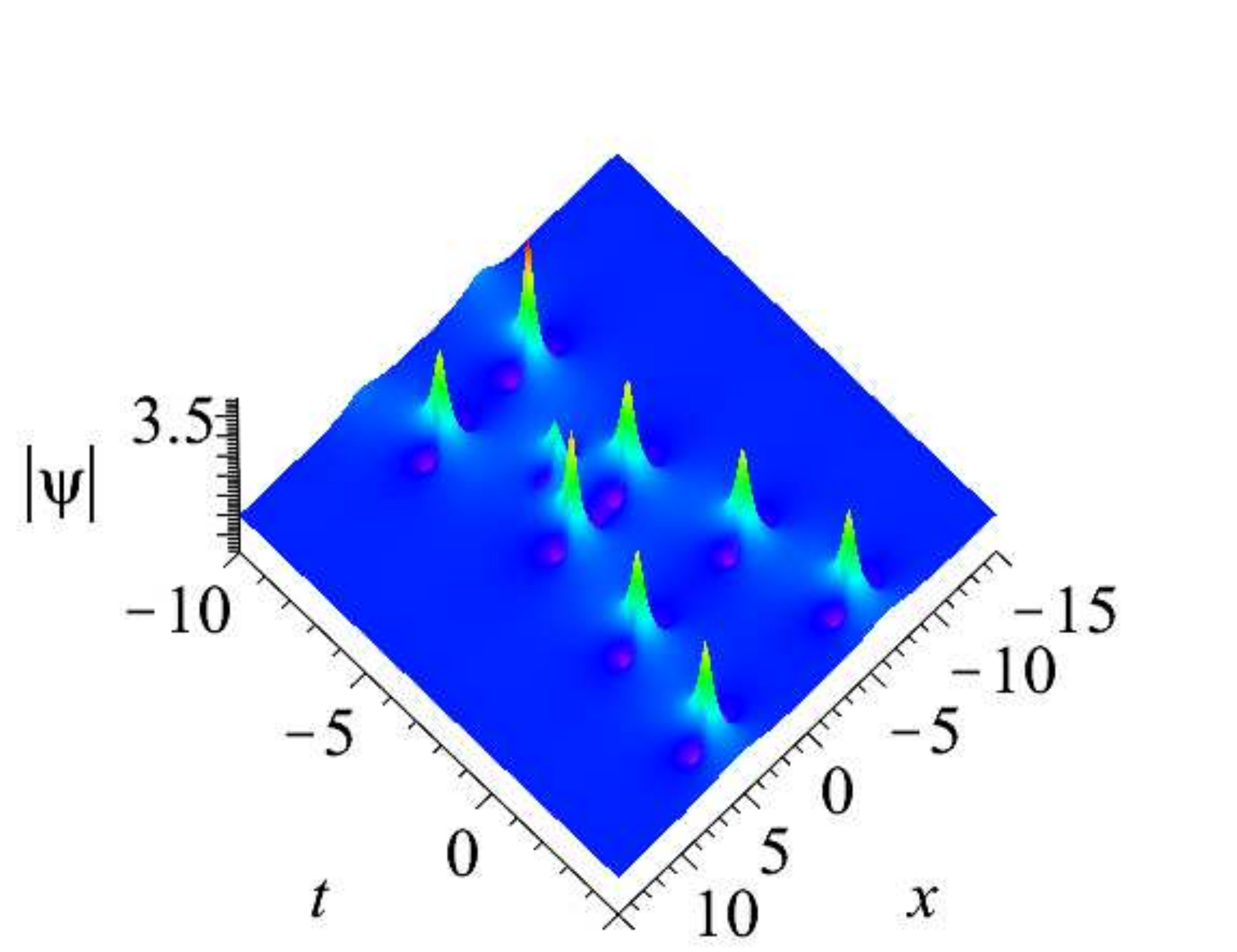}}}

$\qquad\qquad\textbf{(a)}\qquad\qquad\qquad\qquad\qquad\textbf{(b)}
\qquad\qquad\qquad\qquad\qquad\textbf{(c)}$\\

%
\noindent { \small \textbf{Figure 6.} (Color online) Breather waves via solution \eqref{JSP-35} with parameters
$N=2, \alpha_{2}=1, \alpha_{3}=0.01, \alpha_{4}=\alpha_{5}=0.001,A_{+}[z_{1}]=A_{+}[z_{2}]=1$:
$\textbf{(a)}$: breather-breather solutions with $z_{1}=0.5i, z_{2}=1.5i$;
$\textbf{(b)}$: breather solutions with $z_{1}=1.5i, z_{2}=1.5i,$;
$\textbf{(c)}$: breather-breather solutions with $z_{1}=0.1+1.5i, z_{2}=-0.1-1.5i$.\\}

$~~~$
{\rotatebox{0}{\includegraphics[width=5.2cm,height=3.6cm,angle=0]{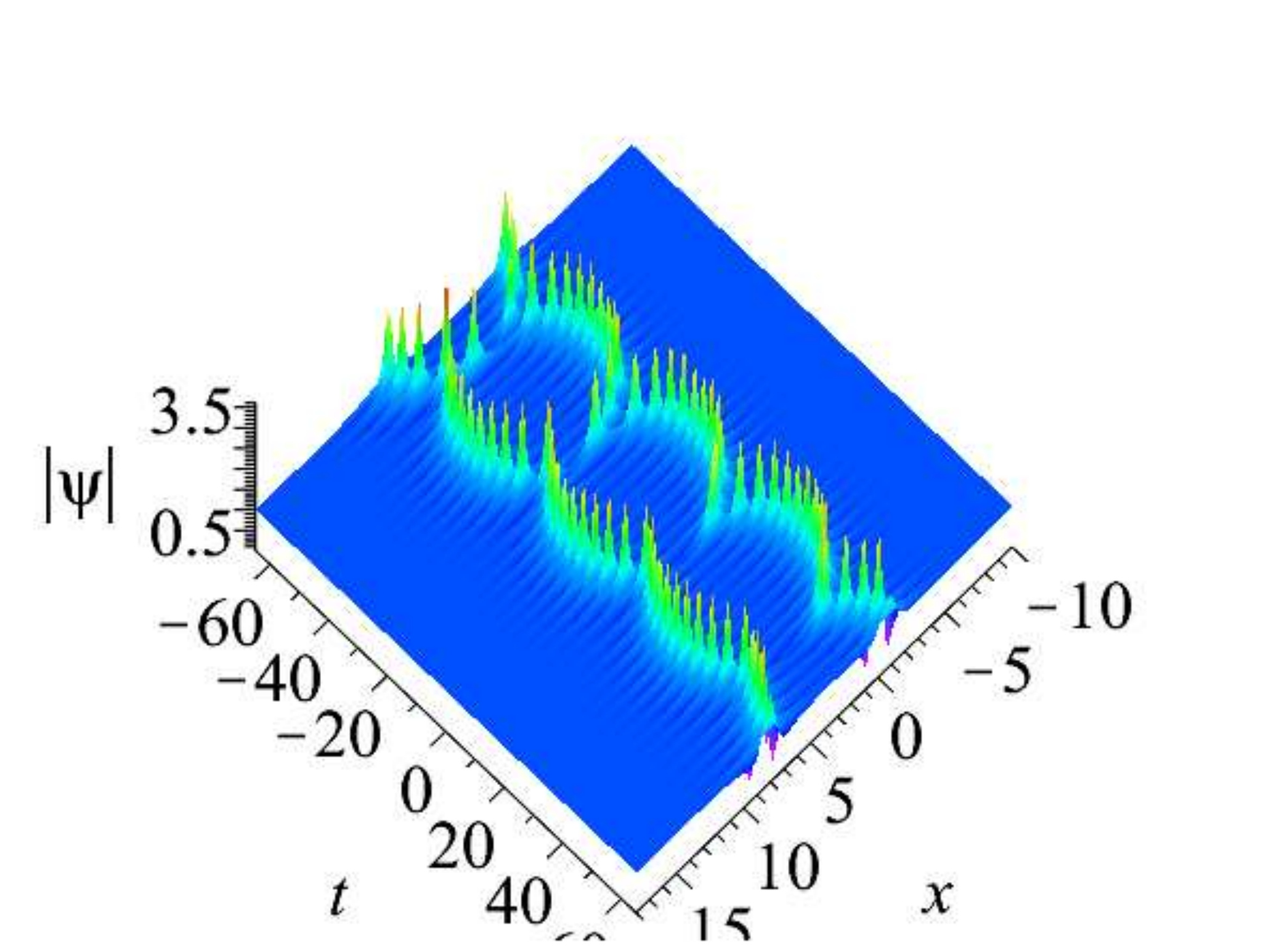}}}
~~~~~~~~~~~~~~~
{\rotatebox{0}{\includegraphics[width=5.2cm,height=3.6cm,angle=0]{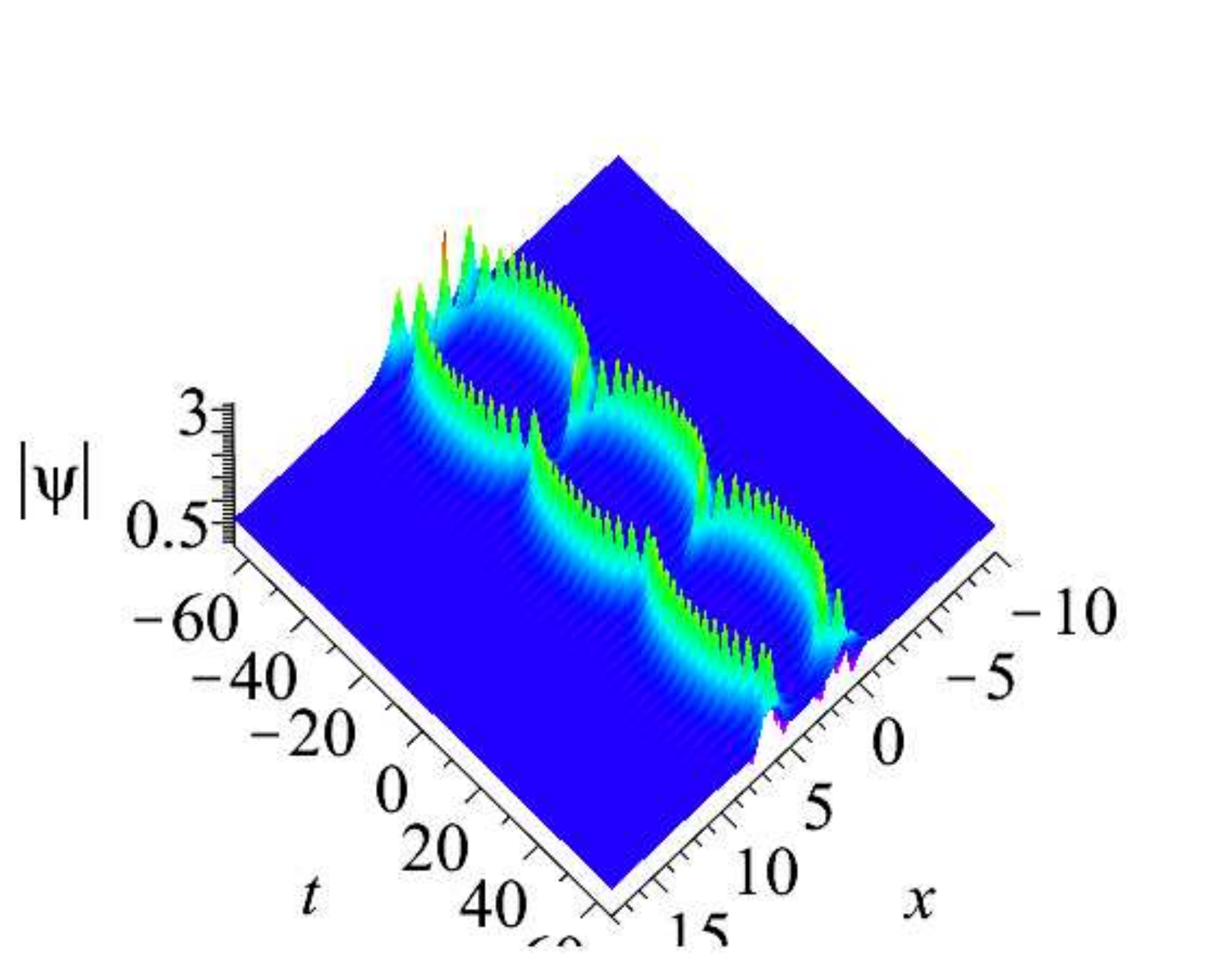}}}

$\qquad\qquad\qquad\qquad\textbf{(a)}
\qquad\qquad\qquad\qquad\qquad\qquad\qquad\textbf{(b)}
$\\

$~~~$
{\rotatebox{0}{\includegraphics[width=5.2cm,height=3.6cm,angle=0]{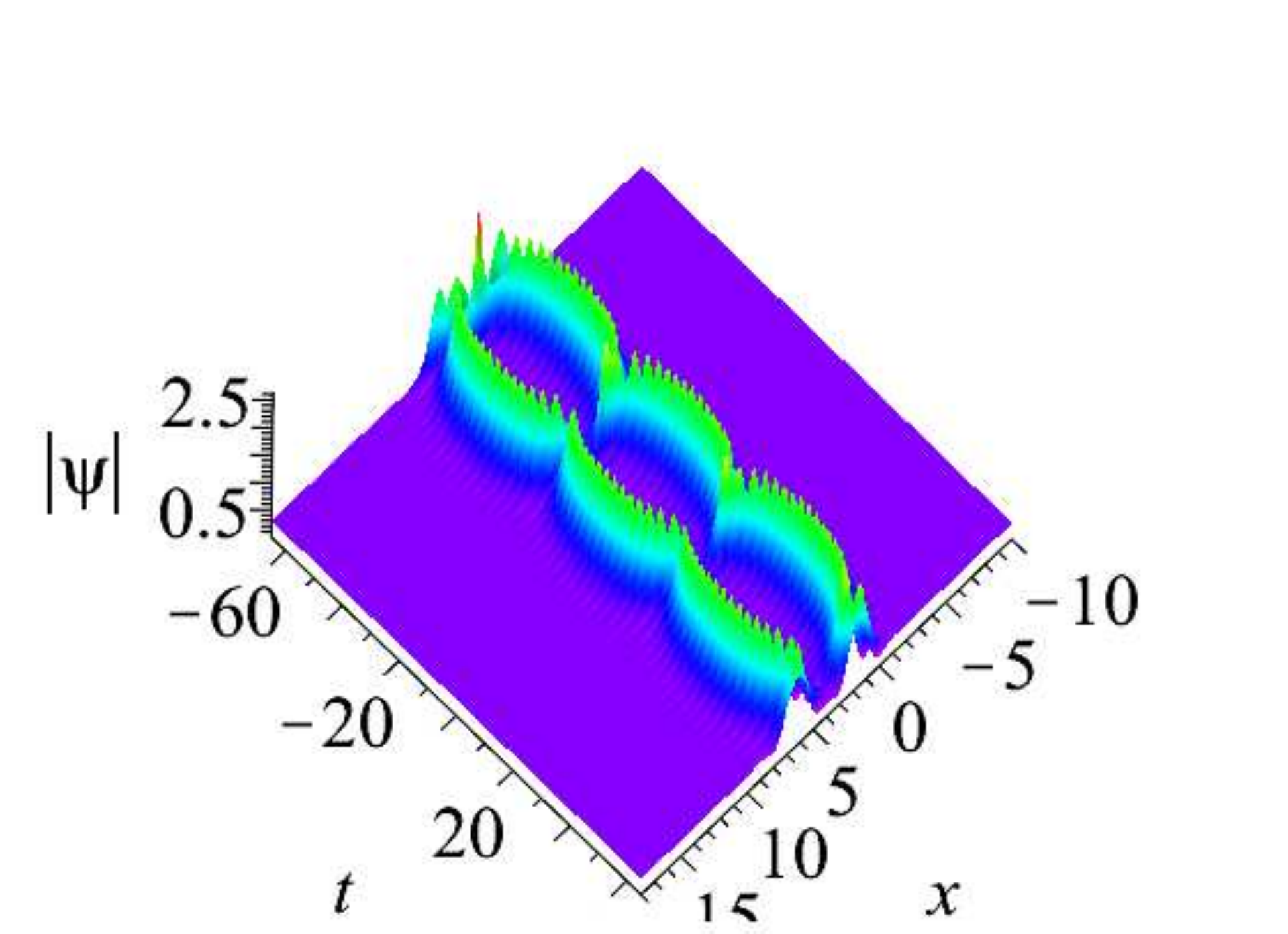}}}
~~~~~~~~~~~~~~~
{\rotatebox{0}{\includegraphics[width=5.2cm,height=3.6cm,angle=0]{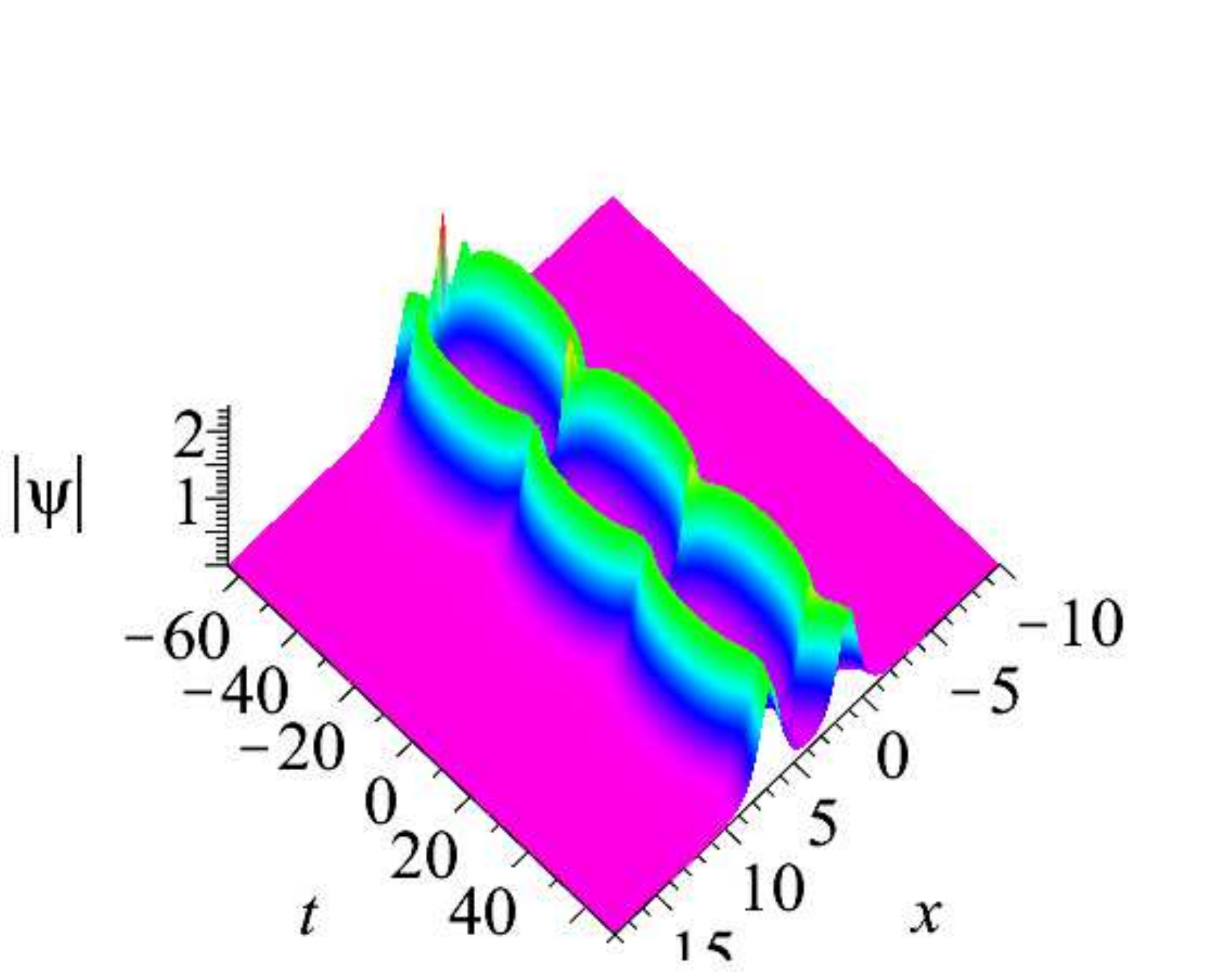}}}

$\qquad\qquad\qquad\qquad\textbf{(c)}
\qquad\qquad\qquad\qquad\qquad\qquad\qquad\textbf{(d)}
$\\
\noindent { \small \textbf{Figure 7.} (Color online) Breather waves via solution \eqref{JSP-35} with parameters
$N=2, z_{1}=1/65+1.05i, z_{2}=-1/65+1.05i, \alpha_{2}=1, \alpha_{3}=\alpha_{4}=\alpha_{5}=0.01,A_{+}[z_{1}]=A_{+}[z_{2}]=1$:
$\textbf{(a)}$: breather-breather solutions with $\psi_{-}=1$;
$\textbf{(b)}$: breather-breather solutions with $\psi_{-}=0.6$;
$\textbf{(c)}$: breather-breather solutions with $\psi_{-}=0.3$;
$\textbf{(d)}$: bright-bright solitons with $\psi_{-}\rightarrow0$.\\}

\section{The GFONLS equation with NZBCs: Double poles}
In the section, we assume that the discrete spectral points $Z^{f}$ are double zeros of the scattering coefficients
$s_{11}(z)$ and $s_{22}(z)$, i.e., $s_{11}(z_{0})=s'_{11}(z_{0})\neq 0$ for $\forall z_{0}\in Z^{f}\bigcap D_{+}^{f}$,
and $s_{22}(z_{0})=s'_{22}(z_{0})=0$, $s''_{22}(z_{0})\neq0$ for $\forall z_{0}\in Z^{f}\bigcap D_{-}^{f}$.
For convenience's sake, we present a simple proposition \cite{MP-2014}: If $f(z)$ and $g(z)$ are analytic in some complex domain $\Omega$,
and $z_{0}\in\Omega$ represents a double zero of $g(z)$ and $f(z_{0})\neq0$,
then the function $f(z)/g(z)$ admits the double pole $z=z_{0}$,
and its residue $\mbox{Res}[f/g]$ and the coefficient $P_{-2}[f/g]$ of $(z-z_{0})^{-2}$ in the Laurent series are expressed by
\begin{align}\label{DP-1}
\mathop{P_{-2}}\limits_{z=z_{0}}\left[\frac{f}{g}\right]=2f(z_{0})/g''(z_{0}),
~~\mathop{\mbox{Res}}\limits_{z=z_{0}}\left[\frac{f}{g}\right]
=2\left(\frac{f'(z_{0})}{g''(z_{0})}-\frac{f(z_{0}g'''(z_{0}))}{3[g''(z_{0}]^2}\right).
\end{align}
Similar to \cite{MP-2014}, for $s_{11}(z_{0})=s'_{11}(z_{0})=0$, $s'_{11}(z_{0})\neq0$ in $\forall z_{0}\in Z^{f}\bigcap D_{+}^{f}$.
Eq.\eqref{ISP-6} still holds. The first expression of \eqref{Lax-15} can be rewritten as
\begin{equation}\label{DP-2}
s_{11}(z)\gamma_{f}(z)=\left|\phi_{+1}(x,t,z),\phi_{-2}(x,t,z)\right|,
\end{equation}
whose first-order partial derivative with respect to $z$ reads
\begin{equation}\label{DP-3}
[s_{11}(z)\gamma_{f}(z)]'=\left|\phi'_{+1}(x,t,z),\phi_{-2}(x,t,z)\right|+\left|\phi_{+1}(x,t,z),\phi'_{-2}(x,t,z)\right|.
\end{equation}
Choosing $z=z_{0}\in Z^{f}\bigcap D_{+}^{f}$ in \eqref{DP-3} and utilizing $s_{11}(z_{0})=s'_{11}(z_{0})=0$ and \eqref{ISP-8} can lead to
\begin{equation}\label{DP-4}
\left|\phi'_{+1}(x,t,z_{0})-b_{+}(z_{0})\phi'_{-2}(x,t,z_{0}),\phi_{-2}(x,t,z_{0})\right|=0,
\end{equation}
which means that there exists another constant $c_{+}(z_{0})$ such that
\begin{equation}\label{DP-5}
\phi'_{+1}(x,t,z_{0})=c_{+}(z_{0})\phi_{-2}(x,t,z_{0})+b_{+}(z_{0})\phi'_{-2}(x,t,z_{0}).
\end{equation}
It follows from \eqref{ISP-8}, \eqref{DP-1} and \eqref{DP-5} that
\begin{align}\label{DP-6}
&\mathop{P_{-2}}\limits_{z=z_{0}}\left[\frac{\phi_{+}(x,t,z)}{s_{11}(z)}\right]
=\frac{2\phi_{+1}(x,t,z_{0})}{s''_{11}(z_{0})}=\frac{2b_{+}(z_{0})}{s''_{11}(z_{0})}\phi_{-2}(x,t,z_{0})=A_{+}\left[z_{0}\right]\phi_{-2}(x,t,z_{0}),
\notag\\
&\mathop{\mbox{Res}}\limits_{z=z_{0}}\left[\frac{\phi_{+1}(x,t,z)}{s_{11}(z)}\right]=\frac{2\phi'_{+1}(x,t,z_{0})}{s''_{11}(z_{0})}
-\frac{2\phi_{+1}(x,t,z_{0})s'''_{11}(z_{0})}{3(s''_{11}(z_{0}))^{2}}\notag\\
&\qquad\qquad\qquad\qquad=A_{+}\left[z_{0}\right]\left[\phi'_{-2}(x,t,z_{0})+B_{+}[z_{0}]\phi_{-2}(x,t,z_{0})\right].
\end{align}

Following a similar way, for $s_{22}(z^{*}_{0})=s'_{22}(z^{*}_{0})=0$, $s''_{22}(z^{*}_{0})\neq0$ in $\forall z^{*}_{0}\in Z^{f}\bigcap D_{-}^{f}$,
equation \eqref{ISP-8} holds.
According to the second one of \eqref{Lax-15} and \eqref{ISP-8}, we have
\begin{equation}\label{DP-7}
\phi'_{+2}(x,t,z^{*}_{0})=c_{-}(z^{*}_{0})\phi_{-1}(x,t,z^{*}_{0})+b_{-}(z^{*}_{0})\phi'_{-1}(x,t,z^{*}_{0})
\end{equation}
for $c_{-}(z^{*}_{0})$.

It follows from \eqref{ISP-8}, \eqref{DP-1} and \eqref{DP-7} that
\begin{equation}\label{DP-8}
\left\{ \begin{aligned}
&\mathop{P_{-2}}\limits_{z=z^{*}_{0}}\left[\frac{\phi_{+2}(x,t,z)}{s_{22}(z)}\right]=\frac{2\phi_{+2}(x,t,z^{*}_{0})}{s''_{22}(z^{*}_{0})}
=\frac{2b_{-}(z^{*}_{0})}{s''_{22}(z^{*}_{0})}\phi_{-1}(x,t,z^{*}_{0})=A_{-}[z^{*}_{0}]\phi_{-1}(x,t,z^{*}_{0}),\\
&\mathop{\mbox{Res}}\limits_{z=z^{*}_{0}}\left[\frac{\phi_{+2}(x,t,z)}{s_{22}(z)}\right]=A_{-}[z^{*}_{0}]\left[\phi'_{-1}(x,t,z^{*}_{0})
+B_{-}[z^{*}_{0}]\phi_{-1}(x,t,z^{*}_{0})\right].
           \end{aligned} \right.
\end{equation}
We thus know
\begin{equation}\label{DP-9}
\left\{ \begin{aligned}
&A_{+}[z_{0}]=\frac{2b_{+}[z_{0}]}{s''_{11}(z_{0})},~~B_{+}[z_{0}]=\frac{c_{+}[z_{0}]}{b_{+}[z_{0}]}-\frac{s'''_{11}(z_{0})}{3s''_{11}(z_{0})},~~
z_{0}\in Z^{f}\cap D_{+}^{f},\\
&A_{+}[z^{*}_{0}]=\frac{2b_{+}[z^{*}_{0}]}{s''_{11}(z^{*}_{0})},~~B_{+}[z^{*}_{0}]=\frac{c_{+}[z^{*}_{0}]}
{b_{+}[z^{*}_{0}]}-\frac{s'''_{11}(z^{*}_{0})}{3s''_{11}(z^{*}_{0})},~~
z^{*}_{0}\in Z^{f}\cap D_{-}^{f},
           \end{aligned} \right.
\end{equation}
from which we obtain the following symmetries
\begin{align}\label{DP-10}
&A_{+}[z_{0}]=-A^{*}_{-}[z^{*}_{0}]=\frac{z_{0}^4\psi^{*}_{-}}{\psi_{0}^4\psi_{-}}A_{-}\left[-\frac{\psi_{0}^2}{z_{0}}\right],\notag\\
&B_{+}[z_{0}]=B^{*}_{-}[z^{*}_{0}]=\frac{\psi_{0}^2}{z_{0}^2}B_{-}\left[-\frac{\psi_{0}^2}{z_{0}}\right]+\frac{2}{z_{0}},~~z_{0}\in Z^{f}\cap D_{+}^{f},
\end{align}
which arrive at
\begin{equation}\label{DP-11}
\left\{ \begin{aligned}
&A_{+}[z_{n}]=-A^{*}_{-}[z^{*}_{n}]=-\frac{z_{n}^{4}\psi^{*}_{-}}{\psi_{0}^4\psi_{-}}A_{+}\left[-\frac{\psi_{0}^2}{z^{*}_{n}}\right],~~~~z_{n}\in Z^{f}\cap D_{+}^{f},\\
&B_{+}[z_{n}]=B^{*}_{-}[z^{*}_{n}]=\frac{\psi_{0}^2}{z_{n}^2}B^{*}_{+}\left[-\frac{\psi_{0}^2}{z^{*}_{n}}\right]+\frac{2}{z_{n}},~~~~~z^{*}_{n}\in Z^{f}\cap D_{-}^{f}.
           \end{aligned} \right.
\end{equation}
To sum up, we have
\begin{equation}\label{DP-12}
\left\{ \begin{aligned}
&\mathop{P_{-2}}\limits_{z=\xi_{n}} M_{1}^{+}(x,t,z)=P_{-2}\left[\frac{\mu_{+1}(x,t,z)}{s_{11}(z)}\right]=A[\xi_{n}]e^{-2i\theta(x,t,\xi_{n})}\mu_{-2}(x,t,\xi_{n}),\\
&\mathop{P_{-2}}\limits_{z=\widehat{\xi}_{n}}M_{2}^{-}(x,t,z)=P_{-2}\left[\frac{\mu_{+2}(x,t,z)}{s_{22}(z)}\right]=A\left[\widehat{\xi}_{n}\right]
e^{2i\theta(x,t,\widehat{\xi}_{n})}\mu_{-1}(x,t,\xi_{n}),\\
&\mathop{\mbox{Res}}\limits_{z=\xi_{n}}M_{1}^{+}(x,t,z)=\mbox{Res}\left[\frac{\mu_{+1}(x,t,z)}{s_{11}(z)}\right]\\
&=A\left[\xi_{n}\right]e^{-2i\theta(x,t,\xi_{n})}\left\{\mu'_{-2}(x,t,\xi_{n})
+\left[B\left[\xi_{n}\right]-2i\theta'(x,t,\xi_{n})\right]\mu_{-2}(x,t,\xi_{n})\right\},\\
&\mathop{\mbox{Res}}\limits_{z=\widehat{\xi}_{n}}M_{2}^{-}(x,t,z)=\mbox{Res}\left[\frac{\mu_{+2}(x,t,z)}{s_{22}(z)}\right]\\
&=A\left[\widehat{\xi}_{n}\right]e^{2i\theta\left(x,t,\widehat{\xi}_{n}\right)}\left\{\mu'_{-1}\left(x,t,\widehat{\xi}_{n}\right)
+\left[B\left[\widehat{\xi}_{n}\right]
+2i\theta'\left(x,t,\widehat{\xi}_{n}\right)\right]\mu_{-1}\left(x,t,\widehat{\xi}_{n}\right)\right\}.
           \end{aligned} \right.
\end{equation}
The RHP in Proposition 3.1 still holds in the case of double poles. To solve this kind of RHP, we must subtract out the asymptotic values as $z\rightarrow\infty$ and $z\rightarrow0$ and the singularity contributions
\begin{align}\label{DP-13}
&M_{dp}(x,t,z)=I+\frac{i}{z}\sigma_{3}Q_{-}+\sum_{n=1}^{2N}M_{dp}^{n},\notag\\
&M_{dp}^{n}=\frac{\mathop{P_{-2}}\limits_{z=\xi_{n}}M^{+}}{(z-\xi_{n})^2}
+\frac{\mathop{P_{-2}}\limits_{z=\widehat{\xi}_{n}}M^{-}}{\left(z-\widehat{\xi}_{n}\right)^2}
+\frac{\mathop{\mbox{Res}}\limits_{z=\xi_{n}} M^{+}}{z-\xi_{n}}+\frac{\mathop{\mbox{Res}}\limits_{z=\widehat{\xi}_{n}} M^{+}}{z-\widehat{\xi}_{n}}.
\end{align}
Then it follows from the jump condition $M^{-}=M^{+}(I-J)$ that
\begin{equation}\label{DP-14}
M^{-}(x,t,z)-M_{dp}(x,t,z)=M^{+}(x,t,z)-M_{dp}(x,t,z)-M^{+}(x,t,z)J,
\end{equation}
where $M^{\pm}(x,t,z)-M_{dp}(x,t,z)$ are analytic in $D_{\pm}^{f}$.
Additionally, their asymptotics are both $O(1/z)$ as $z\rightarrow\infty$ and $O(1)$ as $z\rightarrow0$,
and $J(x,t,z)$ is $O(1/z)$ as $z\rightarrow\infty$, and $O(z)$ as $z\rightarrow0$.
Consequently, the Cauchy projectors and Plemelj's formulae are used to solve
\eqref{DP-14} to generate
\begin{equation}\label{DP-15}
M(x,t,z)=M_{dp}(x,t,z)+\frac{1}{2\pi i}\int_{\Sigma^{f}}\frac{M^{+}(x,t,\zeta)J(x,t,\zeta)}{\zeta-z}d\zeta,~~z\in\mathbb{C}\backslash\Sigma^{f},
\end{equation}
where $\int_{\Sigma^{f}}$ stands for the integral along the oriented contours seen in Fig.1 (right).

Thus, by using \eqref{DP-12}, the parts of $P_{-2}(\cdot)$  and $\mbox{Res}(\cdot)$ in \eqref{DP-15} can be expressed as
\begin{align}\label{DP-16}
&M_{dp}^{n}=\notag\\
&\left(C_{n}(z)\left[\mu'_{-2}(\xi_{n})+\left(D_{n}+\frac{1}{z-\xi_{n}}\right)\mu_{-2}(\xi_{n})\right],
\widehat{C}_{n}(z)\left[\mu'_{-1}\left(\widehat{\xi}_{n}\right)+\left(D_{n}+\frac{1}{z-\widehat{\xi}_{n}}\right)\mu_{-2}(\widehat{\xi}_{n})\right]\right),
\end{align}
where
\begin{align}\label{DP-17}
&C_{n}(z)=\frac{A_{+}\left[\xi_{n}\right]}{z-\xi_{n}}e^{-2i\theta\left(\xi_{n}\right)},~~D_{n}=B_{+}\left[\xi_{n}\right]-2i\theta'\left(\xi_{n}\right),\\
&\widehat{C}_{n}(z)=\frac{A_{-}\left[\widehat{\xi}_{n}\right]}{z-\widehat{\xi}_{n}}e^{2i\theta\left(\widehat{\xi}_{n}\right)},
~~\widehat{D}_{n}=B_{-}\left(\widehat{\xi}_{n}\right)+2i\theta'\left(\widehat{\xi}_{n}\right).
\end{align}
To obtain $\mu'_{-2}(\xi_{n})$, $\mu_{-2}(\xi_{n})$, $\mu'_{-1}$, and $\mu_{-1}(\xi_{n})$ in \eqref{DP-16},
for $z=\xi_{s}(s=1,2,\ldots,2N)$.
According to the second column of $M(x,t,\lambda)$ presented by \eqref{DP-15} and \eqref{DP-16}, we have
\begin{align}\label{DP-18}
&\mu_{-2}(z)=\notag\\&\left(
              \begin{array}{c}
                \frac{i\psi_{-}}{z} \\
                1 \\
              \end{array}
            \right)+\sum_{n=1}^{2N}\widehat{C}_{n}(z)\left[\mu'_{-1}\left(\widehat{\xi}_{n}\right)
            +\left(\widehat{D}_{n}+\frac{1}{z-\widehat{\xi}_{n}}\right)\mu_{-1}\left(\widehat{\xi}_{n}\right)\right]+\frac{1}{2\pi i}\int_{\Sigma^{f}}\frac{(M^{+}J)_{2}(\zeta)}{\zeta-z}d\zeta,
\end{align}
whose first-order derivative with respect to $z$ can lead to
\begin{align}\label{DP-19}
&\mu'_{-2}(z)=\notag\\&\left(
              \begin{array}{c}
                -\frac{i\psi_{-}}{z^2} \\
                1 \\
              \end{array}
            \right)-\sum_{n=1}^{2N}\frac{\widehat{C}_{n}(z)}{z-\widehat{\xi}_{n}}\left[\mu'_{-1}\left(\widehat{\xi}_{n}\right)
            +\left(\widehat{D}_{n}+\frac{1}{z-\widehat{\xi}_{n}}\right)\mu_{-1}\left(\widehat{\xi}_{n}\right)\right]+\frac{1}{2\pi i}\int_{\Sigma^{f}}\frac{(M^{+}J)_{2}(\zeta)}{(\zeta-z)^2}d\zeta.
\end{align}
Additionally, in view of \eqref{ISP-1}, we obtain
\begin{equation}\label{DP-20}
\mu'_{-2}(z)=-\frac{i\psi_{-}}{z^2}\mu_{-}\left(\frac{-\psi_{0}^2}{z}\right)+\frac{i\psi_{0}^2}{z}\frac{\psi_{-}}{z^2}\mu'_{-1}\left(-\frac{\psi_{0}^2}{z}\right).
\end{equation}
Then plugging \eqref{DP-20} into \eqref{DP-18} and \eqref{DP-19} yields
\begin{align}\label{DP-21}
&\sum_{n=1}^{2N}\widehat{C}_{n}(\xi_{s})\mu'_{-1}\left(\widehat{\xi}_{n}\right)+\left[\widehat{C}_{n}(\xi_{k})\left(\widehat{D}_{n}
+\frac{1}{\xi_{s}-\widehat{\xi}_{n}}\right)-\frac{i\psi_{-}}{\xi_{s}}\delta_{sn}\right]\mu_{-1}\left(\widehat{\xi}_{n}\right)\notag\\
&~~~~~~~~~~~~~~=-\left(
    \begin{array}{c}
      \frac{iq_{-}}{\xi_{s}} \\
      1 \\
    \end{array}
  \right)-\frac{1}{2\pi i}\int_{\Sigma^{f}}\frac{\left(M^{+}J\right)_{2}(\zeta)}{\zeta-\xi_{k}}d\zeta,\notag\\
&\sum_{n=1}^{2N}\left(\frac{\widehat{C}_{n}(\xi_{s})}{\xi_{s}-\widehat{\xi}_{n}}+\frac{i\psi_{0}^2\psi_{-}}{\xi_{s}^3\delta_{sn}}\right)
\mu'_{-1}\left(\widehat{\xi}_{n}\right)
+\left[\frac{\widehat{C}_{n}(\xi_{s})}{\xi_{s}-\widehat{\xi}_{n}}\left(\widehat{D}_{n}+\frac{2}{\xi_{s}-\widehat{\xi}_{n}}\right)
-\frac{i\psi_{-}}{\xi_{s}^2}\delta_{sn}\right]
\mu_{-1}\left(\widehat{\xi}_{n}\right)\notag\\&~~~~~~~~~~~~~~=\left(
                               \begin{array}{c}
                                 -\frac{\psi_{-}}{\xi_{s}^2} \\
                                 0 \\
                               \end{array}
                             \right)+\int_{\Sigma^{f}}\frac{(M^{+}J)_{2}(\zeta)}{2\pi i(\zeta-\xi_{k})^2}d\zeta,
\end{align}
which can lead to $\mu_{-}\left(x,t,\widehat{\xi}_{n}\right)$, $\mu'_{-1}\left(x,t,\widehat{\xi}_{n}\right), n=1,2,\ldots,2N$
such that one can also find $\mu_{-2}(x,t,\xi_{n})$, $\mu'_{-2}(x,t,\xi_{n}), n=1,2,\ldots,2N$ from \eqref{DP-20}.
As a result, substituting them into \eqref{DP-16}, and then substituting \eqref{DP-16} into \eqref{DP-15}
can lead to the $M(x,t,z)$ in view of the scattering data.

From \eqref{DP-15} and \eqref{DP-16}, we see that the asymptotics of $M(x,t,z)$ is still of the form \eqref{JSP-27}.
However, $M^{(1)}(x,t)$ should be replaced by
\begin{align}\label{DP-22}
&M^{(1)}(x,t)=i\sigma_{3}Q_{-}-\frac{1}{2\pi i}\int_{\Sigma^{f}}(M^{+}J)(\zeta)d\zeta\notag\\
&+\sum_{n=1}^{2N}\left[A_{+}\left[\xi_{n}\right]e^{-2i\theta\left(\xi_{n}\right)}\left(\mu'_{-2}(\xi_{n})+D_{n}\mu_{-2}(\xi_{n})\right),
A_{-}\left[\widehat{\xi}_{n}\right]e^{2i\theta\left(\widehat{\xi}_{n}\right)}\left(\mu'_{-2}\left(\widehat{\xi}_{n}\right)
+\widehat{D}_{n}\mu_{-1}\left(\widehat{\xi}_{n}\right)\right)\right].
\end{align}

Summarizing the above results, the following proposition 4.1 for the potential $u(x,t)$ for the case of double poles holds.\\

\noindent
\textbf{Proposition 4.1.} The potential with double poles of the GFONLS equation \eqref{gtc-NLS} with NZBCs is expressed by
\begin{equation}\label{DP-16}
\psi(x,t)=\psi_{-}-i\sum_{n=1}^{2N}A_{-}\left[\widehat{\xi}_{n}\right]e^{2i\theta(\xi_{n})}
\left(\mu'_{-11}\left(\widehat{\xi}_{n}\right)+\widehat{D}_{n}\mu_{-11}\left(\widehat{\xi}_{n}\right)\right)
+\frac{1}{2\pi}\int_{\Sigma'}\left(M^{+}J\right)_{12}(\zeta)d\zeta,
\end{equation}
where $\widehat{C}_{n}(z)=\frac{A\left[\widehat{\xi}_{n}\right]}{z-\xi_{n}}e^{2i\theta\left(\widehat{\xi}_{n}\right)}$,
$\widehat{D}_{n}=B\left[\widehat{\xi}_{n}\right]+2i\theta'\left(\xi_{n}\right)$,
and $\mu_{-11}\left(\widehat{\xi}_{n}\right)$ and $\mu'_{-11}\left(\widehat{\xi}_{n}\right)$ are given by \eqref{DP-21}.

Similar to the case of simple poles, we also have the trace formulae in the case of double poles as
\begin{equation}\label{DP-17}
s_{11}(z)=e^{s(z)}s_{0}(z)~~\mbox{for}~~z\in D_{+}^{f},~~s_{22}(z)=e^{-s(z)}/s_{0}(z)~~\mbox{for}~~z\in D_{-}^{f},
\end{equation}
where
\begin{equation}\label{DP-18}
s(z)=-\frac{1}{2\pi i}\int_{\Sigma^{f}}\frac{\log\left[1+\rho(\zeta)\rho^{*}(\zeta^{*})\right]}{\zeta-z}d\zeta,~~
s_{0}(z)=\prod_{n=1}^{N}\frac{(z-z_{n})^2\left(z+\psi_{0}^2/z^{*}_{n}\right)^2}{(z-z^{*}_{n})^2\left(z+\psi_{0}^2/z_{n}\right)^2}.
\end{equation}
In terms of the limit $z\rightarrow0$ of $s_{11}(z)$ in \eqref{DP-17}, one can write the theta condition as
\begin{equation}\label{DP-19}
\mbox{arg}\left(\frac{\psi_{+}}{\psi_{-}}\right)=8\sum_{n=1}^{N}\mbox{arg}(z_{n})
+\int_{\Sigma^{f}}\frac{\log\left[1+\rho(\zeta)\rho^{*}(\zeta^{*})\right]}{\zeta-z}d\zeta.
\end{equation}
Particularly, for the reflectionless case, i.e, $\rho(z)=\hat{\rho}(z)=0$, the Theorem 1.2 holds.

Thus, the trace formulae and theta condition become
\begin{align}\label{DP-22}
&s_{11}(z)=\prod_{n=1}^{N}\frac{(z-z_{n})^2\left(z+\psi_{0}^2/z^{*}_{n}\right)^2}{(z-z^{*}_{n})^2\left(z+\psi_{0}^2/z_{n}\right)^2},~~z\in D_{+}^{f},\\
&s_{22}(z)=\prod_{n=1}^{N}\frac{(z-z^{*}_{n})^2\left(z+\psi_{0}^2/z_{n}\right)^2}{(z-z_{n})^2\left(z+\psi_{0}^2/z^{*}_{n}\right)^2},~~z\in D_{-}^{f},
\end{align}
and
\begin{equation}\label{DP-23}
\mbox{arg}\left(\frac{\psi_{+}}{\psi_{-}}\right)=\mbox{arg}(\psi_{+})-\mbox{arg}(\psi_{-})=8\sum_{n=1}^{N}\mbox{arg}(z_{n}),
\end{equation}
respectively.

In the following, the double-pole breather-breather solutions of the GFONLS equation \eqref{gtc-NLS} with NZBCs are shown in Figs.8-10,
which can help us understand the propagation properties of the breather solutions.
More importantly, when $\psi_{-}\rightarrow0$, we have the double-pole bright-bright soliton solutions of the GFONLS equation \eqref{gtc-NLS}
(see Figs.9(d) and 10(d)).

\noindent
{\rotatebox{0}{\includegraphics[width=4.0cm,height=3.2cm,angle=0]{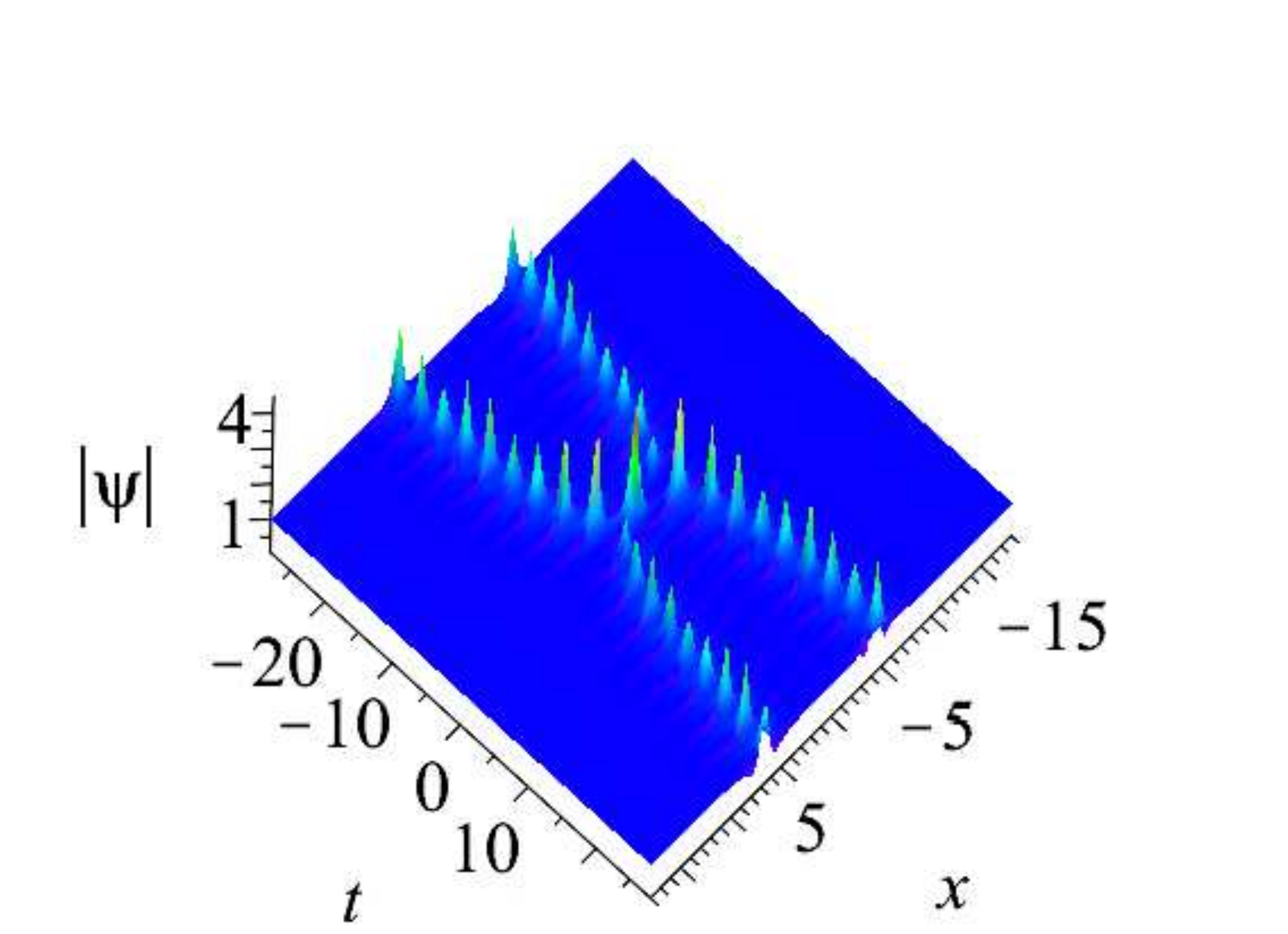}}}
{\rotatebox{0}{\includegraphics[width=4.0cm,height=3.2cm,angle=0]{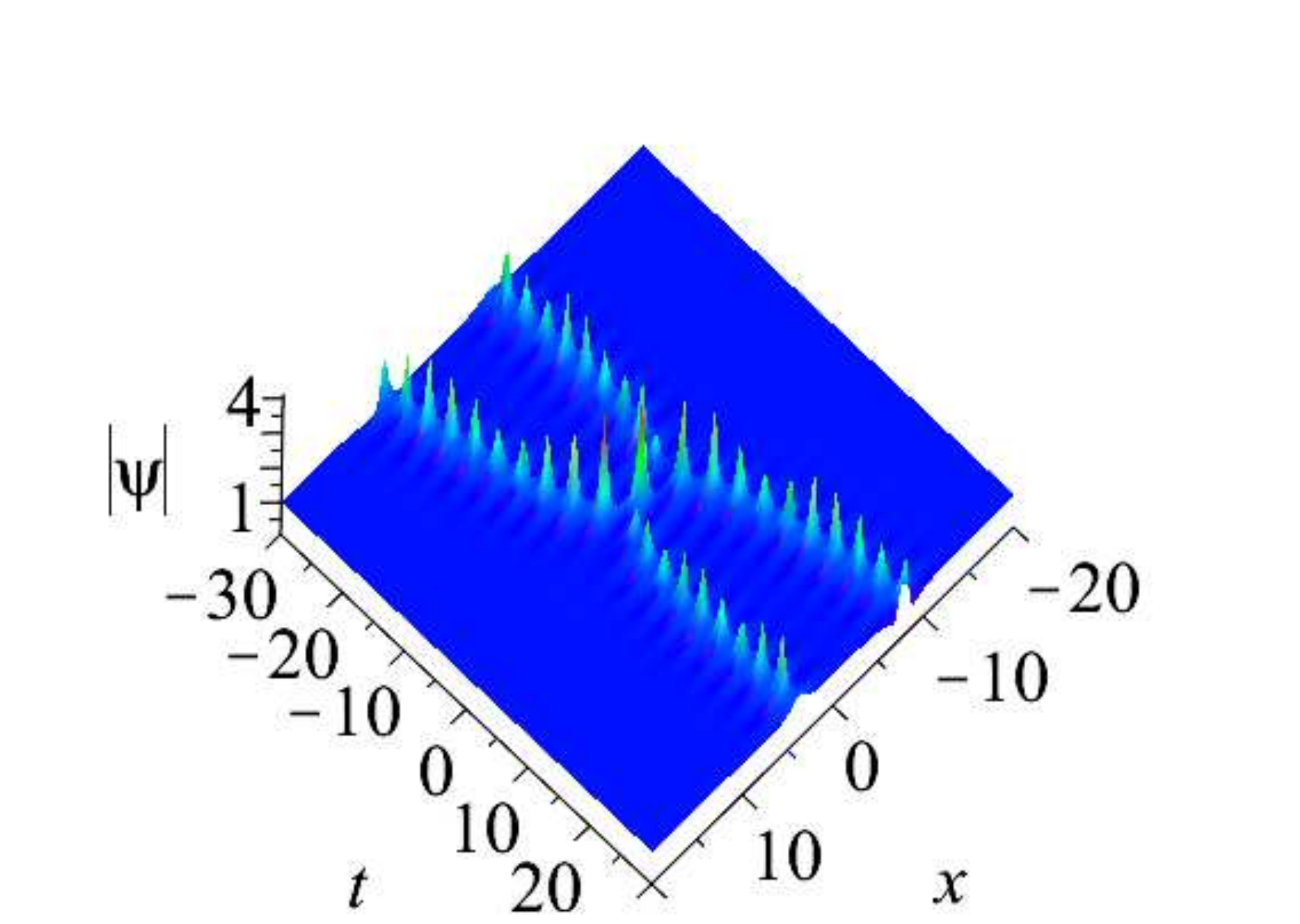}}}
{\rotatebox{0}{\includegraphics[width=4.0cm,height=3.2cm,angle=0]{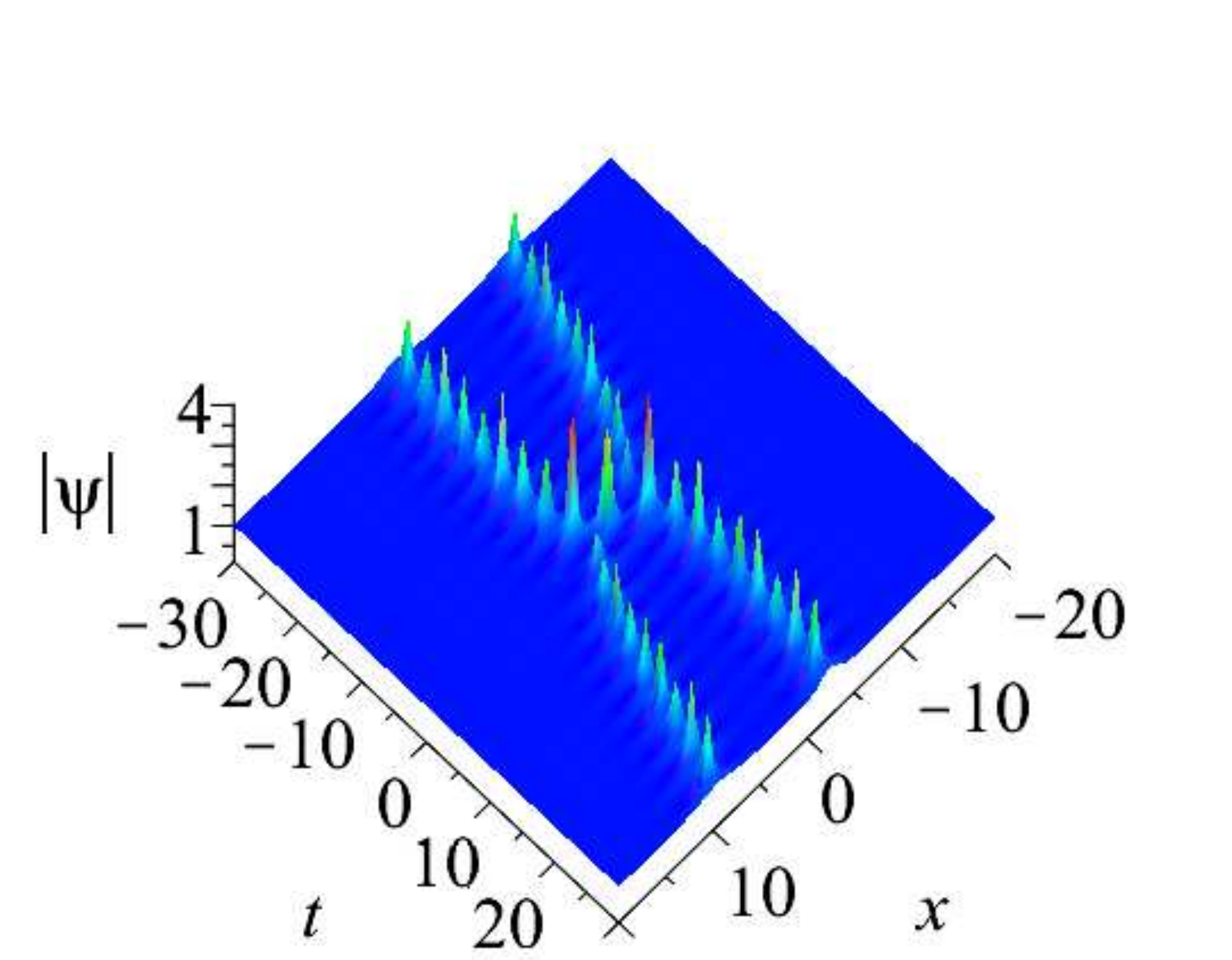}}}

$\qquad\qquad\textbf{(a)}\qquad\qquad\qquad\qquad\qquad\textbf{(b)}
\qquad\qquad\qquad\qquad\qquad\textbf{(c)}$\\

\noindent
{\rotatebox{0}{\includegraphics[width=4.0cm,height=3.2cm,angle=0]{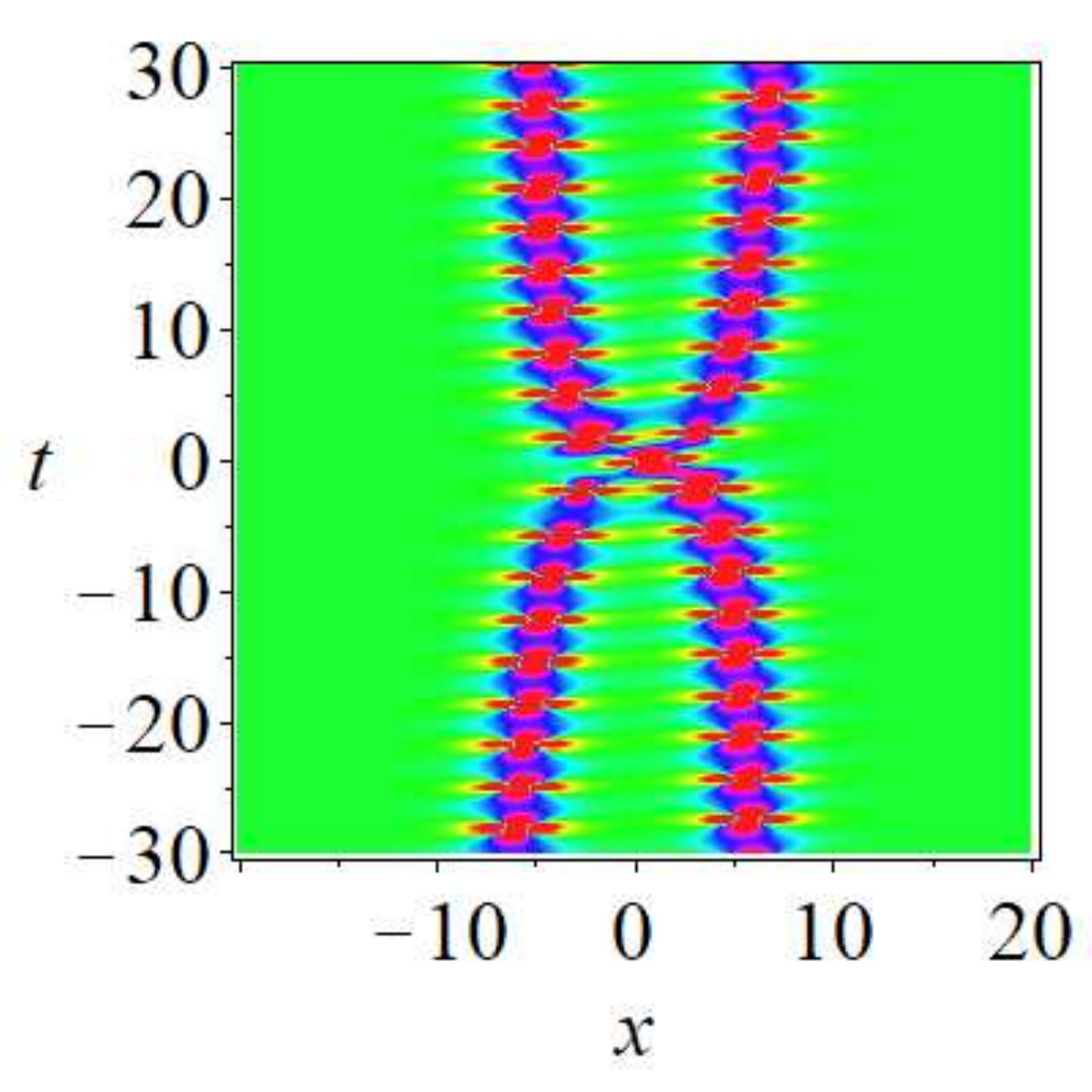}}}
{\rotatebox{0}{\includegraphics[width=4.0cm,height=3.2cm,angle=0]{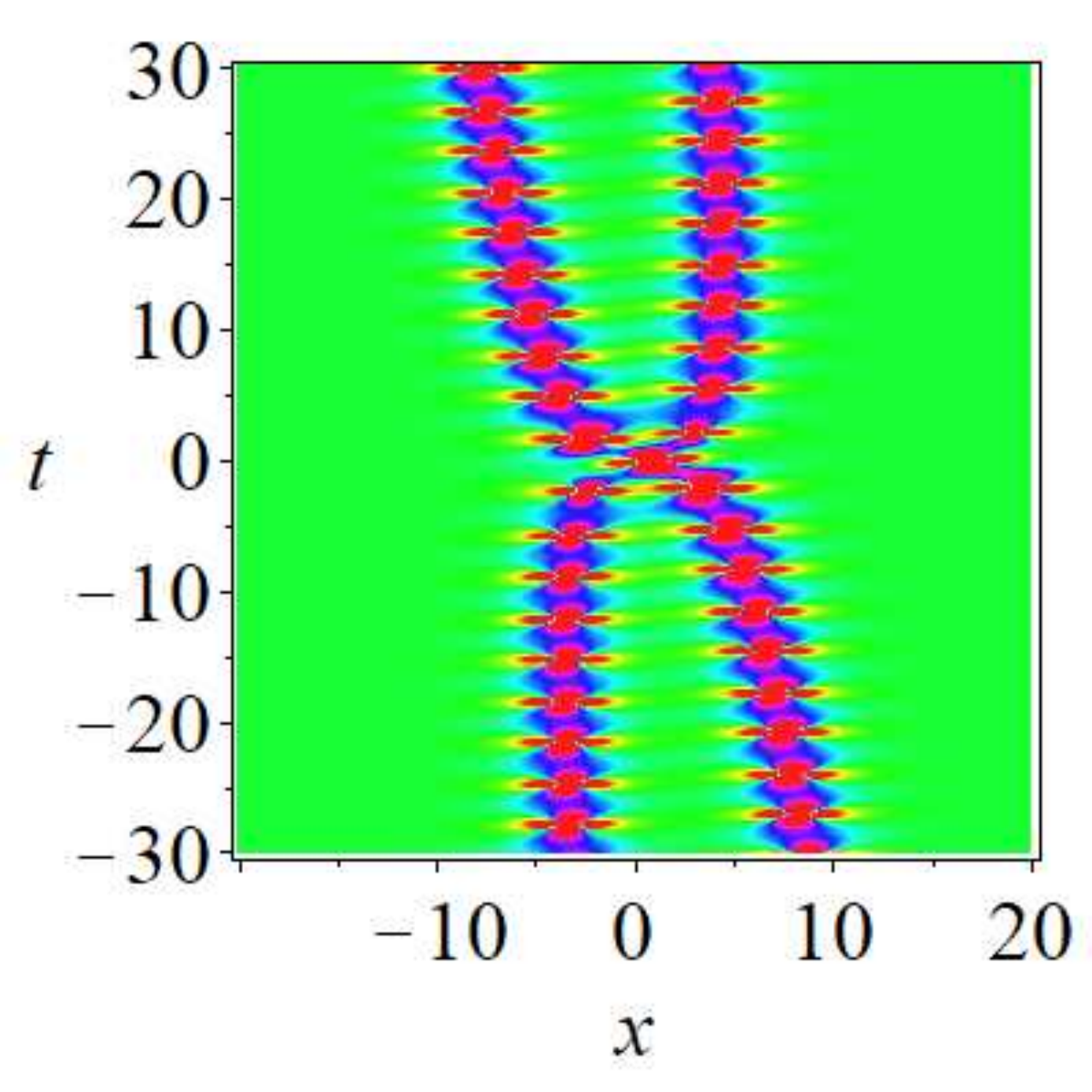}}}
{\rotatebox{0}{\includegraphics[width=4.0cm,height=3.2cm,angle=0]{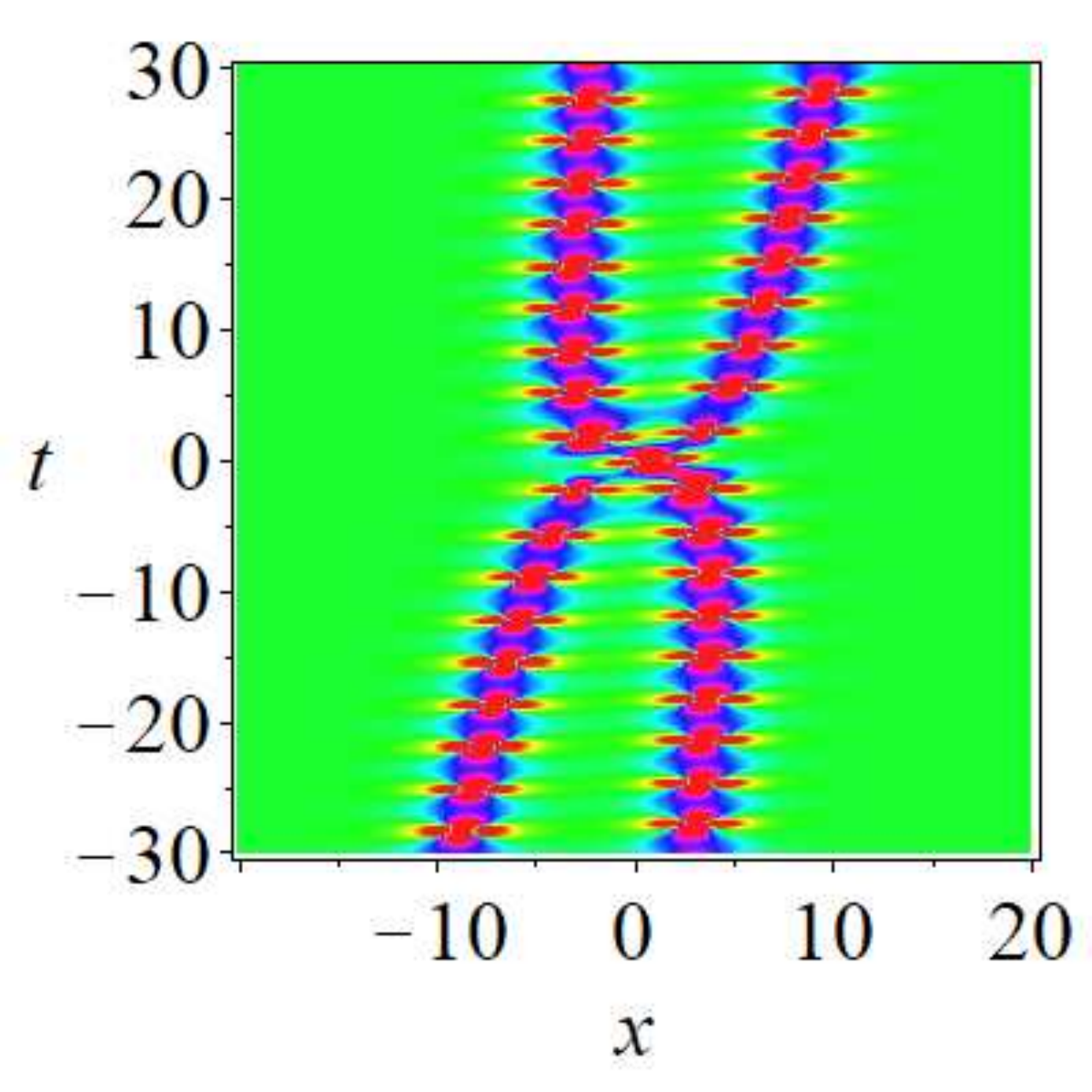}}}

$\qquad\qquad\textbf{(d)}\qquad\qquad\qquad\qquad\qquad\textbf{(e)}
\qquad\qquad\qquad\qquad\qquad\textbf{(f)}$\\
\noindent { \small \textbf{Figure 8.} (Color online) Breather waves via solution \eqref{DP-20} with parameters
$N=1, \psi_{-}=1,\alpha_{2}=1, \alpha_{3}=\alpha_{4}=\alpha_{5}=0.01, A_{+}[z_{1}]=B_{+}[z_{1}]=1$:
$\textbf{(a,d)}$: $z_{1}=-0.08+1.5i$;
$\textbf{(b,e)}$: $z_{1}=-0.08+1.5i$;
$\textbf{(c,f)}$: $z_{1}=-0.06+1.5i$.\\}

$~~~$
{\rotatebox{0}{\includegraphics[width=5.2cm,height=3.6cm,angle=0]{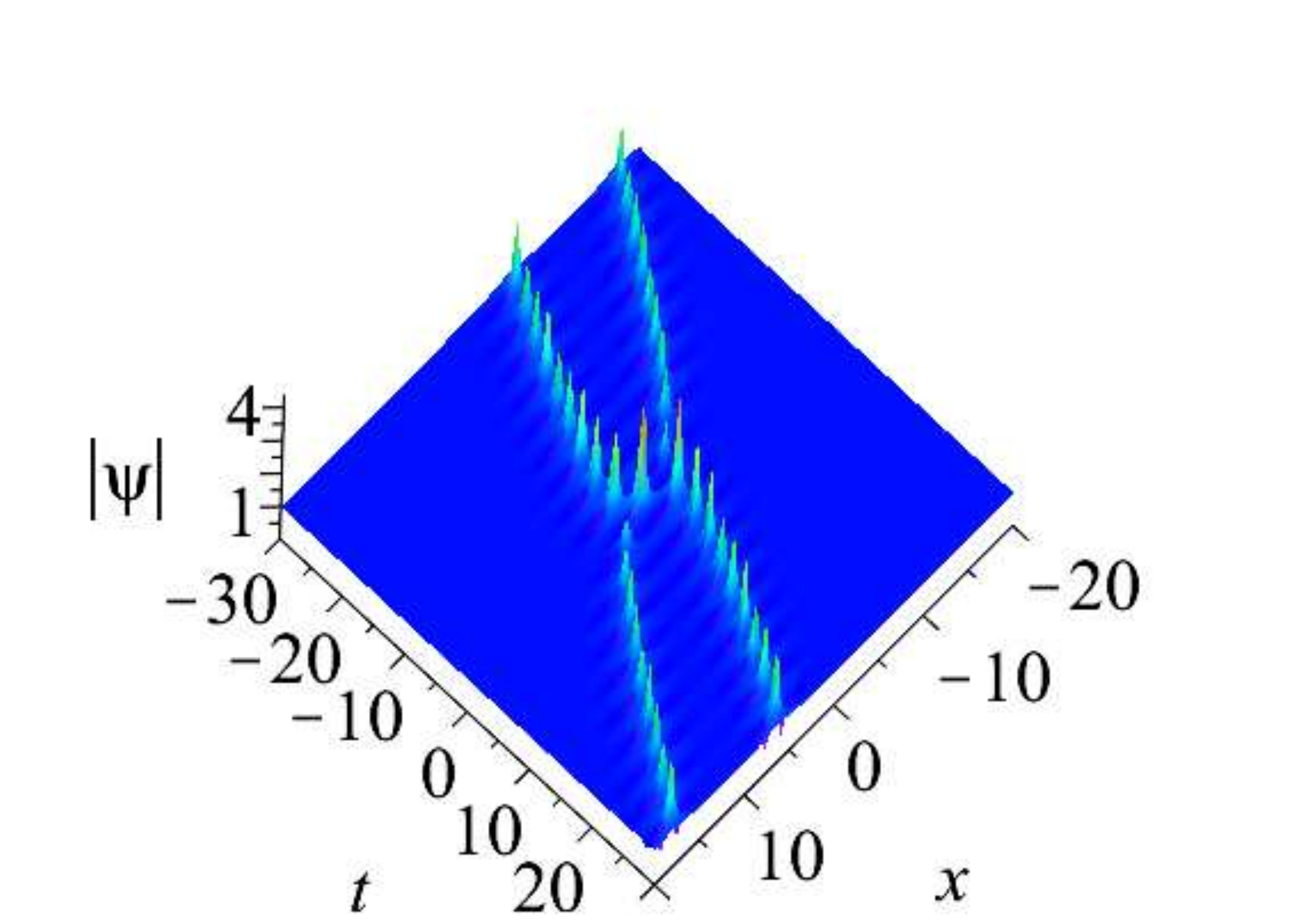}}}
~~~~~~~~~~~~~~~
{\rotatebox{0}{\includegraphics[width=5.2cm,height=3.6cm,angle=0]{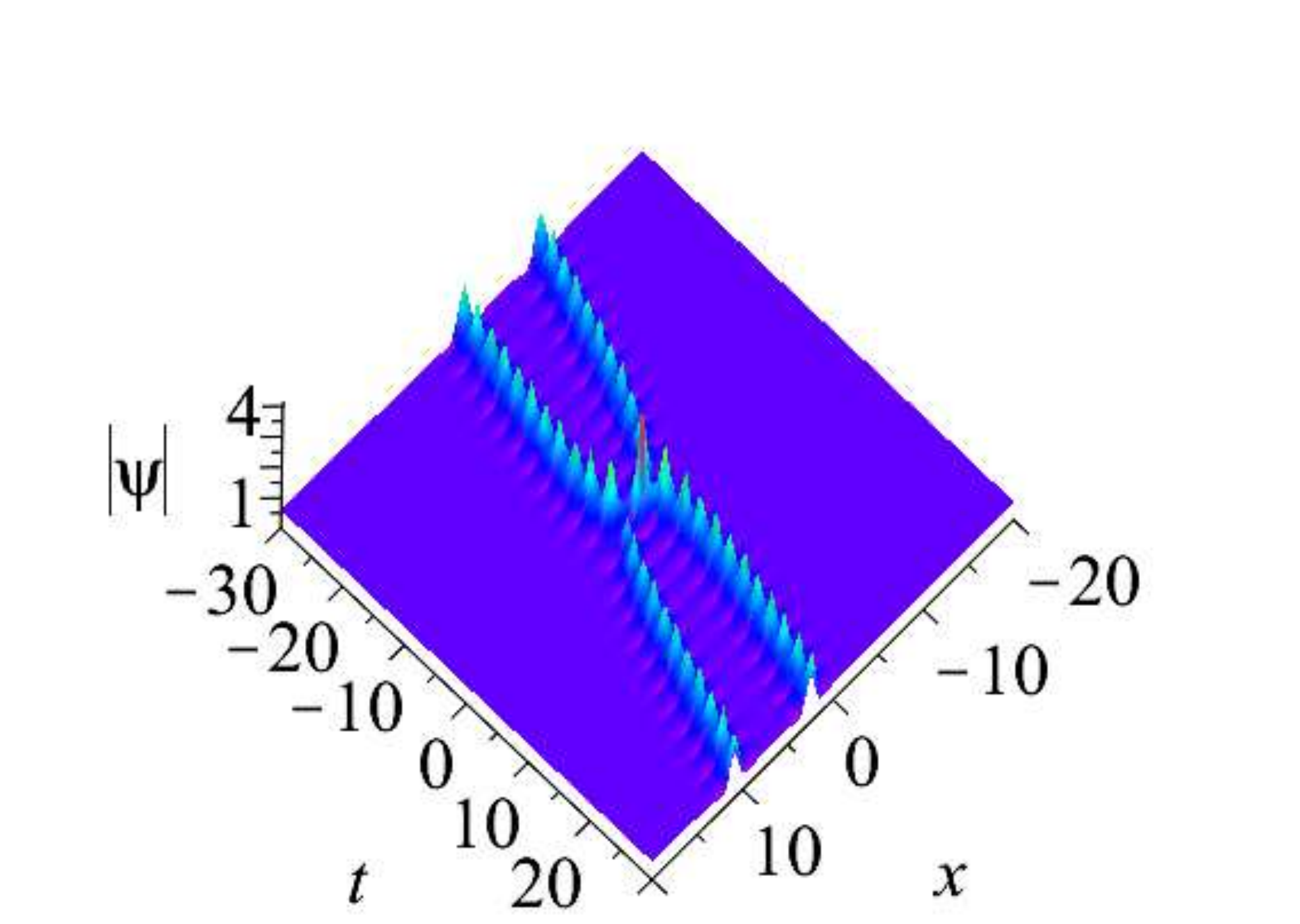}}}

$\qquad\qquad\qquad\qquad\textbf{(a)}
\qquad\qquad\qquad\qquad\qquad\qquad\qquad\textbf{(b)}
$\\

$~~~$
{\rotatebox{0}{\includegraphics[width=5.2cm,height=3.6cm,angle=0]{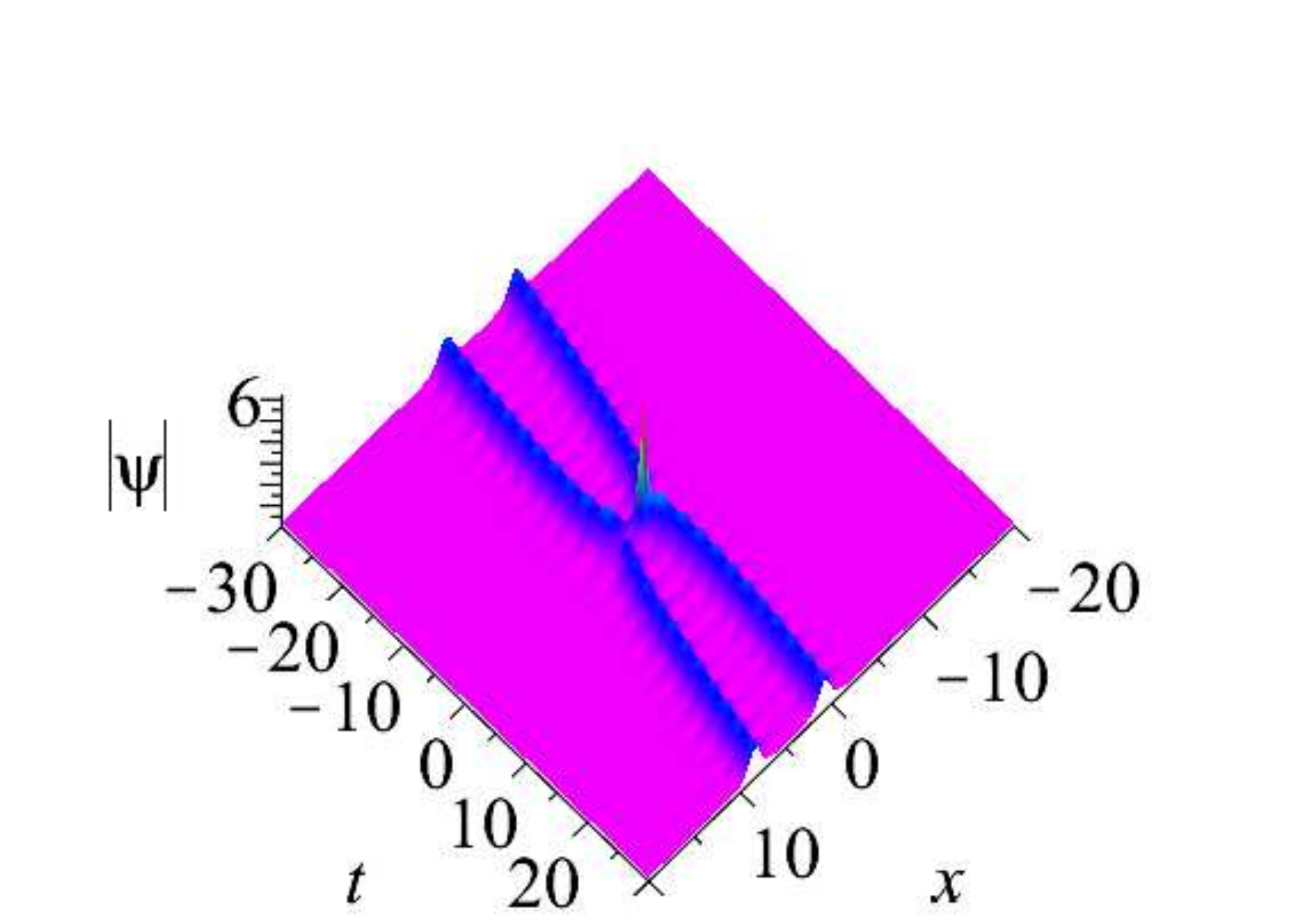}}}
~~~~~~~~~~~~~~~
{\rotatebox{0}{\includegraphics[width=5.2cm,height=3.6cm,angle=0]{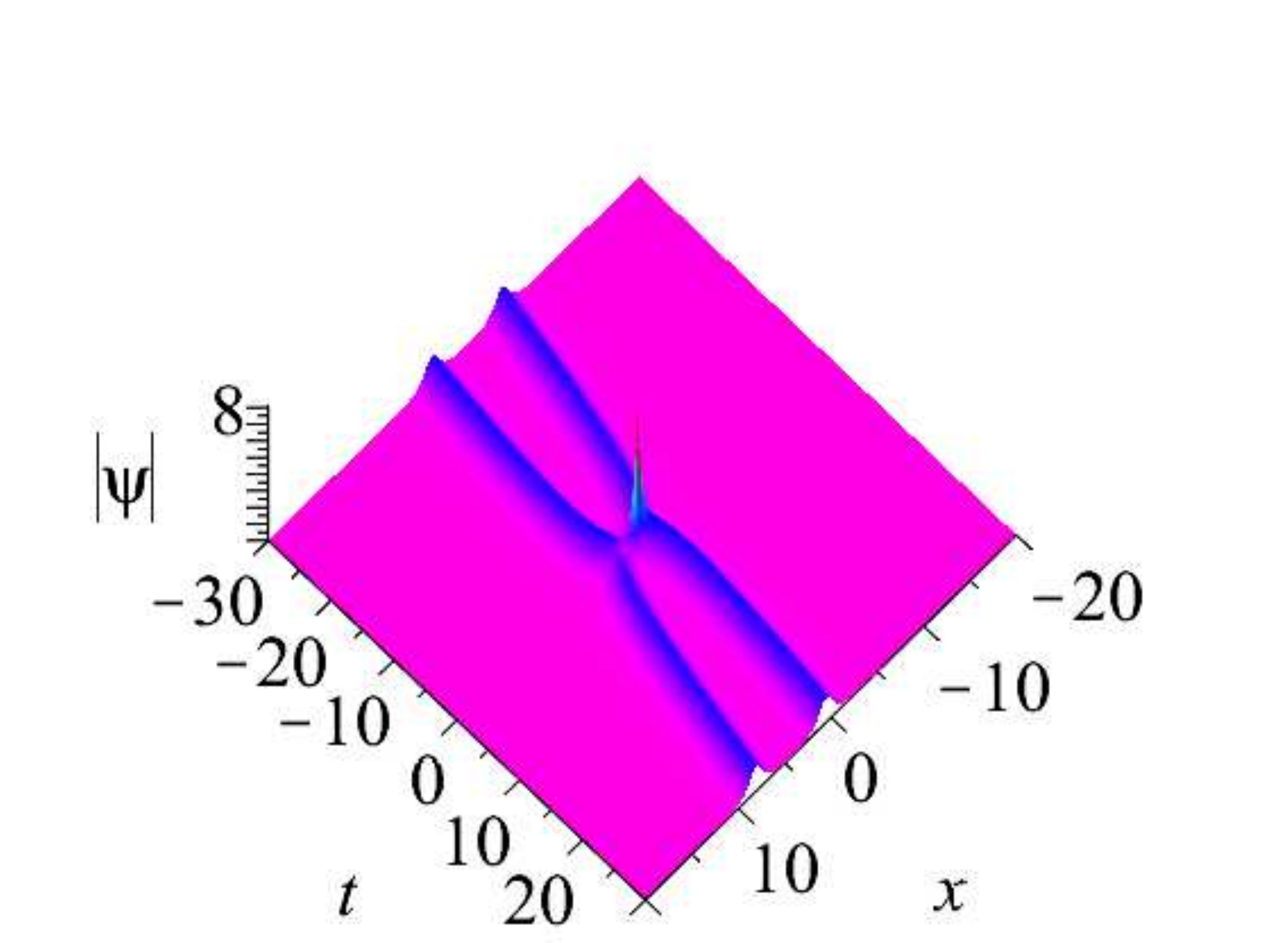}}}

$\qquad\qquad\qquad\qquad\textbf{(c)}
\qquad\qquad\qquad\qquad\qquad\qquad\qquad\textbf{(d)}
$\\
\noindent { \small \textbf{Figure 9.} (Color online) Breather waves via solution \eqref{DP-20} with parameters
$N=1, z_{1}=1.5i, \alpha_{2}=1, \alpha_{3}=\alpha_{4}=\alpha_{5}=0.01, A_{+}[z_{1}]=B_{+}[z_{1}]=1$:
$\textbf{(a)}$: breather-breather solutions with $\psi_{-}=1$; $\textbf{(b)}$: breather-breather solutions with $\psi_{-}=0.5$;
$\textbf{(c)}$: breather-breather solutions with $\psi_{-}=0.3$;
$\textbf{(d)}$: bright-bright solitons with $\psi_{-}\rightarrow0$.\\}

$~~~$
{\rotatebox{0}{\includegraphics[width=5.2cm,height=3.6cm,angle=0]{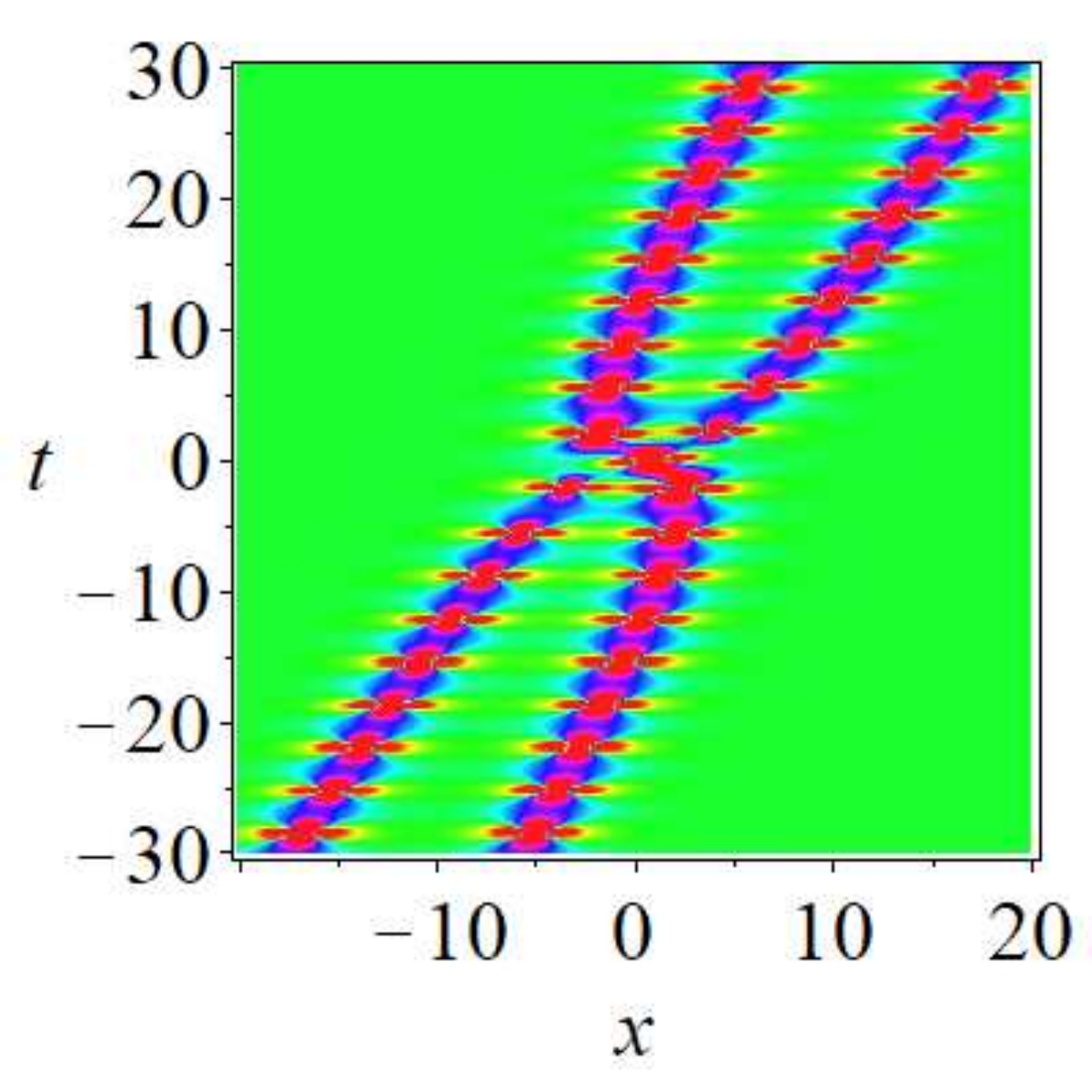}}}
~~~~~~~~~~~~~~~
{\rotatebox{0}{\includegraphics[width=5.2cm,height=3.6cm,angle=0]{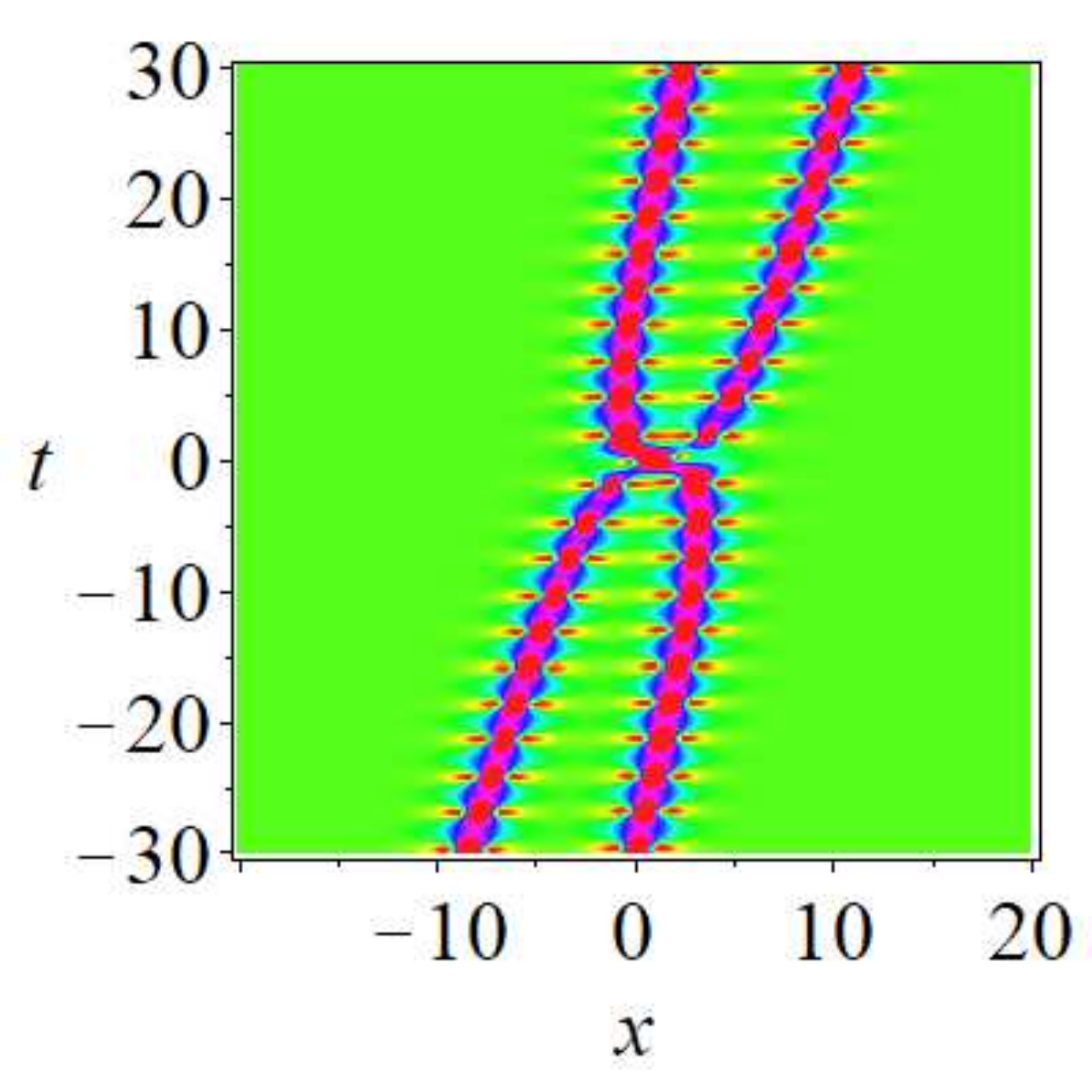}}}

$\qquad\qquad\qquad\qquad\textbf{(a)}
\qquad\qquad\qquad\qquad\qquad\qquad\qquad\textbf{(b)}
$\\

$~~~$
{\rotatebox{0}{\includegraphics[width=5.2cm,height=3.6cm,angle=0]{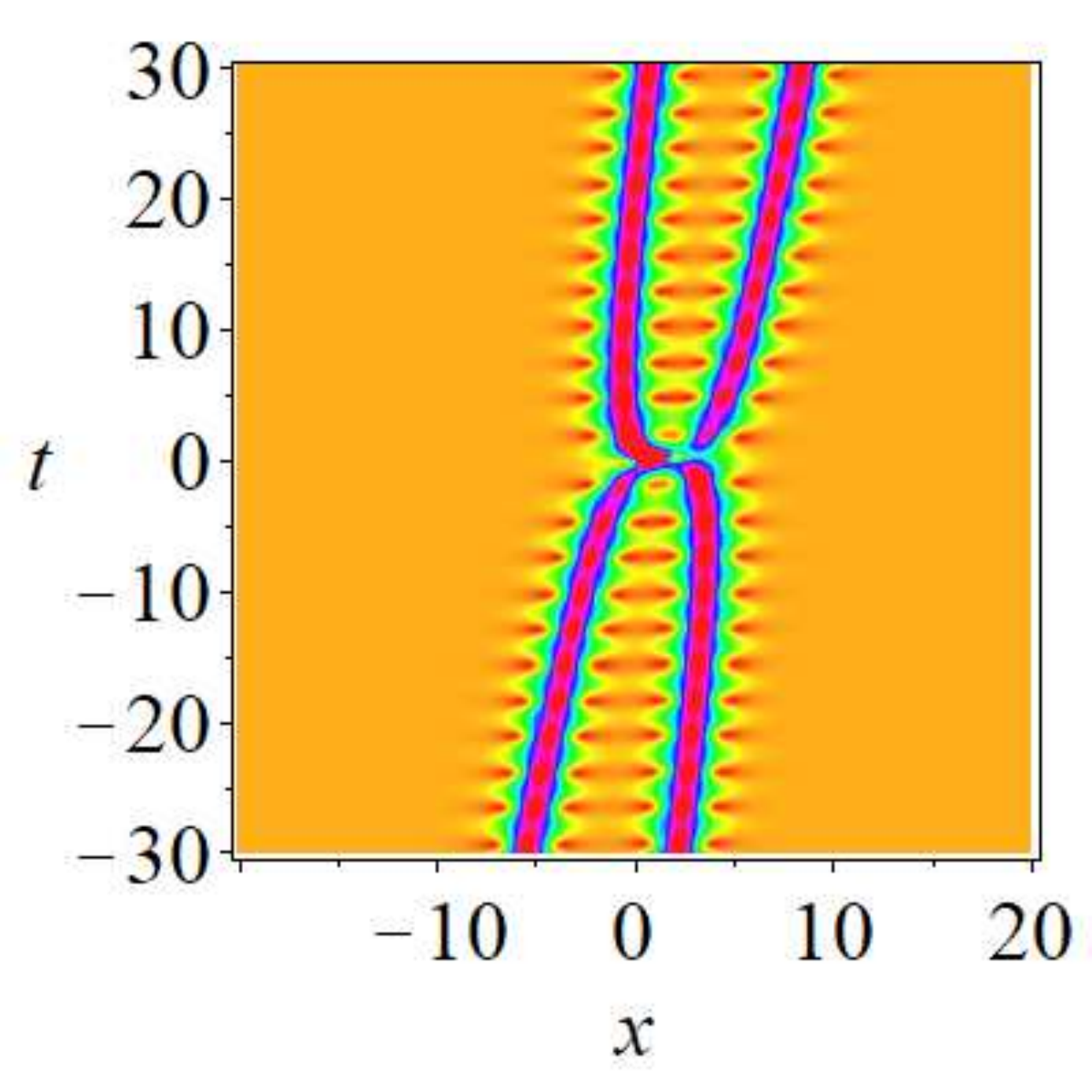}}}
~~~~~~~~~~~~~~~
{\rotatebox{0}{\includegraphics[width=5.2cm,height=3.6cm,angle=0]{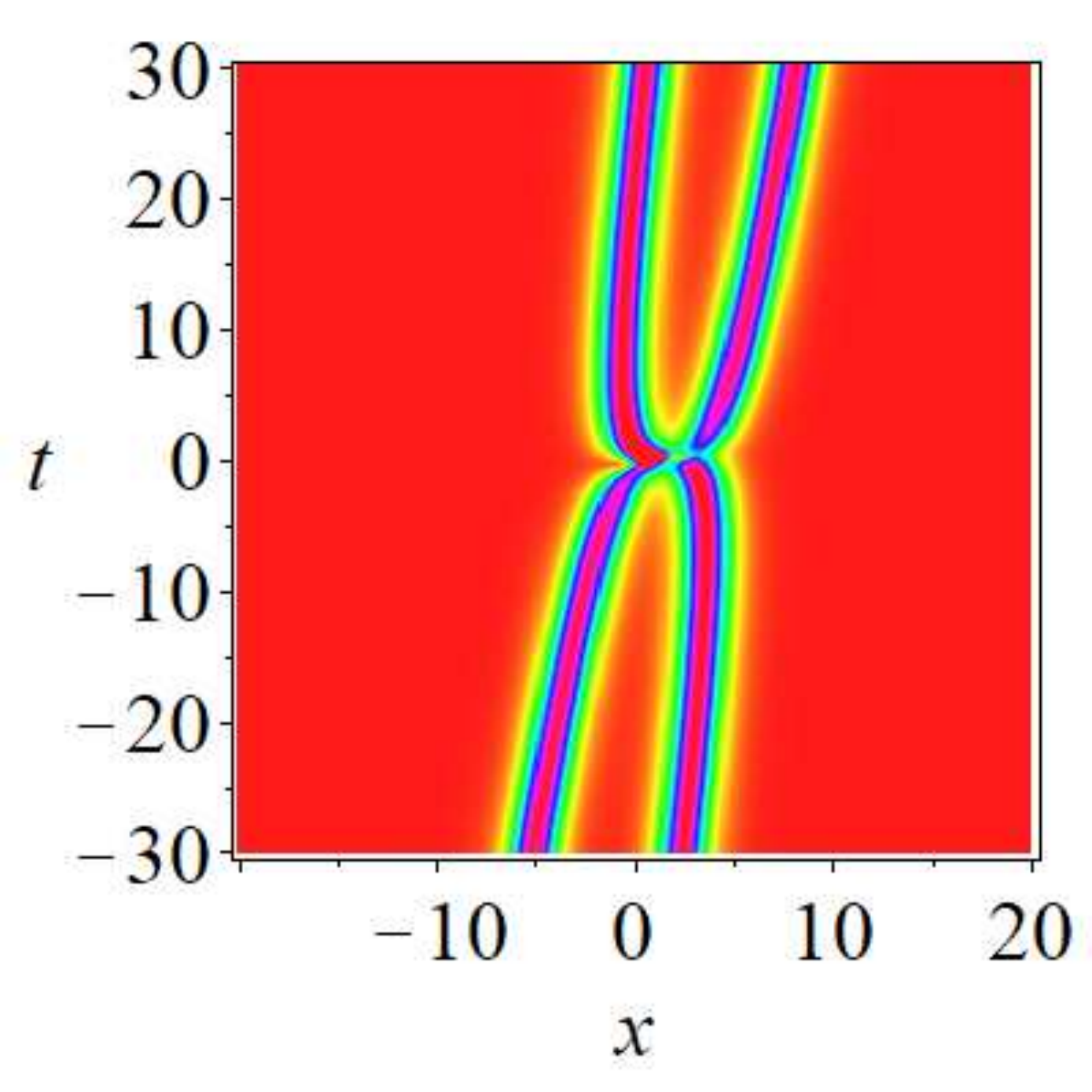}}}

$\qquad\qquad\qquad\qquad\textbf{(c)}
\qquad\qquad\qquad\qquad\qquad\qquad\qquad\textbf{(d)}
$\\
\noindent { \small \textbf{Figure 10.} (On line version in colour) Density plot of figure 9.\\}

\section{Conclusion and Discussion}

In this work, the GFONLS equation \eqref{gtc-NLS} with NZBCs has been systematically investigated,
which can be reduced to some classical integrable equations such as the NLS equation with NZBCs \eqref{Case1}, the mKdV equation \eqref{mKdV} and Hirota equation with NZBCs \eqref{Hirota} etc.
We have discussed the IST and soliton solutions of the GFONLS equation \eqref{gtc-NLS} with NZBCs.
Then its simple-pole and double-pole solutions are found via solving the matrix RHP with the reflectionless potentials.
Some representative solitons are well constructed.
Moreover, in order to help the readers understand the solutions better,
Figs.2-9 of the breather waves, bright-soliton,  breather-breather waves and bright-bright solitons, respectively, are plotted
by seeking appropriate parameters.
More importantly, the GFONLS equation \eqref{gtc-NLS} we investigate in this article are fairly more general as it involves four
real constant $\alpha_{2}$, $\alpha_{3}$, $\alpha_{4}$, $\alpha_{5}$.
The celebrated NLS equation \eqref{Case1} with NZBCs, an important model in fiber
optics, is its special case. Another important reduction of \eqref{gtc-NLS} is the complex mKdV equation \eqref{mKdV}.
Consequently, multisoliton
solutions of the NLS equation \eqref{Case1} \eqref{NLS11} with NZBCs, the mKdV equation \eqref{mKdV} and Hirota equation \eqref{Hirota} with NZBCs can be respectively derived by reducing the
multisoliton solutions of \eqref{gtc-NLS}.
We think that the effective method can help readers to know the diversity and
integrability of nonlinear wave equations well, which should be suitable to study other models in mathematical physics and
engineering.


\subsection*{Acknowledgment}
This work is supported by the National Natural Science Foundation of China under Grant No.11871180.

\end{document}